\documentclass{article}
\usepackage{emulateapj}
\usepackage{float,epsfig,psfig}

\newcommand{\LA}{\mbox{\raisebox{-0.6ex}{$\stackrel{\textstyle<}{\sim}$}}}
\newcommand{\GA}{\mbox{\raisebox{-0.6ex}{$\stackrel{\textstyle>}{\sim}$}}}
\newcommand{\cxo}{{\sl Chandra}}
\newcommand{\cha}{{\sl Chandra}}
\newcommand{\ein}{{\sl Einstein}}
\newcommand{\xmm}{{\sl XMM-Newton}}
\newcommand{\hst}{{\sl Hubble}}
\newcommand{\msun}{M$_{\odot}$}
\newcommand{\ergl}{ergs~s$^{-1}$}

\newcommand{\ergcms}{ergs~cm$^{-2}$~s$^{-1}$}

\newcommand{\ros}{{\sl ROSAT}}
\newcommand{\asca}{{\sl ASCA}}
\newcommand{\etal}{et al.}
\slugcomment{Submitted to Astrophysical Journal}

\begin{document}

\title{
{\sl Chandra} X-ray Observations of the Spiral Galaxy M81}

\author{
Douglas~A.~Swartz\altaffilmark{1},
Kajal~K.~Ghosh\altaffilmark{1},
Michael~L.~McCollough\altaffilmark{1},
Thomas~G.~Pannuti\altaffilmark{2},
Allyn~F.~Tennant\altaffilmark{3},
Kinwah~Wu\altaffilmark{4}
} 
\altaffiltext{1}{Universities Space Research Association, 
NASA Marshall Space Flight Center, SD50, Huntsville, AL, USA}
\altaffiltext{2}{MIT Center for Space Research, 77 Massachusetts Ave.,
NE80-6015, Cambridge, MA, USA}
\altaffiltext{3}{Space Science Department, 
NASA Marshall Space Flight Center, SD50, Huntsville, AL, USA} 
\altaffiltext{4}{MSSL, University College London, Holmbury St. Mary, Surrey,
RH5 6NT, UK}

\begin{abstract}

A {\sl Chandra X-Ray Observatory} ACIS-S imaging observation is used to 
study the population of X-ray sources in the nearby Sab galaxy M81 (NGC 3031). 
A total of 177 sources are detected with 124 located within the $D_{25}$
isophote to a limiting X-ray luminosity of $\sim$$3\times 10^{36}$ \ergl.
Source positions, count rates, luminosities in the 0.3~--~8.0~keV band,
limiting optical magnitudes, and potential counterpart identifications are
tabulated.
Spectral and timing analysis of the 36 brightest sources are reported
including
the low-luminosity active galactic nucleus,
SN~1993J, and the \ein-discovered
ultra-luminous X-ray source X6.
The nucleus accounts for $\sim$86\%, or $5\times$$10^{40}$~\ergl, of the total
X-ray emission from M81. Its spectrum is well-fit by an absorbed power law with
photon index 1.98$\pm$0.08 consistent with previous observations (average
index 1.9). SN~1993J has softened and faded since its discovery.
At an age of 2594 days, SN~1993J displayed a complex thermal spectrum from a
reverse shock rich in Fe~L and highly-ionized Mg, Si, and S but lacking O. A
hard X-ray component, emitted by a forward shock, is also present. X6 is
spatially-coincident with a stellar object with optical brightness and colors
consistent with an O9~--~B1 main sequence star. It is also coincident with a
weak radio source with a flux density of $\sim$95~$\mu$Jy at $\lambda=3.6$~cm.
The continuum-dominated X-ray spectrum of X6 is most closely reproduced by a
blackbody disk model suggesting the X-ray source is an $\sim$18~\msun\ object
accreting at nearly its Eddington limit.

The non-nuclear point source population of M81 accounts for 88\%
of the non-nuclear X-ray luminosity of $8.1\times10^{39}$~\ergl.
The remaining (unresolved) X-ray emission is confined
within $\sim$2~kpc of the galactic center.
The spatial distribution of this emission and of the resolved X-ray bulge
sources closely follows that of the bulge optical light.
In particular, there is no evidence for an X-ray signature 
accompanying the filamentary H$\alpha$ or excess UV emission 
seen in the central \LA1.0~kpc of the galaxy.
The shape of the luminosity function of the bulge sources is a power law
with a break at $\sim4\times10^{37}$~\ergl; suggesting the presence of an
aging ($\sim$400~Myr) population of low-mass X-ray binaries. Extrapolating this
luminosity function to lower luminosities  accounts for only
$\sim$10\% of the unresolved X-ray emission.
Spectroscopically, the unresolved emission can be represented as a combination
of soft, $kT$$\sim$0.3~keV, optically-thin plasma emission and of a $\Gamma=1.6$
power law. The unresolved bulge X-ray emission is therefore most likely a
combination of hot gas and of one or more large and distinct populations of
low-luminosity X-ray sources confined in the gravitational potential and tracing
the old population of bulge stars.
The distribution of disk sources shows a remarkably strong
correlation with spiral arms with
the brightest disk sources located closest to spiral arms.
The luminosity function of sources near the spiral arms
is a pure power law (slope $-0.48\pm0.03$) while that of sources further away
exhibits a break or cut-off in the power law distribution with no
high-luminosity members. This is interpreted as a natural consequence of the
passage of spiral density waves that leave the brightest (when averaged over
their lifetimes) and shortest-lived X-ray sources immediately downstream of the
spiral arms. Consistent with model predictions, we conclude that the shapes of 
the X-ray luminosity functions of the different galactic components of M81 are
most likely governed by the birth rates and lifespans of their constituent X-ray
source populations and that the luminosity functions can be used as a
measure of the star formation histories of their environments.

\end{abstract}

\keywords{galaxies: individual (M81) --- X-rays: galaxies --- X-rays: binaries
 --- X-rays: stars --- supernovae: individual (SN 1993J)}

\section{Introduction}

Systematic investigations of the X-ray properties of normal galaxies began in
earnest with the \ein\ observatory over two decades ago.
The picture that emerged for spiral galaxies
(see the early reviews by Long \& van~Speybroeck 1983; Helfand 1984; and
Fabbiano 1989) 
is that the bulk of the X-ray emission 
takes place in two distinct physical
environments:
the star-forming
disks of late-type spiral and irregular galaxies and among
the old stellar population in dense globular clusters 
and compact bulges at the centers of early-type spiral galaxies.
In addition to these trends along the Hubble sequence, variations
were sometimes found among spiral
galaxies of similar morphological type suggesting a dependence on
star formation histories.
In particular, the brightest X-ray emissions are
associated with starbursts in merging and interacting galaxies (David, Jones, \&
Forman 1992). Thus, by
tracing the endpoints of stellar evolution, the X-ray source populations of
external galaxies provide important clues to the physical nature and 
evolutionary history of their hosts.

The contemporary view for spiral galaxies
is rapidly being refined following
the launch of the \cha\ and \xmm\ X-ray Observatories.
Moderately-deep images reveal point sources to limiting
X-ray luminosities of order $10^{37}$~\ergl\ 
in galaxies out to Virgo cluster distances. 
While this samples only the high luminosity end of the distribution of X-ray
sources, of order 100 sources are routinely detected in normal
galaxies similar to our own. Reliable spectral analysis
is usually limited to an even smaller subset of the brightest individual
sources. Nevertheless, using probabilistic methods, the observed sample of X-ray
sources can help us understand current-epoch galaxy evolution in its broader
context.

A formal expression of the relationship between the star formation history of
local galaxies and their observed X-ray source
populations has recently been put forth by
Wu (2001; see also Wu \etal\ 2002a,b; Kilgard \etal\ 2002;
Dalton \& Sarazin 1995). There it was shown that the basic shape of the
observed X-ray luminosity function 
is governed simply by the birth and death rates of the source population 
under the assumption that the more luminous X-ray sources are
shorter-lived. Thus, in the absence of ongoing star formation, the
luminosity function will develop a cutoff at high luminosity that 
evolves toward lower luminosity. Conversely,
if the population of X-ray sources is replenished through star formation
processes, such as is found in spiral arms, 
then a power law shaped luminosity function can be sustained.

Certain complications arise when applying this basic interpretation
to X-ray populations in individual galaxies (Wu \etal\ 2002a,b).
Among these are the presence of different classes of X-ray sources, such as
supernova remnants and accreting compact objects, which evolve on differing
timescales; alternative source-formation mechanisms uncorrelated with 
stellar evolution such as binary captures in globular clusters; and non-steady
or luminosity-limited emission characteristics 
such as those associated with X-ray transients and novae and in
Eddington-limited neutron star binaries, respectively.
Therefore, only when specific counterparts to individual X-ray sources
are identified and their multiwavelength properties assessed can
the full power of the hypothesis of Wu \etal\ (2002a,b)
be applied to address the nature and evolution
of X-ray sources in different environments.

The nearby Sab galaxy M81 (NGC 3031) is ideal for such a study
in that it contains both a
strong two-arm grand-design spiral pattern and a well-defined circumnuclear
bulge. The distance to M81, 3.6~Mpc, has been well-established from Cepheid
measurements (Freeman \etal\ 1994) which are in good agreement with
other distance estimates (Ferrarese \etal\ 2000). Populations of several classes
of objects in M81 have been investigated and catalogued including globular
clusters (Perelmuter \& Racine 1995; Chandar, Ford, \& Tsvetanov 2001),
supernova remnants (Matonick \& Fesen 1997), H~II regions (Hodge \& Kennicutt
 1983; Petit, Sivan \& Karachentsev 1988), and stars and star clusters (Zickgraf
\& Humphreys 1991; Ivanova 1992; Sholukhova \etal\ 1998).
In addition,
the plane of the galaxy is oriented 32$^{\circ}$ from face-on allowing
detailed mapping of the velocity field
(Goad 1976; Rots \& Shane 1975; Adler \& Westphal 1996)
for dynamical studies and testing
spiral density wave models (Visser 1980; Roberts \& Hausman 1984).

The center of M81 contains a
compact radio core (Bartel \etal\ 1982) surrounded by a region of
enhanced far-infrared (Rice 1993; Davidge \& Courteau 1999),
H$\alpha$ (Devereux, Jacoby \& Ciardullo 1995),
and ultraviolet (Hill \etal\ 1992; Reichen \etal\ 1994)
emission extending to $\sim50\arcsec$  ($\sim$900 parsecs). This emission
probably comes from an old population of hot, low-mass stars rather than
from young massive stars (O'Connell \etal\ 1992; Devereux, Ford \& Jacoby 1997).

In contrast to the bulge,
HI velocity contours show a sharp discontinuity beyond the bulge identified as 
a spiral velocity shock (Visser 1980). Downstream of this 
shock are regions of star formation in the spiral arms. 
The distributions of these components are consistent (Kaufman \etal\ 1989) with
density wave models predicting a broad spiral density enhancement (e.g., 
Roberts \& Hausman 1984).

Beyond the visible disk of M81 is an envelope of neutral hydrogen
(Roberts 1972) enclosing
M81 and nearby group members. A bridge of gas, a relic of tidal interaction 
(Cottrell 1977), connects M81 and the starburst galaxy M82.

The hypothesis of Wu (2001) and Wu \etal\ (2002a,b) was motivated in large part 
by the initial results from our \cha\ observation of M81 presented in Tennant
\etal\ (2001). There it was shown that the X-ray luminosity function of the
population of bulge sources displays a break at $\sim$4$\times$$10^{37}$ \ergl\
similar to that observed in M31 (e.g., Shirey \etal\ 2001).
The X-ray luminosity function of the disk
sources, on the other hand, follows a single power law slope over three 
decades in flux. This is what is expected if an impulsive episode of
star formation occurred in the bulge in the past, ostensibly
during an encounter between M81 and one of its companion galaxies, while 
continuous star formation in the disk is being driven by the passage of spiral 
density waves.

Here we build upon the earlier work of Tennant \etal\ (2001). After
presenting detailed information on the individual X-ray sources in 
\S\ref{s:discrete_src}
and in-depth analysis of the brightest objects in \S\ref{s:bright_src},
the properties of the bulge (\S\ref{s:bulge}) and disk (\S\ref{s:disk}) regions 
are addressed separately then discussed (\S\ref{s:discussion}) 
within the common framework of galaxy evolution. 

\section{Observations} \label{s:observations}

The primary X-ray dataset is a 49926 second observation of M81
obtained on 2000 May 7 with the \cha\ Advanced CCD Imaging Spectrometer (ACIS) 
spectroscopy array operating in imaging mode. Unless otherwise noted, references
to X-ray data will refer to this dataset. The X-ray data was reprocessed by the
\cha\ X-ray Center (CXC) on 2001 January 4. This reprocessed data is used in
this work. There are no significant differences between the reprocessed data and
the originally-distributed data analyzed by Tennant \etal\ (2001). The
observation was taken in faint timed exposure mode at 3.241 s-frame$^{-1}$ at a
focal plane temperature of $-120^{\circ}$C. Standard CXC
processing has applied aspect corrections and compensated for spacecraft dither.

The primary target, SN 1993J, was located near the nominal aimpoint on the
back-illuminated (BI) device S3.
The nucleus of M81 lies 2.$\arcmin$79 from SN 1993J towards the center of
S3 in this observation. Accurate positions of these two objects and two G0 stars
located on device S2 were used to 
identify any offset and to determine absolute 
locations of the remaining \cha\ sources as well as objects in
other X-ray images
and those obtained at other wavelengths. Table~1 shows that the positions are
accurate to within 1.4$\arcsec$. No offset correction was applied to the
\cxo\ X-ray positions.

\begin{center}
\small{
\begin{tabular}{lllll}
\multicolumn{5}{c}{{\sc TABLE 1}} \\
\multicolumn{5}{c}{{\sc M81 Astrometry}} \\ 
\hline \hline
Object &\multicolumn{2}{c}{\cha\ Position}&\multicolumn{2}{c}{Catalogued
Position}\\
       & \multicolumn{1}{c}{RA} & \multicolumn{1}{c}{DEC} &
         \multicolumn{1}{c}{RA} & \multicolumn{1}{c}{DEC} \\ \hline
SN 1993J   & 9 55 24.77 & 69 1 13.4 & 9 55 24.77 & 69 1 13.7$^1$ \\
Nucleus    & 9 55 33.19 & 69 3 55.1 & 9 55 33.17 & 69 3 55.1$^2$ \\
PPM 17242  & 9 55 1.00 & 68 56 22.1 & 9 55 1.00 & 68 56 22.2$^3$ \\
PPM 17243  & 9 55 2.57 & 68 56 21.2 & 9 55 2.76 & 68 56 22.1$^3$ \\
\hline
\multicolumn{5}{l}{{\sc References.}--(1) Marcaide \etal\ 1993; (2) Ma \etal\
 1998}\\
\multicolumn{5}{l}{(3) Positions and Proper Motions catalog} \\
\end{tabular}
} 
\end{center}

A charge transfer inefficiency (CTI) corrector algorithm 
(Townsley \etal\ 2000)
was then applied to the Level~1 event list to partially correct for 
the charge loss and charge smearing effects of CTI in the ACIS detectors.
Matching response matrices were also provided by L. Townsley. 
After correction, a single response matrix is adequate for sources on the 
S3 device since the spectral resolution does not exhibit a 
strong spatial dependence. This is particularly advantageous for analysis of the
 unresolved emission extending over the $\sim$8$\arcmin$ diameter
bulge of M81 which is located entirely within the S3 device. 

The corrector algorithm was applied to all but the front-illuminated (FI) device
S4 because no correction was available for that device. Instead,
 the Level~2 event list was used and the {\tt destreak}
algorithm\footnote[5]{
http://asc.harvard.edu/ciao2.1/downloads/scripts/destreak.ps}
was applied to remove charge randomly deposited along pixel rows during read
out.

The entire dataset was then cleaned of bad pixels and columns and
the standard grade set and events in pulse 
invariant (PI) channels corresponding to $\sim 0.2$ to $8.0$ keV 
were selected for source detection.
The range $0.3$ to $8.0$ keV is used for spectral analysis.

No periods of high particle background occurred during the 
observation. The 0.3~--~8.0~keV background in the BI device S3 is
$\sim$0.04~cts~pixel$^{-1}$ and that in the FI devices is
$\sim$0.01~cts~pixel$^{-1}$ for the observation. 
Separate background spectra were
extracted from large source-free regions of each device. The background 
spectrum for S3 was chosen far from the nucleus because excess X-ray
emission was detected near the nucleus (\S\ref{s:bulge}). The
background spectra appear similar to deep quiescent blank sky compilations
available from the CXC.

The detector viewing area covers 57\% of the 
optical extent of the galaxy, 
defined as the ellipse of major diameter 26.$\arcmin$9 corresponding to the
$D_{25}$ isophote as tabulated in de~Vaucouleurs \etal\ (1991),
oriented at position angle 149$^{\circ}$,
and with major-to-minor axis ratio 1.94:1 corresponding to the 58$^{\circ}$
inclination angle of M81. This area includes all of the
S3 device, approximately half of each of the
S2 and S4 devices, and the outer corner of I3.
The data from each device is analyzed independently owing to differing energy
resolutions, low-energy responses, and background signals.

In addition to this primary dataset, a 2.4-ks ACIS-S image taken 2000 Mar 21
and numerous \ros\ PSPC and HRI datasets were used to construct 
long-term light curves of the brightest 
sources (see Immler \& Wang 2001 for their analysis of the \ros\ observations).

\section{The Discrete X-ray Source Population} \label{s:discrete_src}

Table~2 lists the 177 discrete X-ray sources detected in the 
primary \cha\ observation. The table lists the
source positions (in order of increasing right ascension), 
the aperture-corrected number of source counts, 
the signal-to-noise ratio for the count rate,
apparent visual magnitudes derived from either \hst\ WFPC2 images or
from the Perelmutter \& Racine (1995) catalogue of bright objects, ACIS
CCD device identification, global environment (where b~denotes bulge, d~denotes
disk, and $D_{25}$ denotes source is outside the $D_{25}$ isophote),
and the unabsorbed luminosity in the 0.3~--~8.0~keV energy range. 
The table also lists corresponding X-ray detections from \ein\
(denoted "X", Fabbiano 1988) and \ros\ ("P" and "H" denoting PSPC- and
HRI-identified sources, respectively, Immler \& Wang 2001), and potential source
type based on spatial correlations with catalogued objects or on other
information. Explanations of the quantitative entries
are given in the following subsections. References to \cxo\ X-ray source numbers
in this work refer to the source numbering adopted for Table~2.

 \subsection{X-ray Source Detection}

A source-finding method was used that assumes a source is located at a
given position and compares the distribution of detected events to a known point
spread function (PSF). The algorithm first calculates the fraction
of the PSF within each pixel within a detection region. Then, using the PSF
fraction as the independent variable, it calculates an unweighted least
squares fit of a straight line to the counts detected in the pixels in the
region.  If a source is present, then the slope of the line will be
positive and will represent the total number of counts from the source
(integrated over the PSF). The line intercept is the background per pixel.  A
key value is the uncertainty in the slope and hence the number of source counts.
The uncertainty is determined by applying the standard propagation of
errors directly to the sums.

The algorithm then calculates the estimated source counts (slope) and error
at every pixel in the image.  The estimated source counts divided
by the uncertainty is the signal to noise ratio (S$/$N).  If the S$/$N
exceeds some threshold then there is a source in the neighborhood.
However, since both source counts and error increase near a source,
the S$/$N is not as sharply peaked as either component individually.
To best separate sources in confused regions, a source is defined
to be a peak in the estimated source count spatial distribution and must also
exceed a minimum S$/$N threshold.  The threshold S$/$N is best defined by
constructing the S$/$N histogram.  This histogram is roughly a Gaussian
core centered near zero (due to noise) with exponential wings (due to sources).
By fitting the core to a Gaussian, the S$/$N value for
which the Gaussian will contribute less than one source in the
field can be estimated.  This threshold is as low as 2.5 for a source on
axis and as high as 3.0 for a source far off-axis.
For this paper a constant value of 2.8 is used as the threshold.

The PSF used in the search can have any shape and the source-finding algorithm
allows either a mathematical model PSF or a high-fidelity
simulated PSF available from the CXC PSF
library\footnote[6]{available from
http://cxc.harvard.edu/caldb/download.html}.
A circular Gaussian approximation to the PSF does a good job of locating
sources, in comparison to simulated PSFs, assuming the width of the Gaussian
increased quadratically off-axis so that the size of the Gaussian roughly
matches the observed off-axis broadening of a point source image. This is not
unexpected since bright sources are easily detected in any method and, for faint
sources, Poisson noise removes the importance of the exact PSF shape. Assuming a
circular Gaussian PSF gives higher weight to sources with a central
concentration of events. This is superior to a cell detect method which only
looks for an excess of counts in an arbitrary source region and does not depend
on the distribution of events within the region.

The source detection process was repeated using the
CXC source detection tool {\tt wavdetect} (Freeman \etal\ 2002).
The {\tt wavedetect} tool is similar to our method in
that it is more likely to detect sources with a central condensation. {\tt
wavdetect} was applied on spatial scales from 1 to 16 pixels in logarithmic
steps using a significance threshold for source detection corresponding to
less than a $10^{-6}$ chance probability of detection due to local background
fluctuations. The results were consistent with our method at the equivalent
significance level.

Point-source counts and spectra were extracted from within 
the 95\% encircled-energy aperture of the model PSF.
Background regions were typically chosen from annular regions 
surrounding the source regions except in crowded regions 
of the field where background regions adjacent to the source were used.
The background-subtracted counts within the source regions were scaled to
obtain the aperture-corrected count values. The background-subtracted point
source detection limit is 12 counts for the 2.8 minimum S$/$N
threshold and a minimum 5$\sigma$ above background. The resulting source
positions, count rates, and S$/$N ratios are listed in columns 2~--~5,
respectively, of Table~2.


 \subsection{Counterparts \& Source Identifications} \label{s:counterparts}

Optical properties of the X-ray sources were determined using archival
\hst\ WFPC2 images, the catalogue (complete to $V\le21$) of Perelmuter
\& Racine (1995), and Digitized Sky Survey (DSS) images. Additionally,
compilations of supernova remanant (SNR) candidates (Matonick \& Fesen 1997),
HII regions (Petit \etal\ 1988; Kaufman \etal\ 1987), globular
clusters (Perelmuter, Brodie \& Huchra 1995; Chandar \etal\ 2001),
novae (Shara, Sandage, \& Zurek 1999), and stellar objects (Zickgraf \&
Humphreys 1991; Ivanova 1992; Zickgraf, Szeifert, \& Humphreys 1996; Sholukhova
\etal\ 1998) were queried for spatial correlations to the X-ray sources.
Objects within the  3$\sigma$ uncertainty of the X-ray
source positions are considered potential counterparts to the X-ray
sources.

Archival \hst\ WFPC2 images of portions of the \cxo\ field were searched for
potential optical counterparts to the X-ray sources based on spatial
coincidence.  For most of the images there were $V$-type filters (F555W, F606W,
and F547M) available.  For a subset of these observations there were images
with filters corresponding approximately to standard $U$ (filter F336W), $B$
(F439W), $V$ (F555W), $R$ (F675W), and $I$ (F814W) for each field.  When
there were multiple images, using the same filter, they were combined using the
{\tt IRAF} task {\tt crrej} to remove cosmic ray contamination.  For a few
images in which there was a position offset the task {\tt xregister} was used
to create matching images and the task {\tt gcombine} was used to combine
the images and to remove cosmic rays.
A list of point sources detected in each field in a $V$ filter image was
generated using the star-finding algorithm {\tt daofind} in {\tt DAOPHOT}
(Stetson 1987).  Photometry was performed for each source using a $0.\arcsec3$
radius aperture and concentric background annuli of inner radius $1.\arcsec0$
and $0.\arcsec5$ width.  For extended sources, a radius as large as
$1.\arcsec0$ was needed for the source aperture.  Where multiple optical
objects fall within the  3$\sigma$ width of the \cxo\ model PSF at the location
of an X-ray source,
we selected the most probable candidate optical source based
on (in order of preference) optical brightness, colors, or distance from
the X-ray source centroid.

A total of 66 X-ray sources lie within one or more of the \hst\ imaging fields.
Of these sources, 34 have potential optical counterparts based on spatial
coincidence with an average separation between X-ray source and optical
candidate of $\sim$1.5$\sigma$. Forty-three of the 66 sources lie in the bulge
of M81 and 13 of these have optical candidates. In contrast, 21 disk
sources are within the \hst\ fields and 19 of these have optical
candidates based on spatial coincidence. The search for optical candidates
was repeated using an artificial distribution of X-ray source positions and
position uncertainties (while preserving the radial distribution of sources).
Of 89 artifical X-ray sources, 51 fell within one or more of the \hst\ imaging
fields and 16 of these had optical candidates (5 candidates in the bulge of
 36 possible and 11 candidates in the disk of 15 possible). Thus, we expect
about $20\pm4$ of the $34$ X-ray sources with potential optical counterparts are
simply chance coincidence with a greater probability of chance coincidence for
sources in the disk. Nevertheless, the \hst\ observations place valid upper
limits to the optical luminosity of the X-ray sources located within their
fields.

The apparent visual magnitudes of the optical candidates (from
F555W, F606W, or F547M filter measurements) are listed in column~6 of Table~2.
The values have not been corrected for charge transfer efficiency (CTE) effects
nor color-corrected to obtain true Johnson $V$ magnitudes. CTE effects are
estimated to be of order 0.01~magnitude while the color-correction depends on
the intrinsic properties of the source and on interstellar reddening. The
correction is typically less than 0.3 magnitudes. For two of the sources,
numbers 52 and 157, only F814W ($I$-band) images were available and they are
denoted as such in column~6 of Table~2. Potential optical counterparts to 7 
X-ray sources appear extended in the \hst\ images. Three of these are listed as
globular clusters in the catalogue of Chandar \etal\ (2001, see below). Two
others, source numbers 82 and 127, are probably also globular clusters (see
below). The remaining 2 sources that appear extended in the \hst\ images are
denoted "Extended?" in column 10 of Table~2. X-ray sources located within \hst\
fields but without optical candidates are designated as "$>$HST" in column~6 of
Table~2 to indicate they are fainter than the \hst\ limiting magnitude of
$\sim$27~magnitudes. Visual magnitudes of objects in the Perelmuter \& Racine 
(1995) catalogue within 3$\sigma$ of X-ray sources are listed as upper limits in
column~6 of Table~2 unless superseded by \hst-derived values.

Archived radio observations were used to search for potential radio counterparts 
to the X-ray sources. Radio data
 obtained using the Very Large Array (VLA) of the National Radio Astronomy
Observatory (NRAO)\footnote[7]{The NRAO is a facility of the National
Science Foundation operated under a cooperative agreement by Associated
Universities, Inc.} as part of an on-going study of SN~1993J were kindly
provided by N. Bartel and M. Beitenholz for our use. For the present
work, we analyzed a radio map made from a 1999 November 23 observation at a
wavelength of 6~cm (4985 MHz) in the B configuration. The angular resolution is
approximately  1.$\arcsec$5, and the root-mean-square noise level is 25~$\mu$Jy.
The image was prepared using standard radio data reduction techniques using the
NRAO software package AIPS (Astronomical Image Processing System). The field of
view of the radio map is approximately 8.$\arcmin$5$\times$8.$\arcmin$5 centered
on the optical nucleus of the galaxy. Of the 177 \cxo-detected X-ray sources, 91
fell within this field of view. The two most luminous radio sources
in the field of view are the nucleus and SN 1993J, and
both of these sources are known to be time-variable in
their radio emission. Only two other X-ray sources were
clearly detected above the 3$\sigma$ level in this radio data, source numbers 79
(55$\pm$17~$\mu$Jy)and 155 (95$\pm$26~$\mu$Jy). They are designated "Radio" in
column~10 of Table~2. A source at the location of \ein\ source X6 was detected
at 3.6~cm, as discussed in \S\ref{s:x6}. Further analysis of the radio
properties of the X-ray sources will be the subject of a separate paper (T.
Pannuti et al., in preparation).

There are 5 SNRs in the tabulation of Matonick \& Fesen (1997) coincident with
X-ray sources including \ein\ source X6. As discussed in \S\ref{s:x6}, the X-ray
emission from X6 is not from a SNR. All four remaining SNR candidates are
located along spiral arms, as expected for core-collapse supernovae from young
massive stars, and are listed with the designation "SNR" in column~10, Table~2.

There are no X-ray sources coincident with any of the 25 globular clusters
identified by Perelmuter \etal\ (1995). The positions of four globular clusters
in the tabulation by Chandar \etal\ (2001) are within the 3$\sigma$ error circle
of X-ray sources as listed in column~10, Table~2, with the designation "GC". 
Three of these, corresponding to X-ray source numbers 141, 146, and 158, are
located near a prominent spiral arm yet the colors reported by Chandar \etal\
(2001) place the optical candidates among the typical population of old
(\GA1~Gyr) clusters. The remaining candidate, corresponding to number 148 of
Table~2, is also an old globular cluster but its projected position is between
spiral arms. In addition, one of the bright objects in the catalogue of
Perelmuter \& Racine (1995) that is coincident with X-ray sources has an optical
magnitude  18$\le$$V$$\le$21 and colors in the ranges  0.5$\le$$(B-V)$$\le$1.1
and  0.3$\le$$(V-R)$$\le$0.7. This object, source number 127, meets the criteria
given by Perelmuter \etal\ (1995) for globular cluster candidates and is
designated "GC?" in column~10 of Table~2. Also listed with the same designation
are X-ray source numbers 67 and  82. Although the optical candidates of both
objects exceed the Perelmuter \& Racine (1995) brightness criteria, Ghosh \etal\
(2001) have argued that they are very likely globular clusters (see also
\S\ref{s:others}).

Three hundred and ninety of the 492 HII regions tabulated by Petit, Sivan, \&
Karachentsev (1988) are within the ACIS imaging field of view.
Twelve of these are coincident with \cxo\ x-ray sources. All are located along
 M81 spiral arms. They are designated "HII" in column~10 of Table~2.

Three \cxo\ sources are coincident with foreground stars. These are labeled with
a designation "$\star$" in column 10 of Table~2. Two of these, X-ray source 
numbers 56 and 60, are the PPM catalogued stars used to determine absolute X-ray
source positions (Table~1). Inspection of DSS images to search for
uncatalogued bright star-like objects revealed no additional foreground star
candidates. X-ray source number 8, however, is very likely a background galaxy
based on its shape in DSS images.


 \subsection{X-ray Spectral \& Timing Analysis} \label{s:discrete_spec}

Statistically-constrained model fits could be achieved for a total
of 39 sources in the M81 field.
Detailed X-ray properties of SN~1993J, \ein\ source X6, and the nucleus are
presented in \S\ref{s:bright_src} and of the three brightest supersoft sources
in Swartz \etal\ (2002). Best-fit spectral model parameters for the remaining 33
bright sources are listed in Table~3.

The X-ray spectra of these 33 sources were fit to absorbed power law models
characterized by the photon index $\Gamma$; to absorbed
Raymond-Smith spectral models representing emission from low density,
optically-thin plasma characterized by the temperature $kT$; 
and to absorbed blackbody models.
The metal abundances in the thermal plasma models were assumed to
be 3\% of their solar value consistent with
the results of Kong \etal\ (2000) for the galaxy as a whole. Spectra were
grouped to contain a minimum of 20 counts per spectral bin and fit to models
using the XSPEC spectral fitting package (Arnaud 1996).

The power law model provides a significantly better fit to all but two of
the X-ray sources. The two exceptions, X-ray source numbers 160 and 161, are
best fit by blackbody spectral models with effective temperatures 0.2 and
 0.57~keV, respectively. The best-fit absorption column density, power-law index
(or blackbody temperature), fit statistic, and unabsorbed 0.3~--~8.0~keV
luminosities are listed in columns 2~--~5 of Table~3. Quoted errors are 90\%
confidence limits for a single interesting parameter based on the $\chi^2$ fit
statistic. Absorption column densities were constrained in the fitting
procedure to be at least as large as the Galactic column density along the line
of sight to M81. This resulted in a best-fit value of $N_H$ equal to
this lower limit for 8 sources. Therefore, the column densities are listed as
$N_H=4.0\pm0.0$ in column~2 of Table~3 for these sources.

The luminosities for the remaining sources in Table~2 were estimated
assuming an absorbed power law spectrum with photon index $\Gamma=1.5$
and hydrogen column density $N_{20} = N_H/10^{20} = 11.0$ cm$^{-2}$.
These are the average values for the 31 sources in Table~3 whose spectra are
best fit with a power law model. For comparison, the Galactic hydrogen column
density in the direction of M81 is $N_{20}=4.0$~cm$^{-2}$ (Stark \etal\ 1992).
The source detection limit of 12 counts corresponds to an observed flux of
$F_{0.3-8 keV} = 1.9 \times 10^{-15}$~\ergcms\ for sources on device S3
($F_{0.3-8 keV} = 2.5 \times 10^{-15}$~\ergcms\ on FI devices) or an
unabsorbed 0.3~--~8 keV luminosity of
$L_{\rm X} = 3.4 (4.5) \times 10^{36}$ \ergl\
for sources on the BI (FI) devices. The luminosities of all
sources are listed in column~9 of Table~2.

The Kolmogorov-Smirnov statistic was used to test these sources
(including background) for constant
count rates over the duration of the primary
observation. The Kolmogorov-Smirnov statistic results are listed in
column~6 of Table~3. Sources with a low value of $P_{KS}$ have a high
probability of being variable.

The 0.3~--~8.0~keV light curves were used to estimate the overall X-ray power
density spectrum (PDS) of the X-ray sources listed in Table~3.
Light curves were sampled at the 3.24104~s nominal ACIS frame time and
used to compute the Leahy-normalized power spectra of each source.
The average power in each
normalized-PDS was 2.0, which is consistent with what is expected from Poison
noise (Leahy et al. 1983). Fluctuations up to 10~--~15 are commonly seen
in the normalized power spectra with a maximum power in one or two frequency
bins typically between 15 and 20 for all the sources with one
exception. Fluctuations of this order are due to noise. For the one exception,
X-ray source number 69, the average normalized-PDS power
was 2.0 and the power of all the frequencies was less than 15
except at 0.0245 Hz (corresponding to a 40.8 s period), which has a power
of 28. Given that there are 4096 frequency bins in the PDS for each of 33
sources, the probability of seeing one peak with power 28 is 12\%. Therefore we
conclude that the detection of the peak with power 28 could be a statistical
fluctuation and that no X-ray pulsations are detected in any of the
sources with the existing data.

\section{Individual M81 Sources} \label{s:bright_src}

The eleven brightest sources in the \cxo\ field appear to exceed the Eddington
limit luminosity for a 1.5 \msun\ spherically-accreting object. It should be noted that
many accreting compact sources in our Galaxy and in
the Magellenic Clouds that exceed the Eddington limit are in fact neutron stars
with episodes of high X-ray luminosity (see, for example, the compilation in
Grimm, Gilfanov, \& Sunyaev  2001). The four brightest sources in M81 are
SN~1993J, \ein\ source X6, the nucleus, and the brightest supersoft source
candidate. The supersoft source is discussed in detail by Swartz \etal\ (2002)
and no further analysis  is given here. Details of the other three sources are
given in the following subsections followed by a brief discussion of the
remaining bright sources (\S\ref{s:others}).

 \subsection{SN 1993J} \label{s:sn1993j}

Supernova (SN) 1993J
was discovered 1993 March 28 and was observed by \cxo\ 2594~days after outburst.
The pre-supernova star was likely a $\sim$17~\msun\ star that lost all but
$\sim$0.2--0.4~\msun\ of its hydrogen envelope prior to explosion (see Wheeler
\& Filippenko 1996 for an early review). This mass was lost through a
combination of stellar winds and mass transfer to a binary companion, possibly
involving a common envelope phase (Podsiadlowski \etal\ 1993; Woosley \etal\
 1994; Nomoto, Iwamoto, \& Suzuki 1995). A circumstellar medium (CSM), a relic
of this mass loss, surrounded the system at the time of explosion as evidenced
by its early radio (Van Dyk \etal\ 1994) and X-ray signatures (e.g., Zimmerman
\etal\ 1994).

Monitoring of this emission has continued since discovery. X-ray observations by
\ros\ (Immler, Ashenbach, \& Wang 2001) and \asca\ (Uno \etal\ 2002) began 6
and 8 days after the explosion and continued intermittently for 1817 and 564
days, respectively. Radio monitoring continues (e.g., Marcaide \etal\ 1997;
Bietenholz, Bartel, \& Rupen 2001). The evolution of optical line profiles also
shows an increasing contribution from SN debris interacting with surrounding gas
(Patat, Chugai, \& Massali 1995; Houck \& Fransson 1996; Matheson, \etal\ 2000).

The early X-ray and radio data has been successfully explained in terms of the
circumstellar interaction model (Fransson, Lundqvist, \& Chevalier 1996). In
this model, the radial flow in free expansion of the SN debris steepens into a
shock wave in the CSM, heating this gas to $T \sim 10^9$ K. This outer shock
cannot be freely expanding so the supernova ejecta interacts with the hot gas
and a second (inner, or reverse) shock front develops. The lower temperature
($T$ \LA\ $10^8$ K) reverse shock gas dominates the soft xX-ray emission with a
flux dependent on the density gradient of the SN ejecta and on absorption in the
cooling, post-shock, gas.

 \subsubsection{The \cxo\ X-ray spectrum of SN 1993J}

The observed X-ray spectrum of SN~1993J, the best-fit model spectrum
($\chi^2=94.6$ for 90 dof), and 
the fit residuals are shown in Figure~\ref{f:93j_spec}. Fits were applied to
the spectrum in the energy range 0.3~--~5.0~keV because of a lack of signal at
higher photon energies, including in the Fe~K band at $\sim6.5$~keV. The model
consists of two low-temperature absorbed thermal emission-line ({\em vmekal})
components at 0.35$\pm$0.06~keV and 1.01$\pm$0.05~keV both with
$N_{20}=40.5\pm0.9$~cm$^{-2}$ and a high-temperature {\em mekal} component with
$kT=6.0\pm0.9$~keV, $N_{20}=4.9\pm0.1$~cm$^{-2}$. Abundances in the
low-temperature components are consistent with solar values with the exception
of N (15 times solar, though see below), Mg (0.2), Si (1.4), Fe (1.6) and He, C,
O, and Ne which are all consistent with an abundance of zero. The
high-temperature component is consistent with subsolar abundances but is not
sensitive to this parameter.

The two low-temperture components are needed to fit the two peaks present in the
observed spectrum (at $0.78\pm 0.02$ and $0.98\pm0.02$~keV, respectively,
Figure~\ref{f:93j_spec}). These peaks are dominated by Fe~L emission from
 different ionization stages of Fe at the different temperatures. In addition,
He-like Si and Mg (from the 0.35~keV component) and He-like S, and H-like
Si and Mg (from the 1.01~keV component) are observed in the
data. Notably absent is the O~VIII Ly$\alpha$ line at 654~eV. This line is
predicted to be strong over a range of plasma temperatures from $\sim$0.1 to
 0.5~keV (e.g., Nahar 1999).
The high-temperature component is needed to account for
X-ray flux above $\sim$1.5~keV. No emission lines are produced by this
component. In fact, replacing the {\em mekal} component with a bremsstrahlung
model at the same temperature provides an equally-acceptable fit to the
spectrum.
The elemental abundances were allowed to vary in the two low-temperature
component model fits (with the abundances and absorbing columns
constrained to be equal between the two components). Beginning with solar
ratios, the abundance of each element was individually varied while holding the
abundances of the remaining elements fixed at their current values until the
model converged. This procedure was repeated for all $\alpha$-chain elements
from He through Ni and for N, Na, and Al. The largest reductions in $\chi^{2}$
occurred for N, O, and Fe. A large N abundance is required to fit a broad
feature near the N VII Ly$\alpha$ line at 0.5~keV. This feature lies just below
the neutral O absorption edge at 0.53~keV and may, instead, be an artifact of a
deeper O edge. This interpretation is also more consistent with evolutionary
models of a massive progenitor star (e.g., Thielemann, Nomoto, \& Hashimoto
 1996) that do not predict an over-abundance of N. The model requires no O
consistent with the lack of observed O VIII Ly$\alpha$. The over-abundance of Fe
needed to fit the spectrum may indicate some of the explosively-synthesized
material has been transported to the outer regions of the SN ejecta which are
presently entering the reverse-shock region.

\begin{center}
\includegraphics[angle=270,width=\columnwidth]{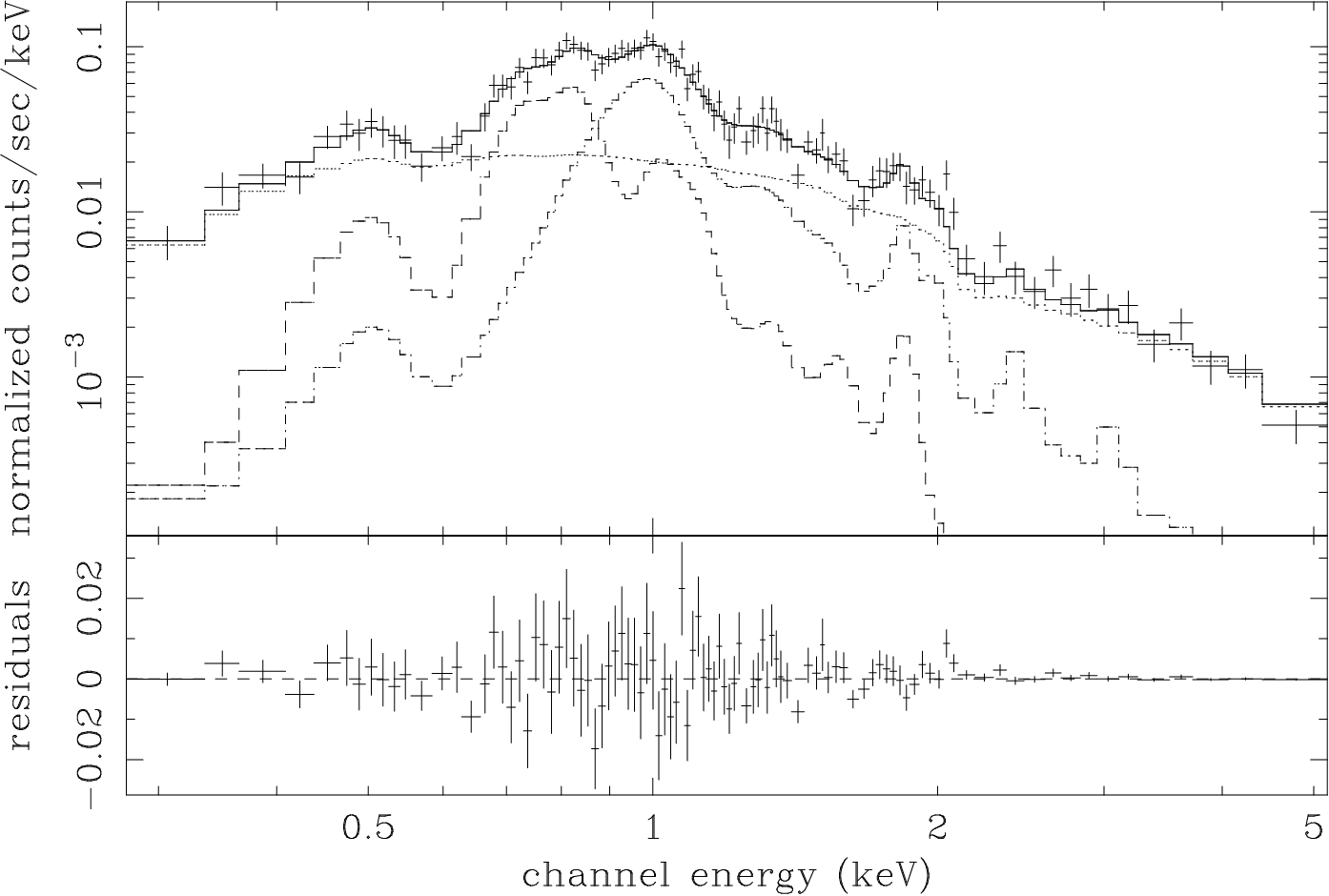}
\vspace{10pt}
\figcaption{Observed X-ray spectrum of SN~1993J ({\em top}) and model fit
residuals ({\em bottom}). Curves trace the full model ({\em solid}), 0.35~keV
({\em dashed}) and 1.01~keV ({\em dot-dashed}) {\em vmekal} components, and the
harder, 6.0~keV, {\em mekal} component ({\em dotted}). \label{f:93j_spec}}
\end{center}

The combination of low-~and high-temperature
components is consistent with the standard CSM interaction model. The
low-temperture emission in this scenario originates in the reverse shock region
while the hard component comes from the much hotter forward shock. The
low-temperature components are more heavily absorbed than the hard component
consistent with an intervening dense cool shell of gas at the contact
discontinuity between the forward and reverse shock fronts. The reverse shock
front is distorted by Rayleigh-Taylor instabilities (Chevalier \& Blondin 1995).
Emission from regions of differing densities (and, hence, cooling rates) will
have different characteristic temperatures. The two low-temperature model
components may therefore only approximate a range of temperatures present in the
reverse shock region.

In contrast, the forward shock is located further from the high-density
contact region. High-resolution VLBI images (e.g., Bietenholz \etal\ 2001) shows
the outer shell to be highly circular. While a single temperature forward
shock may be favored in this case, we note that the temperature of the
high-temperature component is uncertain due to the lack of signal above
$\sim$5~keV. The relatively low temperature of this
component, $\sim$$7\times 10^7$~K, suggests relatively flat ejecta and CSM
density profiles at the time of the \cxo\ observation.

The average densities in the reverse shock region and in the intervening
absorbing shell can be estimated from the model parameters. The radius of the
interaction region at 2594 days was approximately $2\times 10^{17}$~cm based on
extrapolating the observed angular size at 1893 days (Bartel \etal\ 2000) to the
time of the \cxo\ observation and assuming an average velocity of
$\sim$7000~km~s$^{-1}$ (Matheson \etal\ 2000) during the interval. The width of
the shell is of order  10\% of this radius. The number density in the thermal
emission region is then $\sim$$5\times 10^4$~cm$^{-3}$ and that in the cool
absorbing shell is $\sim$$2\times 10^5$~cm$^{-3}$. This implies the mass in the
shell is of order $\sim$0.4 to 2.0~\msun. The lower value is consistent
with the value of $\sim$0.3~\msun\ inferred from the deceleration observed in
radio images of SN~1993J at an age of $\sim$5~yr (Bartel \etal\ 2000).

Conceivably, an underlying neutron star could also contribute to the hard
emission detected from SN~1993J. While the dynamics of the hot,
radioactively-heated, gas at the center of the SN is unknown, accretion at a
rate above a few $10^{-8}$~\msun\ yr$^{-1}$ creates substantial X-ray
emission. However, the emission in the \cxo\ band will be absorbed by the
overlying SN debris. Assuming for simplicity a uniform-density expanding sphere
of gas, the ejecta provides a total column density of order $10^{22}$~cm$^{-2}$
per solar mass of ejecta at the time of the \cxo\ observation. Most of this
material is in the form of metals. Thus the effective hydrogen column is
orders of magnitude higher. Woosley, Pinto, \& Hartmann (1989) predicted no
detectable X-ray flux below $\sim$10~keV from an accreting neutron star in the
center of SN~1987A at an age of 2500 days with the exception of the Fe~K$\alpha$
fluorescence line. A similar conclusion was reached by Xu \etal\ (1988). Though
the amount of material ejected by SN~1993J is nearly an order-of-magnitude less
than in SN~1987A, there is as yet no compelling evidence of a neutron star in
the X-ray spectrum of SN~1993J at an age of $\sim$7~yr.

 \subsubsection{The X-ray light curve of SN 1993J}

The 0.3~--~8.0~keV luminosity of SN~1993J on day 2594 was
$4.8\times10^{38}$~\ergl. The flux in the 0.1~--~2.4~keV \ros\ and 1~--~10~keV
\asca\ energy bands are $\sim3.3\times10^{38}$ and $\sim3.6\times10^{38}$~\ergl,
respectively. \ros\ (Immler, Ashenbach, \& Wang 2001) and \asca\ (Uno,
\etal\  2002) light curves, along with the \cxo\ data, are shown in
Figure~\ref{f:93j_lc}.
Visual inspection shows that the light curves are not simple power laws but that
the rate of decline of the X-ray luminosity increases after $\sim$50 to 100
days. The best-fit broken power law model for the \ros\ light curve has
its break at $\sim$220~days. The luminosity declines as $L \propto
t^{-0.24\pm0.04}$ prior to the break and as $L \propto t^{-0.62\pm0.07}$ at
later times. The break occurs at about 45~days in the \asca\ energy range but
the change in slope is less pronounced, evolving from $L \propto
t^{-0.57\pm0.78}$ to $L \propto t^{-0.84\pm0.24}$.

\begin{center}
\includegraphics[angle=270,width=\columnwidth]{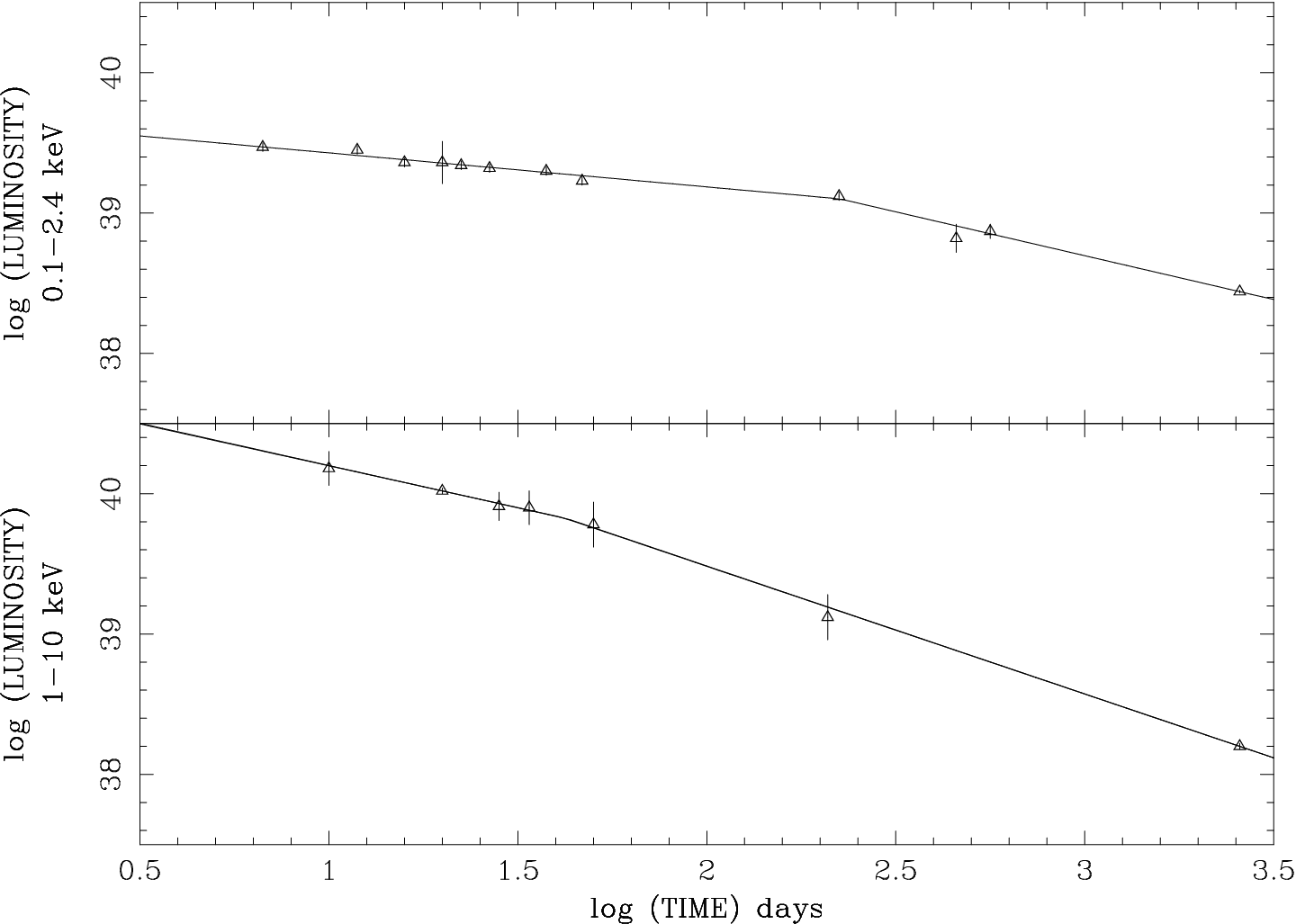}
\vspace{10pt}
\figcaption{Observed X-ray light curve of SN~1993J in the \ros\ ({\em top}) and
\asca\ ({\em bottom}) energy bands. The best-fit broken power law models are
shown as solid lines.
\label{f:93j_lc}} \end{center}

As originally conceived, the self-similar form of the interaction model of
Chevalier (1982; Fransson, Lundqvist, \& Chevalier 1996) applies only to the
early phases of supernova evolution when both the SN ejecta and CSM density
profiles can be represented by single power laws in radius. Extrapolation to
$\sim$7~yr is inappropriate because as the reverse shock progresses through the
ejecta the density profile flattens while the forward shock has traversed some
$10^4$~yr of previous mass-loss history (assuming a wind velocity of order
 10~km~s$^{-1}$ and a shock velocity of order $10^4$~km~s$^{-1}$). Furthermore,
the importance of radiative losses behind the reverse shock and departures from
electron-ion equipartition in the forward shock make numerical calculations
necessary.

Suzuki \& Nomoto (1995) have performed hydrodynamic calculations based on models
of the SN ejecta that accurately reproduce the observed optical light curve of
SN~1993J (Nomoto \etal\ 1995). Suzuki \& Nomoto (1995) followed the evolution
for $\sim$1000~days and find reasonable agreement with X-ray observations for
the first 50 days or more. Beyond that time, Suzuki \& Nomoto (1995) require a
combination of a steeper CSM density gradient and a clumpy CSM morphology to
sustain a level of X-ray flux comparable to that observed. However, their models
predict rapid increases in X-ray flux, followed by declines on timescales of
order months, as the forward shock sweeps through individual clumps. This is in
contrast to the observed steady decline of the X-ray light curve over the entire
monitoring sequence.

While beyond the context and scope of the present work, X-ray observations of
SN~1993J warrant further investigation. As pointed out by Immler \etal\ (2001),
X-rays from the interaction region trace the pre-SN evolution of the progenitor
system and X-rays are a direct means of accessing this evolution in detail as
has been undertaken recently for SN~1987A (Park \etal\ 2002).

 \subsection{Einstein Source X-6} \label{s:x6}

The brightest non-nuclear source in the \cxo\ field is the \ein-discovered
source X6 (Fabbiano 1988) located $\sim$1$\arcmin$ to the southeast of
and along the same prominent spiral arm containing SN~1993J. The X-ray flux in
the \ein\ observation was $9.5\times 10^{-13}$~\ergcms, placing X6 in the class
of Ultra-Luminous X-ray sources (ULXs) whose luminosities (\GA$10^{39}$~\ergl)
far exceed the Eddington limit for spherically-accreting $\sim$1.5~\msun\
objects. The X-ray flux from X6 has remained remarkably steady throughout its
observed history. X6 is coincident with a weak radio source (Fabbiano 1988) and
with an optically-identified SNR candidate (Matonick \& Fesen 1997). \cxo\ can
resolve sources on scales smaller than the 90~pc (5$\arcsec$) diameter reported
for the SNR.

\subsubsection{The \cxo\ X-ray spectrum of X6}

The high X-ray flux from X6 leads to a pileup of events
in the ACIS detector and to statistically-significant detection of events during
frame transfer. Events detected during frame transfer appear as a streak or
trail through the source along the detector readout direction. These events are
not piled up and therefore represent the true count rate (when properly
time-scaled) and source spectrum although spread over a large spatial region.
The 0.3~--~8.0~keV count-rate of X6 is 0.21~c~s$^{-1}$ compared to the
$0.38\pm0.04$~c~s$^{-1}$ rate deduced from the readout trail. The spectrum was
fit using two identical models but with one model convolved with the pileup
model developed by Davis (2001) as implemented in XSPEC v. 11.1.0u. This
combination reflects contributions from pileup and non-pileup spatial regions:
Piled-up events are localized to the central few pixels containing the majority
of the detected counts. The spectral extraction region includes this central
region and additional source counts from surrounding pixels. A
linear energy grid is necessary to apply the convolution model. Therefore, the
spectral analysis is performed with the Level~2 event data using CXC-provided
response and ancillary response files instead of those provided by L. Townsley
which utilize a piece-wise linear energy grid. As a test of our procedures, we
repeated the analysis of S5~0836$+$710 undertaken as a demonstration by Davis
(2001) and derived model parameters consistent with the values presented in
that paper.

Spectral models were applied to photons from X6 in the 0.3 to 10.0~keV range
using a 2$\arcsec$ extraction region.
The extension to 10~keV was made because there are substantial source counts
above 8~keV and the observed spectrum shows a flattening above $\sim$7~keV
characteristic of pileup (Figure~\ref{f:x6_spec}), a feature helpful for
constraining the model parameters. The 2$\arcsec$ extraction region is the same
size region used by Davis (2001) in his analysis of  S5~0836$+$710.
The fitting procedure resulted in a slight adjustment of
the pileup parameters from
the values reported in Davis (2001).
The event grade morphing parameter, $\alpha$, and the
fraction of the extraction region {\em not} experiencing pileup, $1-f$, both
must be increased slightly because the PSF is slightly asymmetric and broadened
at the $\sim$1$\arcmin$ off-axis position of X6 relative to the on-axis location
of S5~0836$+$710. Inspection of the high-resolution image of the model PSF shows
the asymmetry increases the probability that the second photon in a two-photon
event will enter an adjacent pixel rather than a corner pixel and hence will be
registered as a good grade. The broadening of the PSF results in a larger
fraction of the detected photons falling in the wings of the PSF where pileup
does not occur. The resulting pileup parameters are $\alpha = 0.585$ compared to
 0.5 used by Davis (2001) and $1-f = 0.063$ (compared to 0.05) as determined
from the best-fit spectral model.

\begin{center}
\includegraphics[angle=270,width=\columnwidth]{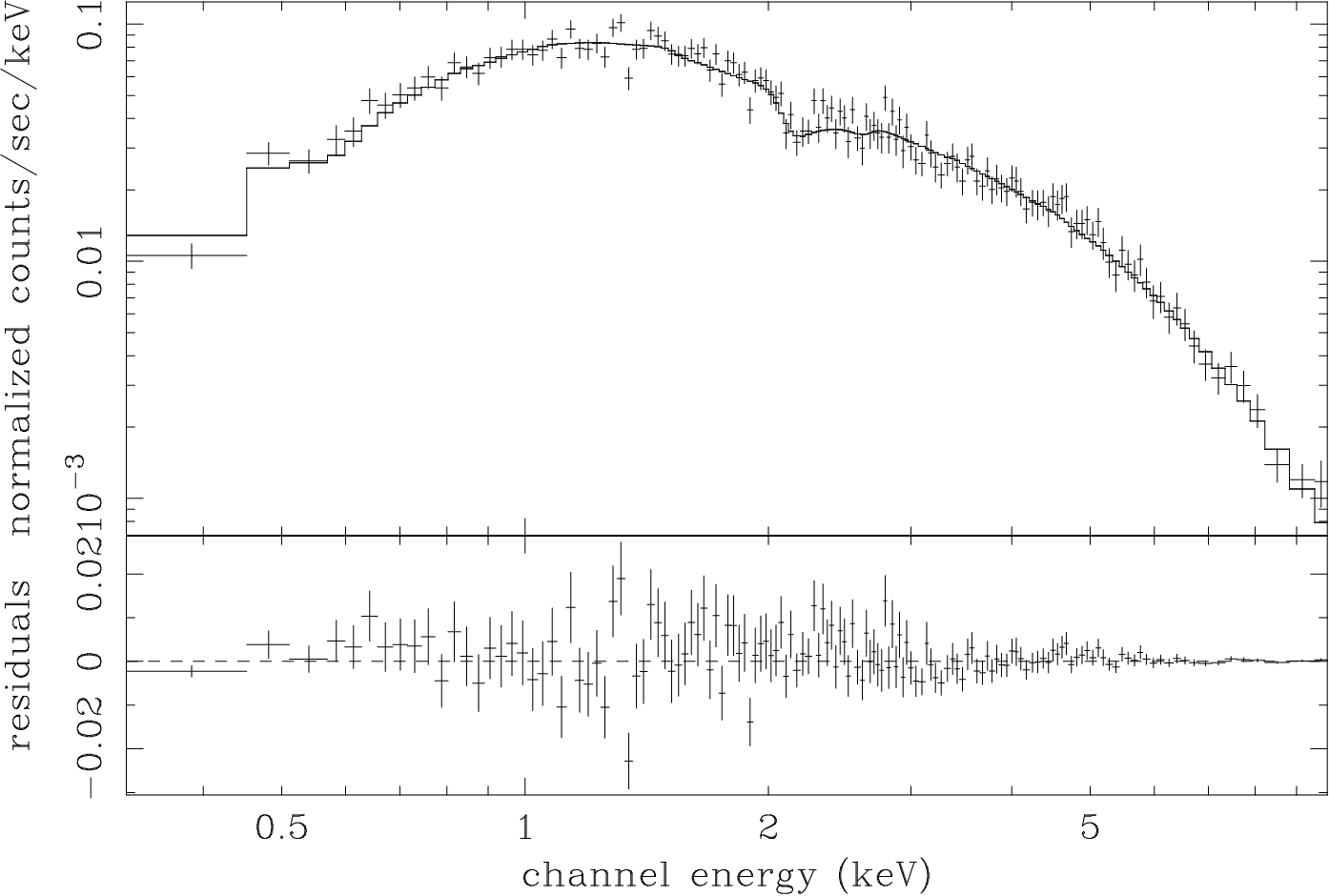}
\vspace{10pt}
\figcaption{Observed X-ray spectrum of Einstein source X6 ({\em top}) and model
fit residuals ({\em bottom}). The curve traces the best-fit disk blackbody
model spectrum with pileup modeled as in Davis (2001). \label{f:x6_spec}}
\end{center}

The best-fit model for the observed spectrum of X6 is an absorbed disk
blackbody.
Makishima \etal\ (2000) have modeled the \asca\ spectrum of X6 and other ULXs
using the disk blackbody model they developed (Mitsuda \etal\ 1984) for modeling
the high soft state of accreting black holes. X-ray emission in this state
originates from an optically thick accretion disk and the model is basically a
superposition of blackbody emission from different disk annuli with local disk
temperatures scaling with the disk radius as $R^{-3/4}$. Thus the model spectrum
in the X-ray regime is dominated by the innermost disk temperature, $T_{in}$,
with a normalization scaling as the disk geometry; $K \propto (R_{in}/D)^2 {\rm
cos}(\theta)$ where $R_{in}$ is the innermost disk radius, $D$ the source
distance, and $\theta$ the disk inclination. This model provides a
fit statistic $\chi^2 = 295.3$ for 283 dof. The best-fit innermost disk
temperature is $T_{in}=1.03\pm 0.11$~keV, the corresponding radius is
$R_{in}=161\pm16$~km, and the absorbing column density is
$N_{20}=21.7\pm1.0$~cm$^{-2}$. The disk blackbody model parameters correspond to
an 18~\msun\ accreting, non-rotating, object and a bolometric luminosity of
$\sim$$2.7\times10^{39}$~\ergl\ according to the relations given by Makishima
\etal\ (2000). This luminosity is equivalent to the Eddington luminosity for the
derived mass. The parameters obtained from analysis of the \asca\ spectra
(Mizuno 2000, Makishima \etal\ 2000) are $T_{in}=1.48\pm 0.08$~keV,
$R_{in}=83\pm8$~km, and $N_{20}=21\pm3$~cm$^{-2}$. The higher temperature
implies a lower mass since $T_{in} \propto M^{-1/4}$ (eq.~12, Makishima \etal\
 2000) but a higher luminosity, $L \propto R_{in}^2 T_{in}^4$, implying the
luminosity of X6 exceeds the Eddington limit. Makishima \etal\
(2000) argue that, if the compact object is a Kerr black hole, then the inner
radius can be reduced by as much as a factor of 6 thereby reducing the derived
bolometric luminosity to values below the Eddington limit for the
temperature-estimated mass. In contrast, the fit parameters to the \cxo\ data
leads to self-consistent values of bolometric luminosity, mass of the compact
object, and the associated Eddington limit luminosity.

Mizuno (2000) analyzed 7 individual
\asca\ observations of X6 and found a temperature $T_{in} =1.3\pm0.1$~keV in
two of the observations and $1.6\pm0.1$~keV in the remaining observations. The
value derived here, $T_{in} =1.0\pm0.1$~keV, is significantly lower than these
values. It is unclear if this is a real effect or is an artifact of the pileup
model.

A power law model fit to the \cxo\ spectrum was statistically less acceptable.
The best-fit result with $\alpha$ and $1-f$ fixed as above is $\chi^2 = 342.8$
for 284 dof; for an absorbing column $N_{20} = 39.5\pm 2.0$~cm$^{-2}$, and a
photon index $\Gamma=2.1\pm0.1$. Across the 90\%-confidence range of $\alpha$
determined from the disk blackbody model, 0.56$\le$$\alpha$$\le$1, the range of
acceptable power law indices is 2.0$\le$$\Gamma$$\le$2.3. The largest systematic
contributions to $\chi^2$ for this model occur just above the 2~keV Ir-M edge.
This is precisely where contributions from pileup from photons at the peak of
the energy distribution (at $\sim$1.0~--~1.5~keV) occurs. Adding a broad
Gaussian line to the model significantly improved the fit to $\chi^2 = 298.0$
for 281 dof. The resulting absorbing column is $N_{20} = 31.2\pm 2.0$~cm$^{-2}$
and the photon index is $\Gamma=1.71\pm0.09$. The aperture-corrected model
predicted flux in the 0.3~--~8.0 keV band is 3.9$\times$$10^{-12}$~\ergcms\
corresponding to a luminosity of $L=6.0 \times 10^{39}$~\ergl.

An optically-thin thermal plasma model can also produce an acceptable fit 
provided the abundance is kept low to reproduce the observed continuum-dominated
spectrum. The best-fit parameter values  for X6 are $kT=3.5\pm0.4$~keV,
$N_{20}=30.7\pm1.8$~cm$^{-2}$, for $\chi^2 = 310.0$, 283 dof, and the metal
abundance constrained to 0.03 of the solar value. This model produces few strong
spectral lines with the exception of Fe~K$\alpha$ at 6.7~keV. Adding a line at
this energy to either the disk blackbody or power law models is not
statistically significant.

The results given above are consistent with another test we conducted:
The spectrum of X6 was extracted from an annulus with a 1.$\arcsec$5 inner
radius (compared to the 2$\arcsec$ outer radius used for applying the pileup
model). In this case pileup is not an issue and a simple absorbed disk blackbody
or power law model is sufficient. The resulting model parameters were consistent
with those quoted above though the formal errors are considerably larger because
of the lower number of source photons detected in the extraction region.

\subsubsection{X6 Radial Profile}

A two-dimensional Gaussian model fit to the spatial distribution of X-ray events
places source X6 at 9$^h$55$^m$32.98$\pm$0.08$^s$,
$+$69$^{\circ}$0$\arcmin$33.4$\pm$0.4$\arcsec$ (ignoring the absolute
uncertainty in source positions, see \S\ref{s:observations} and Table~1). Source
X6 is coincident with a large SNR candidate. Wang (1999) identified several
X-ray--bright sources apparently associated with SNRs based on \ros\
observations of M101 and concluded that the blast wave energies of these SNRs
exceed theoretical predictions of supernova explosions by factors of 30 or
greater. In the case of X6, the superb angular resolution of the \cxo\ image can
be used to determine the extent of the source on scales much less than the
$\sim$5$\arcsec$ size of the candidate SNR (No.~22, Table~10, Matonick \& Fesen
1997).

Figure~\ref{f:x6_rp} displays the radial profile of source X6 along with a model
of the radial profile for a point source located at the off-axis position of X6
and the observed radial profile of SN~1993J. The total number of counts in
the model PSF and the SN~1993J profile were scaled to
 1.6 times the total number of counts detected in the X6 profile. This
accounts for the 40\% pileup estimated for source X6. As can be seen, the
majority of the pileup occurs in the innermost two radial bins (with an area of
 9 pixels). The profile of SN~1993J is slightly more concentrated than either
the model PSF or the X6 profile consistent with its closer proximity to the
aimpoint (20$\arcsec$ compared to $\sim$58$\arcsec$ for X6). There is no
evidence that X6 is an extended source; including in the 5$\arcsec$ region
occupied by the SNR.

\begin{center}
\includegraphics[angle=270,width=\columnwidth]{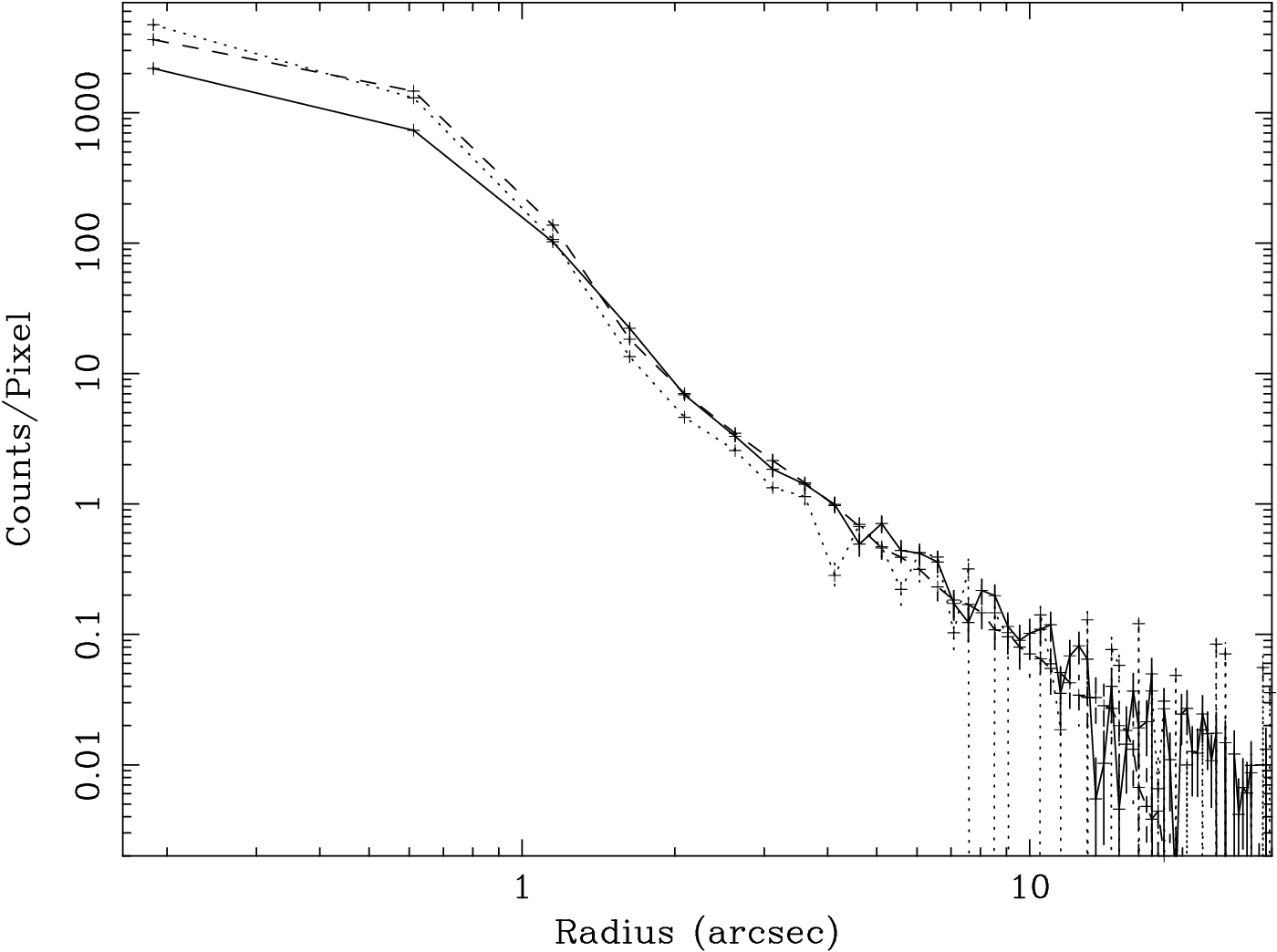}
\vspace{10pt}
\figcaption{Radial profile of source X6 ({\em solid}). X6 is located
$\sim$1$\arcmin$ off-axis. The profiles of SN~1993J ({\em dotted};
$\sim$20$\arcsec$ off-axis) and the 1.5~keV, 1$\arcmin$--off-axis, model PSF
({\em dashed}) are shown for comparison. There is no evidence for source 
extension in the X-ray profile of X6. \label{f:x6_rp}}
\end{center}

\subsubsection{Potential X6 Counterparts}

A weak uncatalogued radio source visible in the 21~cm map of Bash \& Kaufman
(1986) is coincident with the \ein\ High Resolution Imager position of source
X6 according to Fabbiano (1988). The source is not seen in a 6~cm VLA image
taken on 1999 Nov 23 above the $\sim80\mu$Jy 3$\sigma$ limit but is detectable
in a  3.6~cm image obtained 1994 Dec 23 at a flux density of $\sim95\mu$Jy, just
above the $3\sigma$ signal-to-noise limit. The radio source extension
cannot be reliably measured at this low signal level.

Figure~\ref{f:x6_hst} displays an archival \hst\ WFPC2 F555W image of the region
containing X6 with the X-ray source position identified by a 1$\arcsec$-radius
circle. An optical source is clearly present within $\sim0.\arcsec2$ of the X6
location. The observed \hst\ magnitudes are F336W$=22.8\pm0.2$,
F439W$=24.1\pm0.3$, F555W$=24.1\pm0.1$, F675W$=23.9\pm0.2$, and
F814W$=23.7\pm0.5$. Estimating the color excess from the hydrogen column density
obtained from the X-ray spectrum, the optical properties are consistent with an
early-type main sequence star of spectral class O9~--~B1 though the source is
relatively bright in the $R$-band (F675W) image, perhaps due to H$\alpha$
emission. Be stars with circumstellar disks can emit strong H$\alpha$ under
certain circumstances. The equivalent H$\alpha$ flux is
$\sim(5.4\pm0.9)\times10^{-15}$~\ergcms\ assuming all the $R$-band flux is from 
this emission line. Matonick \& Fesen (1997) report the H$\alpha$ flux from the
SNR candidate to be $1.1\times10^{-14}$~\ergcms. The point source can,
therefore, contribute as much as one-half of this amount. Note that the point
source flux estimate is not contaminated by any underlying extended emission
because the source and background extraction regions used in our analysis
(\S\ref{s:counterparts}) lie wholly within the region occupied by the candidate
SNR.

\begin{center}
\includegraphics[angle=270,width=\columnwidth]{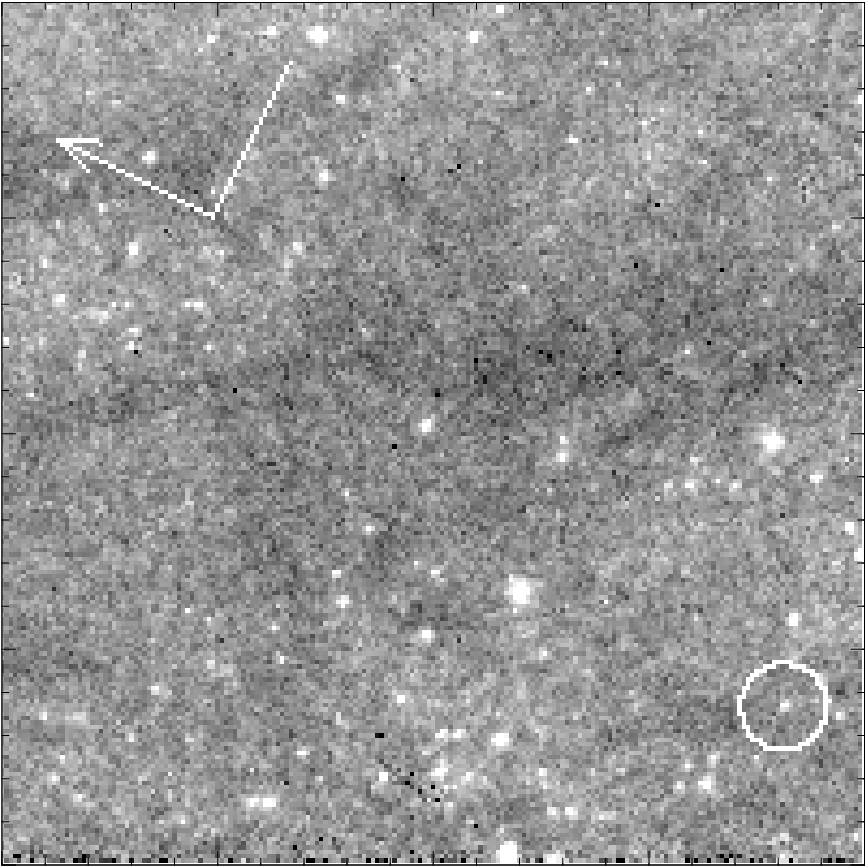}
\vspace{10pt}
\figcaption{\hst/WFPC2 image of the region surrounding source X6 taken with the
F555W filter. The 1$\arcsec$-radius circle at the lower left denotes the X-ray
position in the \cxo\ data. The arrow indicates the direction of north and the
line segment denotes east. The optical properties of the object at the location
of X6 is consistent with a main sequence star of spectral class O9~--~B1.
\label{f:x6_hst}} \end{center}

If the X-ray
emission is from an accretion disk then it is possible that some or all of the
optical emission is also from an accretion disk instead of from a companion
star (or from another object in the field). The color index, $\xi\equiv B_0+2.5
{\rm log}F_X$ where $B_0$ is the reddening-corrected $B$ magnitude and $F_X$ is
the 2~--~10~keV X-ray flux in $\mu$Jy, was introduced by van~Paradijs \&
McClintock (1995) to quantify the observed relationship between X-ray and
optical flux from accretion-powered X-ray binaries. The average value of $\xi$
for systems in which the secondary star does not contribute significantly to the
optical brightness (namely, those with low-mass donor stars) is $21.8\pm1.0$
(errors are $1$ standard deviation). This can be compared to X6 where $\xi
\sim22.5\pm1.5$ (where the error includes an uncertainty of 0.15 magnitudes in
the color excess). The average colors for these systems are $(B-V)=-0.09\pm0.14$
and $(U-B)=-0.97\pm0.17$ compared to $(B-V)=-0.1\pm0.3$ and $(U-B)=-1.4\pm0.3$
for X6.

The observed X-ray-to-optical flux ratio, $\sim$900, is far too large for
typical foreground objects. Normal stars have ratios in the range $10^{-4}$ to
$0.1$ (Maccacaro \etal\ 1988) and cataclysmic variables (CVs) have ratios in the
range 0.1 to  10.0 (Bradt \& McClintock 1983). The only extragalactic sources
with very high ratios of X-ray to optical flux are the BL~Lac sources with
ratios in the range  10 to 50 (Maccacaro \etal\ 1988). To produce the ratio
observed for X6 requires an optical extinction greater than 4 magnitudes, a
possibility excluded by the X-ray-measured hydrogen column density of
$\sim2\times 10^{21}$~cm$^{-2}$.

\subsubsection{X6 Variability} \label{s:x6_timing}

Tests for source flux variability were applied to counts
extracted from both the entire source region and to counts in the inner 3x3
pixel region and to the outer source region. There is no evidence of pulsations
or other variability in any of these regions. However, periods of higher flux
would incur more pileup. This tends to smooth the light curve by
reducing the observed count rate during these high flux periods.

We performed similar analysis of the extensive set of \ros\ observations and
found no large-scale variability on timescales as short as the \ros\ orbital
period. Immler \& Wang (2001) report a factor-of-two change in the X6
\ros\ PSPC count rate over a 6~day period in 1993~Nov. Our analysis of this
data find a modest, maximum 22$\pm$7\%, change in count rate but the data are
statistically consistent with no variation when compared to SN~1993J. Mizuno
(2000) reports that variability of X6 in the \asca\ data cannot be assessed
because of the unavoidable contribution from the bright nucleus.

\subsection{The Galactic Nucleus} \label{s:agn}

The nucleus of M81 has long been an object of study. It is optically
 classified as a low-ionization nuclear emission line region (LINER) (Ho,
Filippenko, \& Sargent 1996). The dominant source of energy in LINERs may be
mechanical heating by shocks,
photoionization by hot stars, or photoionization by a low-luminosity active
galactic nucleus (AGN). The nucleus of M81 shows evidence of a
low-luminosity AGN including a compact radio core (Bietenholz \etal\ 2000),
broad H${\alpha}$ emission (Peimbert \& Torres-Peimbert 1981), a
UV-bright continuum (Devereux, Ford \& Jacoby 1997) with broad, AGN-like
emission lines (Maoz \etal\ 1998), and a power law X-ray continuum (Ishisaki
\etal\ 1996, Pelligrini \etal\ 2000). These are all consistent  with the
presence of a $6 \times 10^7$ \msun\ object, as inferred  from dynamical studies
(Bower \etal\ 2000), at the galactic center.

There have been numerous X-ray studies of the nucleus including \ein\ (Elvis \&
van Speybroeck 1982), {\sl GINGA} (Ohashi \etal\ 1992), {\sl BBXRT} (Petre
\etal\  1993), \ros\ (Radecke 1997), {\sl ASCA} (Ishisaki \etal\ 1996, Iyomoto
\& Makishima 2001), {\sl BeppoSAX} (Pellegrini \etal\ 2000), and \xmm\ (Page
\etal\ 2002). A summary of the current and previous X-ray observations of the
M81 nucleus is given in Table~4.
In addition to a power law with a slope of 1.9, similar to those of luminous
Seyfert~1 nuclei (Turner \& Pounds 1989, Nandra \etal\ 1997, Terashima \etal\
 2002), low-resolution X-ray
observations  suggest a soft thermal component is present in the nucleus.
While both advection-dominated accretion flows (ADAFs) and accretion disk
coronae can provide the necessary Comptonizing medium to produce the observed
power law X-ray spectrum, the source of the thermal component remains an
outstanding issue. This is also true in other wavebands. The UV continuum is
weak relative to X-rays and the ``big blue bump'' is absent in M81; perhaps a
manifestation of a low accretion rate (Ho \etal\ 1996) or the presence of an
ADAF (Quataert \etal\ 1999).

\begin{center}
\small{
{\sc Table 4} \\
{\sc X-Ray History of the M81 Nucleus} \vspace{6pt} \\
\begin{tabular}{clcccc}
\hline \hline
Date  & Obs. & $\Gamma^a$ & $N_H^b$ & $F^c$ & Ref \\ \hline
 1979 Apr  & \ein\   & $3.0_{-1.5}^{+2.0}$ & 68$\pm$58 &$\sim$0.1 & 1\\

 1985 Feb  & {\sl EXOSAT} & 2.1$\pm$0.2 & 19$\pm$9 & 2.65$\pm$0.25 & 2\\

 1987 May  & {\sl GINGA}  & 2.2$\pm$0.2 & 60$\pm$30 & 2.4$\pm$0.3 &  3\\

 1990 Dec  & {\sl BBXRT}  & $2.2_{-0.2}^{+0.3}$ & 41$\pm$10 & 3.6$\pm$0.3 &4\\

 1991~--~94& \ros\    & 2.5$\pm$0.3 & 7.4$\pm$0.4 & 1.24 & 2\\

 1993~--~99& \asca\ & 1.85$\pm$0.04 & $10.0$ & 3.50$\pm$0.12 & 5 \\

 1998 Jun  & {\sl SAX}  & 1.86$\pm$0.03 & $12.0_{-3.0}^{+3.6}$ & 3.8 & 6\\

 2000 May  & \cxo\  & 1.98$\pm$0.08    & 9.4$\pm$2.0 & 3.25$\pm$0.15 & 2\\

 2001 Apr  & {\sl XMM} & 1.94$\pm$0.06 & 3.4$\pm$0.8 & 0.93$\pm$0.06 & 7\\
\hline
\multicolumn{6}{l}{$^a$Power law photon index} \\
\multicolumn{6}{l}{$^b$Solar-abundance absorption column, in units of $10^{20}$
cm$^{-2}$}  \\
\multicolumn{6}{l}{$^c$2~--~10 keV observed flux, in units of $10^{-11}$
\ergcms} \\
\multicolumn{6}{l}{{\sc References.}-- (1) Fabbiano 1988
 (2) this work} \\
\multicolumn{6}{l}{(3) Ohashi \etal\ 1992 (4) Petre \etal\ 1993}\\
\multicolumn{6}{l}{(5) Iyomoto \& Maikshima 2001 (6) Pellegrini \etal\ 2000}\\
\multicolumn{6}{l}{(7) Page \etal\ 2002}\\
\end{tabular}
} 
\end{center}

Though past X-ray observations have been unable to resolve the nucleus from the
surrounding diffuse emission and point-like sources, Tennant \etal\ (2001) have
shown that some $\sim$$10^{39}$~\ergl, or a few percent, of the X-ray
luminosity in the nuclear region actually originates within a rather extended
($\sim2.\arcmin5$) region of the bulge. A similar conclusion has been reached
more recently by Immler \& Wang (2001) from analysis of \ros\ data and by Page
\etal\ (2002) from \xmm\ RGS spectral analysis.

The high X-ray flux from the nucleus leads to a severe pileup of events
in the ACIS detector, making the point-source spectrum unsuitable for study
(although J. Davis has successfully analyzed the nuclear spectrum; J. Davis,
 2002, private communication). In this work, the nuclear spectrum was instead
extracted from two 10$\arcsec$-wide rectangular regions spanning the readout
trail and offset $>$30$\arcsec$ from the nucleus to avoid contamination by
events in the wings of the nuclear PSF. The background was extracted from 4
similar rectangular regions adjacent to the readout trail.

A simple absorbed power law model provides a statistically acceptable fit to the
nuclear spectrum ($\chi^2 = 114$ for 104 dof; Figure~\ref{f:X5_spec}). The
resulting power law fit parameters are listed in Table~4. The addition of a
thermal component did not improve the fit significantly ($\Delta \chi^2 = 0.4$
for one additional parameter with the temperature parameter constrained to
$kT=0.5$~keV). Detection of Fe~K$\alpha$ emission could not be confirmed because
of the lack of source counts above $\sim$5~keV in the extracted spectrum. This
line was previously detected in {\em Ginga} (Ohashi \etal\ 1992), \asca\
(Ishisaki \etal  1996), {\em BeppoSAX} (Pelligrini \etal\ 2000), and,
recently, \xmm\ PN (R. Soria, private communication) observations of M81.

\begin{center}
\includegraphics[angle=-90,width=\columnwidth]{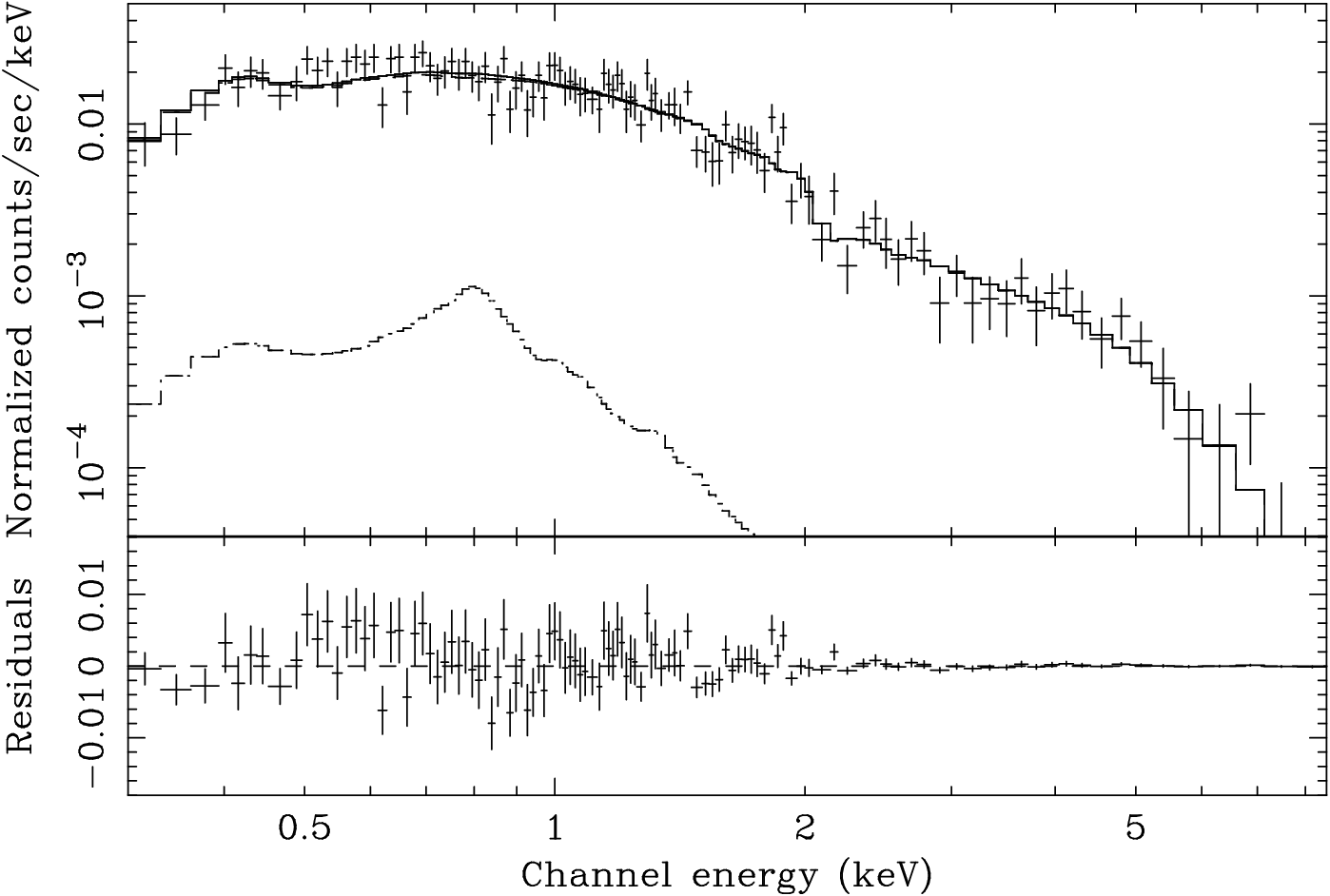}
\vspace{10pt}
\figcaption{Spectrum of the M81 nucleus extracted from the readout trail. Shown
are the contributions from a $\Gamma = 1.98$ power law ({\em solid line}) and a
weak thermal component ({\em dot-dashed line}). The addition of the thermal
component does not improve the fit significantly ($\Delta \chi^2 = 0.4$ for 1
additional parameter). \label{f:X5_spec}} \end{center}

The 90\% upper limit to the thermal model normalization corresponds to a maximum
thermal contribution to the nuclear flux of 2.8\% or a luminosity in the
 0.3~--~8.0~keV band of $\sim$$9\times 10^{38}$~\ergl. While this is only a
small fraction of the nuclear emission it is comparable to the total luminosity
from unresolved emission in the bulge (\S\ref{s:bulge}).
The lack of a significant thermal component in the \cxo\ nuclear spectrum is in
contrast to results based on low spatial resolution observations including
\asca, {\em BeppoSAX}, and \xmm\ that find one or more thermal
components improve the fit significantly. In fact, it is argued in
\S\ref{s:bulge} that any thermal emission from the nuclear region is
consistent with an extrapolation of the emission from the region
beyond $\sim$10$\arcsec$ from the nucleus and therefore is unrelated
to the nucleus.

 \subsection{Other X-ray-bright M81 Sources} \label{s:others}

In addition to SN~1993J, X6, the nucleus, and the brightest supersoft source
(Swartz \etal\  2002), there are seven X-ray point sources with luminosities
exceeding the Eddington limit for a  1.5~\msun\ accretor
in M81. Four of these are the \ein\ sources X7, X4,
X10, and X3 in order of decreasing brightness. These have all also been detected
in the \ros\ data (Table~2). In addition, source number~83 was detected by \ros.
The two remaining bright sources are confused with nearby sources in the \ros\
and \ein\ observations: Source number~116 is only  16$\arcsec$ from the nucleus
and source number~57 is  19$\arcsec$ from (and of comparable luminosity to) X10.
Thus, aside from the two confused sources, the brightest sources must be
relatively steady sources.

The X-ray spectra of these sources are unremarkable with the exception of
source X4 (number 86) which is extremely flat, $\Gamma=1.08\pm0.13$, and source
number 57 which has a relatively high column density, $N_{20}=99.1\pm40$.
Our analysis of the \ros\ PSPC spectra of these two sources is consistent with
these values (though source 57 is confused with source X10, number 52).
All seven sources are within the $D_{25}$ isophote of M81. Three are within the
bulge as is the brightest supersoft source (number 132). There are
no discernable optical or UV counterparts to any of these seven sources with the
exception of source number 146, \ein\ source X7, which is coincident with
 globular cluster number 63 of Chandar \etal\ (2001).

While the majority of these bright sources are persistent sources,
two \cxo\ sources of sub-Eddington luminosity are known to have been much more
luminous in the past. These are \cxo\ source numbers 67 (\ein\ source X2)
and~82. Ghosh \etal\ (2001) analyzed source number 82 in detail and noted that
both it and X2 are spatially-coincident with optically-bright objects which may
be bright M81 globular clusters. The bright transient source number 82 reached a
peak unabsorbed luminosity of $\sim7 \times 10^{38}$~\ergl\ during \ros\
observations but was observed at $1.7 \times 10^{37}$~\ergl\ in the \cxo\ data
(Ghosh \etal\ 2001). Source X2 is not known to have exceeded $\sim2\times
 10^{38}$~\ergl. All \ros\ sources in the tabulation of Immler \& Wang (2001)
falling within the \cxo\ field of view have \cxo-detected counterparts
(Table~2). No other bright transients are present in these two tabulations based
on a comparison of the \cxo\ count rates derived in this work and the \ros\
count rates tabulated by Immler \& Wang (2001).

\section{The M81 Bulge} \label{s:bulge}

The inner Lindblad Resonance, located at an inclination-corrected
radius of 4~kpc, separates the bulge and
inner disk from the spiral arms and outer disk
of M81 (e.g., Kaufman \etal\ 1989, Reichen \etal\ 1994).
This value is larger than the 2.5~kpc radius adopted by Tennant \etal\ (2001)
in their discussion of bulge and disk emission (and based on the observed X-ray
morphology). A 4~kpc radius circle in the plane of the galaxy corresponds to a
$7.\arcmin64 \times 3.\arcmin94$ ellipse on the plane of the sky with major
axis oriented at PA 149$^{\circ}$. 

The \cxo\ image shows the bulge X-ray sources
concentrated towards the galactic center 
and excess or unresolved X-ray emission extending away from the nucleus.
Analysis of this emission is made difficult by the bright
nucleus. 
While the high angular resolution of the \cha\ mirrors concentrates most of
the nuclear X-rays into the central few pixels, 
the small fraction incident in the wings of the PSF
accounts for a large percentage of the total number of X-rays detected at 
larger radii.

In this section, the spatial distribution of the bulge X-ray emission
is analyzed and compared to the distribution observed in other wavebands.
The spectral properties are then assessed to determine the possible
contributions from different sources of X-ray emission as a function of position
within the bulge.

Of particular interest is the central $\sim$30~--~50$\arcsec$ inner region
or core of the galaxy. Unlike the smooth distribution of optical light,
there is filamentary H$\alpha$ emission (Devereux, Jacoby \& Ciardullo 1995)
and excess UV emission (Hill \etal\ 1992, Reichen \etal\ 1994) in the core.
The origin of this emission is uncertain.
It has been attributed to recent star formation activity
(Devereux, Jacoby \& Ciardullo 1995)
but more likely originates from hot evolved post-AGB stars
(O'Connell \etal\ 1992, Devereux, Jacoby \& Ciardullo 1995).
High resolution \hst\ images rule out massive OB stars as a source of
ionization but ionization by shocks originating from nuclear activity
remains a viable alternative (Devereux, Ford, \& Jacoby 1997), a conclusion
also reached by Greenwalt \etal\ (1998).

 \subsection{Spatial Distribution}
  \subsubsection{Broad-band X-ray Surface Brightness}

The unresolved X-ray surface brightness is obtained by removing
the detected sources and the nuclear readout trail from the image.
The radial profile of this
emission is shown in Figure~\ref{f:rp_psf}.
The profile asymptotically approaches the background level of 
 0.04~cts~pixel$^{-1}$ at a radius of $\sim$2$\arcmin$ ($\sim$2~kpc).
Also shown is the estimated contribution from the nuclear point source
and the resulting profile with the nuclear contribution subtracted.
The contribution from the nucleus was estimated using a
model PSF appropriate for the
observed off-axis location of the nucleus and 
for 1.5~keV photon energies. This energy is above  
the $\sim$0.8~keV peak of the observed spectrum of the nucleus (\S\ref{s:agn})
but is representative and avoids interpolation between PSF models at 
 different discrete energies.
The radial profiles show a strong
 nuclear contribution  within the central $\sim$30$\arcsec$ but which
falls much more steeply than the observed excess at larger radii.
Toward the center of the galaxy, the estimated
contribution from the nucleus exceeds the observed surface brightness due to
pileup of nuclear photons. Analysis is therefore confined to the region beyond
$\sim$10$\arcsec$ from the nucleus.

\begin{center}
\includegraphics[angle=-90,width=\columnwidth]{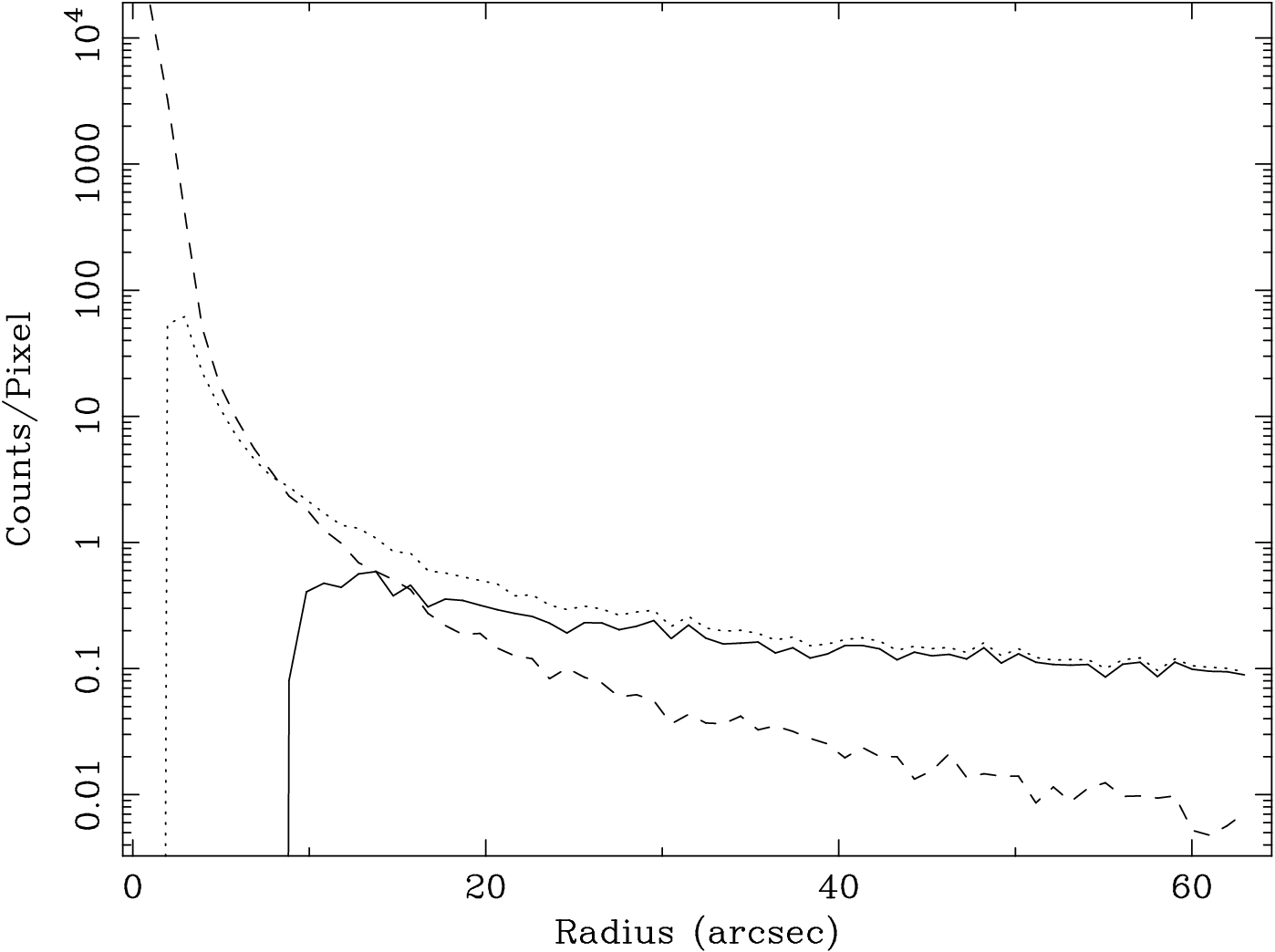}
\vspace{10pt}
\figcaption{Curves tracing the radial profile of the observed
X-ray surface brightness ({\em dotted}), the model point spread function
of the nucleus ({\em dashed}), and the difference of the two ({\em solid}).
Errors have been omitted for clarity. The background level is
$\sim0.04$~c~s$^{-1}$ \label{f:rp_psf}} \end{center}

The excess X-ray emission (with the estimated nuclear PSF contribution
subtracted) was examined to determine the shape of the
emission and to compare to profiles at other wavelengths.
Both the bulge and the underlying galactic disk can contribute to the 
excess X-ray surface brightness. 
If the excess
is confined to the disk, then its distribution
should appear elongated in an elliptical pattern consistent with the 
known inclination of M81. If the emission is from
the bulge, then its distribution on the
plane of the sky should appear azimuthally symmetric about the nucleus.

Azimuthal profiles extracted over a range of spatial scales show no significant
departures from azimuthal symmetry with one exception:
There is a slight ($\sim$30\%) increase in X-ray surface brightness
in a region located 30$\arcsec$ to 45$\arcsec$ to the northeast of the nucleus.
This is also the location of a
nonthermal highly polarized radio arc
that may be a small-scale nuclear radio lobe (Kaufman \etal\ 1996).  
A two-dimensional Gaussian model fit to the
excess X-ray surface brightness 
shows a slight elongation along PA 153$\pm$2.4$^{\circ}$ and an
eccentricity of 0.56. This is much less than expected from the galaxy disk
inclination (eccentricity 0.86) and
slightly smaller than the $\sim$0.66 measured
in the inner 0.1~--~1~kpc from
UBVR isophotes (Tenjes, Haud, \& Einasto 1998).
Qualitatively, this
suggests the excess X-ray surface brightness is emitted predominantly from
the spherical bulge but that the disk also contributes. 

A more quantitative estimate of the disk and bulge contributions can be made
by fitting the
radial profile of the excess X-ray surface brightness with a generalized
exponential of the form $\Sigma(r) = \Sigma_0 {\rm e}^{-(r/h)^{1/n}}$
(e.g., de~Jong 1996).
The PSF-corrected and background-subtracted
X-ray surface brightness profile
is shown in Figure~\ref{f:rp_genexp} along with the best fit curves 
for $n\equiv1$,~2, and~4.
None of the fits are statistically acceptable, with $\chi^2$ values 
ranging from 160.4 to 203.1 for 116 dof. Allowing $n$ to vary improved the
$\chi^2$ value only slightly to 154.6 for the best-fit value $n=1.56\pm0.30$.
This value is intermediate between an exponential disk ($n=1$) and a
de Vaucouleurs bulge ($n=4$ or $R^{1/4}$ law) profile. 
Fitting only the region beyond 1$\arcmin$ of
the nucleus reduced $\chi^2$ substantially for $n=1$ (65.4 for 65 dof) and
results in a best-fit value of $n=1.1\pm0.1$ ($\chi^2=65.3$ for 64 dof).
The X-ray emission at large radii therefore generally follows the exponential
disk profile typically seen in optical light from spiral galaxies
but steepens to a bulge-dominated profile near the nucleus.

\begin{center}
\includegraphics[angle=-90,width=\columnwidth]{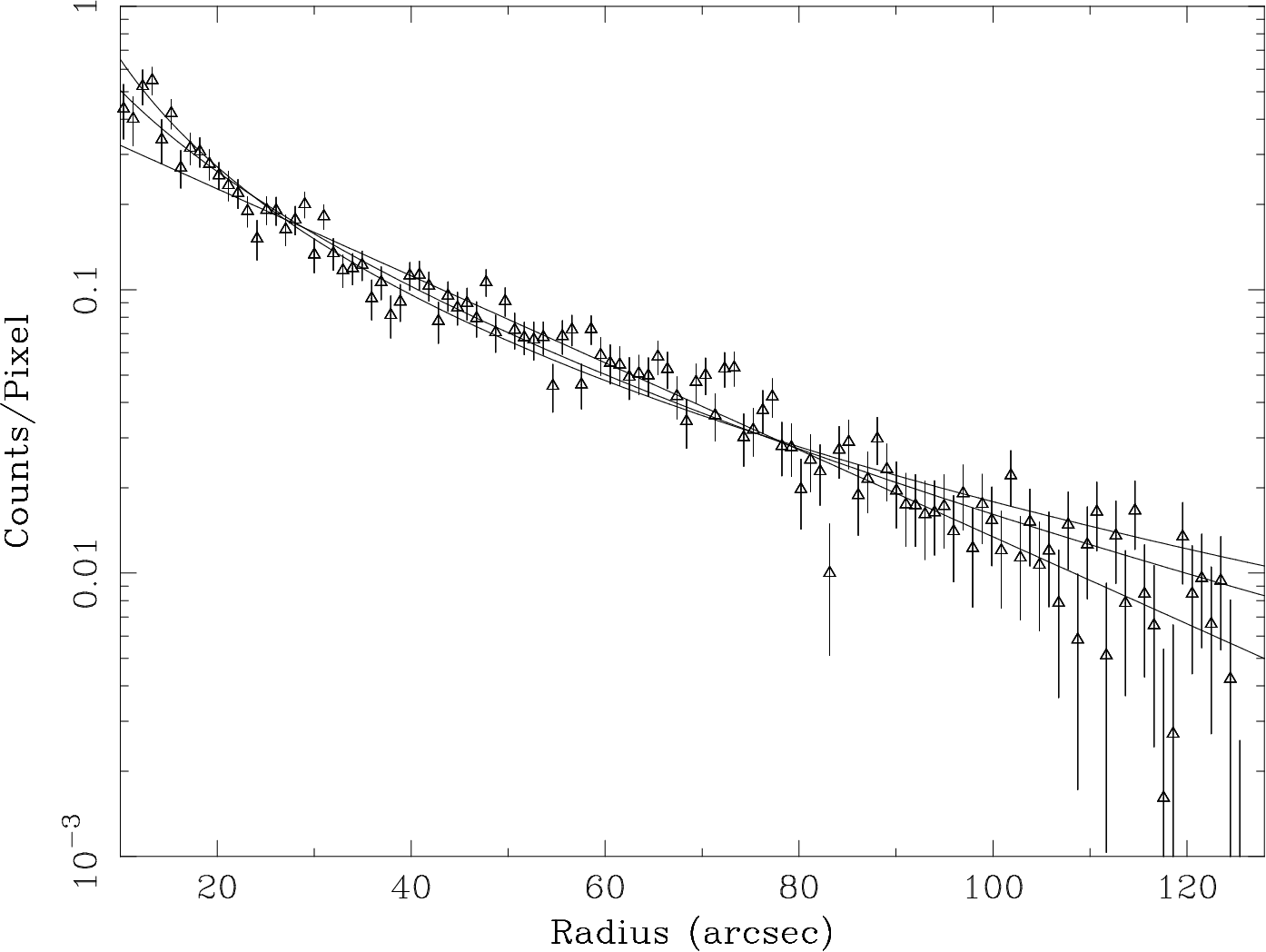}
\vspace{10pt}
\figcaption{Background-subtracted and PSF-corrected excess X-ray surface 
brightness radial profile. Curves represent best fits using a generalized 
exponential function of the form $\Sigma(r) = \Sigma_0 {\rm e}^{-(r/h)^{1/n}}$
with $n=1$, 2, and 4. (Curvature increases with increasing $n$; the
exponential, $n=1$, is the straight line). \label{f:rp_genexp}}
\end{center}

\begin{center}
\includegraphics[angle=-90,width=\columnwidth]{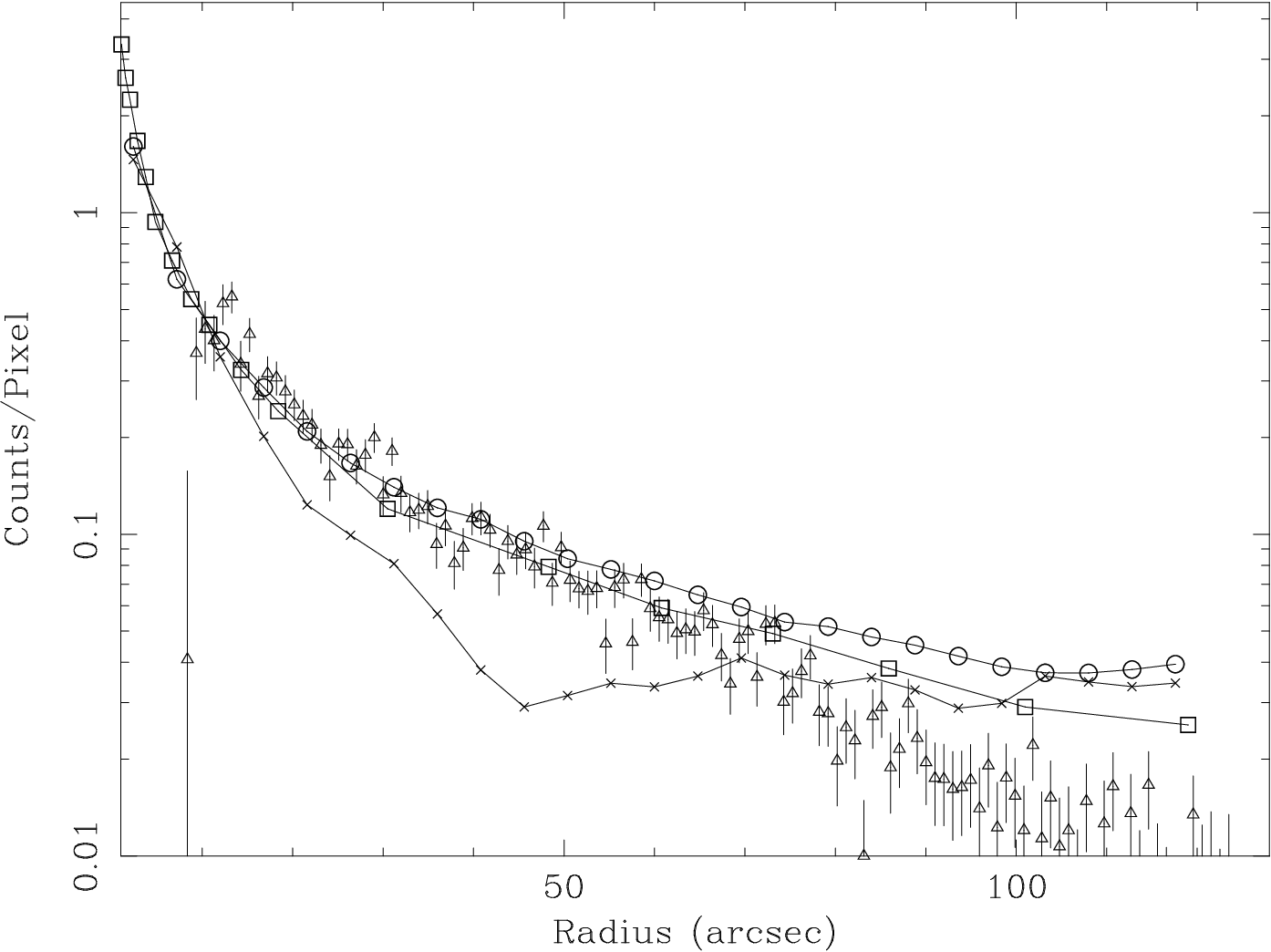}
\vspace{10pt}
\figcaption{Background-subtracted and PSF-corrected excess X-ray surface 
brightness radial profile ({\em triangles}) with optical ({\em squares}),
near-UV ({\em circles}), and far-UV ({\em crosses})
profiles. Non-X-ray profiles have been scaled vertically to match the X-ray
data at a radius of 10$\arcsec$ for comparison. \label{f:rp_optical}}
\end{center}

Figure~\ref{f:rp_optical} compares the X-ray surface brightness profile to
those observed at other wavelengths. 
The X-ray surface brightness follows the optical (courtsey P. Tenjes 2002,
private communication) and
near-UV\footnote[8]{from archival
Ultraviolet Imaging Telescope data
available from http://archive.stsci.edu/uit/index.html} (centered at 2490~\AA)
light from old bulge stars out to $\sim$80$\arcsec$ but declines more rapidly at
larger radii. In contrast, the far-UV (1520~\AA) profile falls rapidly to the
background level beyond 50$\arcsec$ of the nucleus (Hill \etal\ 1992). The
resulting UV color gradient has been interpreted as a gradient in metal
abundance in the inner $\sim$50$\arcsec$ (O'Connell \etal\ 1992). There is no
indication of a similar enhancement in the X-ray data but, again, the morphology
within the inner $\sim$10$\arcsec$ cannot be determined owing to the X-ray
brilliance of the nucleus. 

  \subsubsection{Resolved Sources}

The radial distribution of the observed counts (excluding the nucleus) in the
resolved X-ray population superposed
on the excess X-ray surface brightness profile is displayed in
Figure~\ref{f:rp_srccts}.
The profiles are
remarkably similar with the exception of the single high bin
due to a single bright (supersoft) source (number~132) located $\sim$50$\arcsec$
from the nucleus.

The radial distribution of the 53 resolved bulge X-ray sources (per unit area)
is flatter than the surface brightness profile.
There is, however, a bias in the source
detection efficiency in the center of the galaxy where the excess X-ray
surface brightness is highest and acts as an
increased background for source detection.
This naturally causes a flattening in the resolved source number distribution.

\begin{center}
\includegraphics[angle=-90,width=\columnwidth]{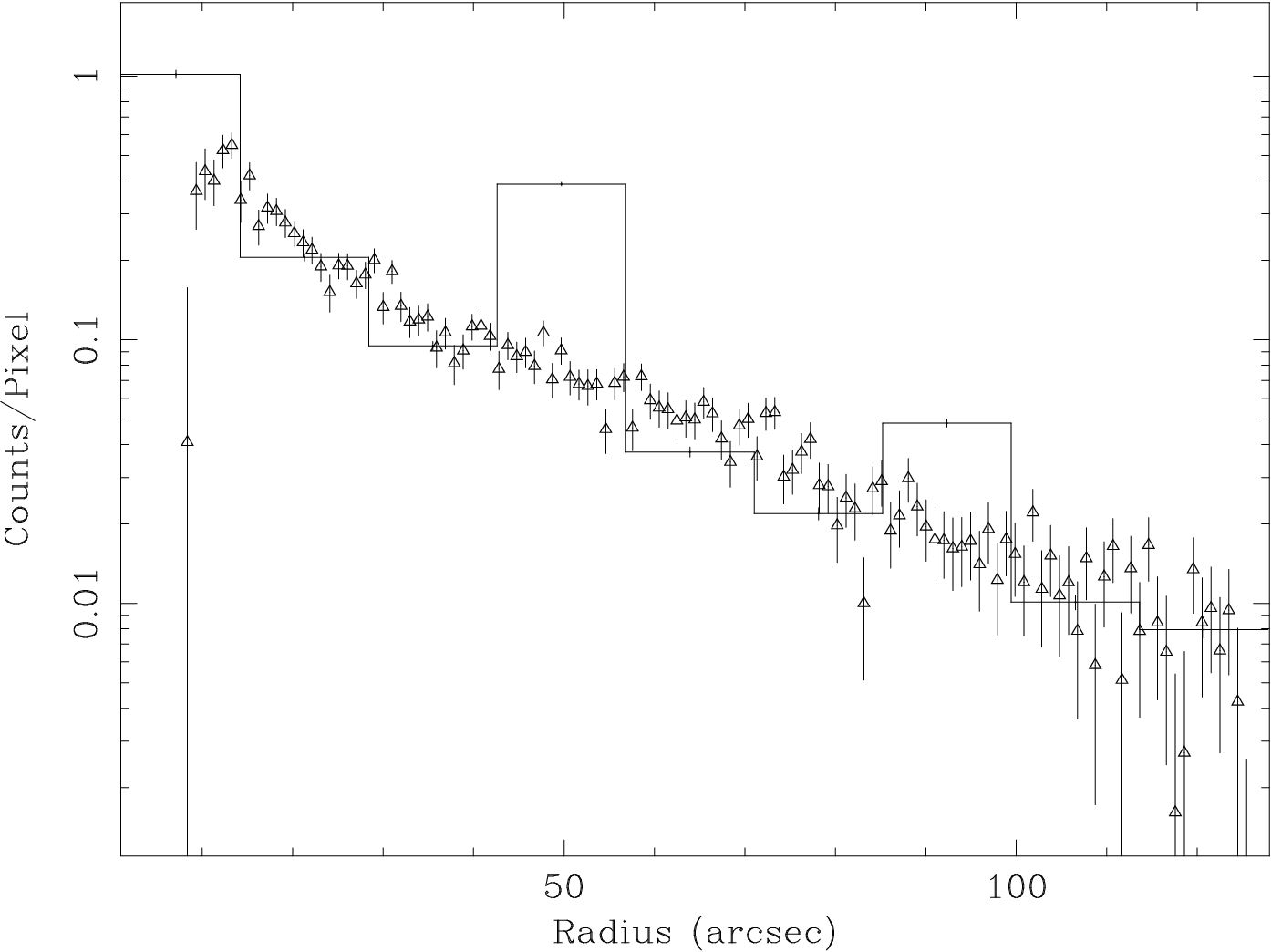}
\vspace{10pt}
\figcaption{Background-subtracted and PSF-corrected excess X-ray surface 
brightness radial profile ({\em triangles}) with histogram distribution of X-ray
counts detected in resolved 
X-ray sources (in units of c~kpc$^{-2}$) overplotted.
Profile of resolved source counts has been scaled vertically by 
$2.5\times10^{-4}$. The high bin at 50$\arcsec$ is due mostly to
a single source contributing $\sim$4000 X-ray counts.
\label{f:rp_srccts}}
\end{center}

 \subsection{X-ray Spectrum of the Unresolved Component}

The unresolved X-ray emission may be composed of
unresolved point sources and of diffuse, shock-heated, gas.
The spectrum of the unresolved X-ray  emission was
extracted from 15$\arcsec$-wide annuli centered on the nucleus with the
inner radius of the first annulus equal to 10$\arcsec$. At least three spectral
components are included in spectral fits for each annulus:
a nuclear contribution, a power law, and a
thermal component.

The contribution from the wings of the PSF of the nucleus can be estimated by
scaling the spectrum obtained from the nuclear readout trail (\S\ref{s:agn}) by 
the fraction of the PSF falling within the chosen annulus.
However, this scaling preserves the shape of the nuclear spectrum and
does not account for the spectral flattening that
occurs because of the energy dependence of the PSF.
The deep calibration observation of the point source LMC X-1 (obsid 1422) was
used to model this effect. The readout trail image of LMC X-1 was extracted
along with spectra from various annuli surrounding the source. The
channel-by-channel ratio of these spectra shows a linear rise with energy
reflecting the PSF energy
dependence. Applying this result to the nucleus of M81 modifies the
power law form of the readout-trail spectrum to a function of the form
$E^{-\Gamma}(aE+b)$ with the constants $a$ and $b$ dependent on the annular
region under study and the power law index $\Gamma$ determined by the shape
of the readout trail spectrum (\ref{s:agn}). Subsequently, all parameters for
the nuclear contribution to the unresolved emission are held constant while
fitting the latter spectra.

The contribution from unresolved sources is assumed to have the same shape
as that of the resolved sources
but with the model normalization left as a free parameter.
The spectra of the resolved bulge sources, with the exception of the
nucleus and the bright supersoft source, were added to obtain the total
bulge source spectrum and then fitted to an absorbed power law. The resulting
column density is $N_{20}=7.7$ and the photon index is $\Gamma=1.6$
($\chi^2=156.6$ for 176 dof). This slope is typical of individual
intermediate-brightness sources for which reliable spectral fits could be made
(\S\ref{s:discrete_spec}). The total flux is $9.2 \times 10^{-13}$ \ergcms\
corresponding to a luminosity of $1.43 \times 10^{39}$~\ergl\ ($1.64 \times
 10^{39}$~\ergl\ unabsorbed).

The third
component is assumed to be thermal because, when the data is fit with only
the first two components, the bulk of the residual lies at low energies
suggestive of a thermal contribution. Models with additional
power law or thermal components were also tested. These
components might represent a contribution from weak unresolved sources
dissimilar in spectral shape to the resolved sources. None
of these added components improved the fit statistic significantly.

A typical spectrum and best-fit model components are displayed in
Figure~\ref{f:diffuse_spec}. For this annulus, the hard X-ray flux above
$\sim$1.5~keV is dominated by the nuclear contribution but the thermal and power
law contributions are clearly present.

\begin{center}
\includegraphics[angle=-90,width=\columnwidth]{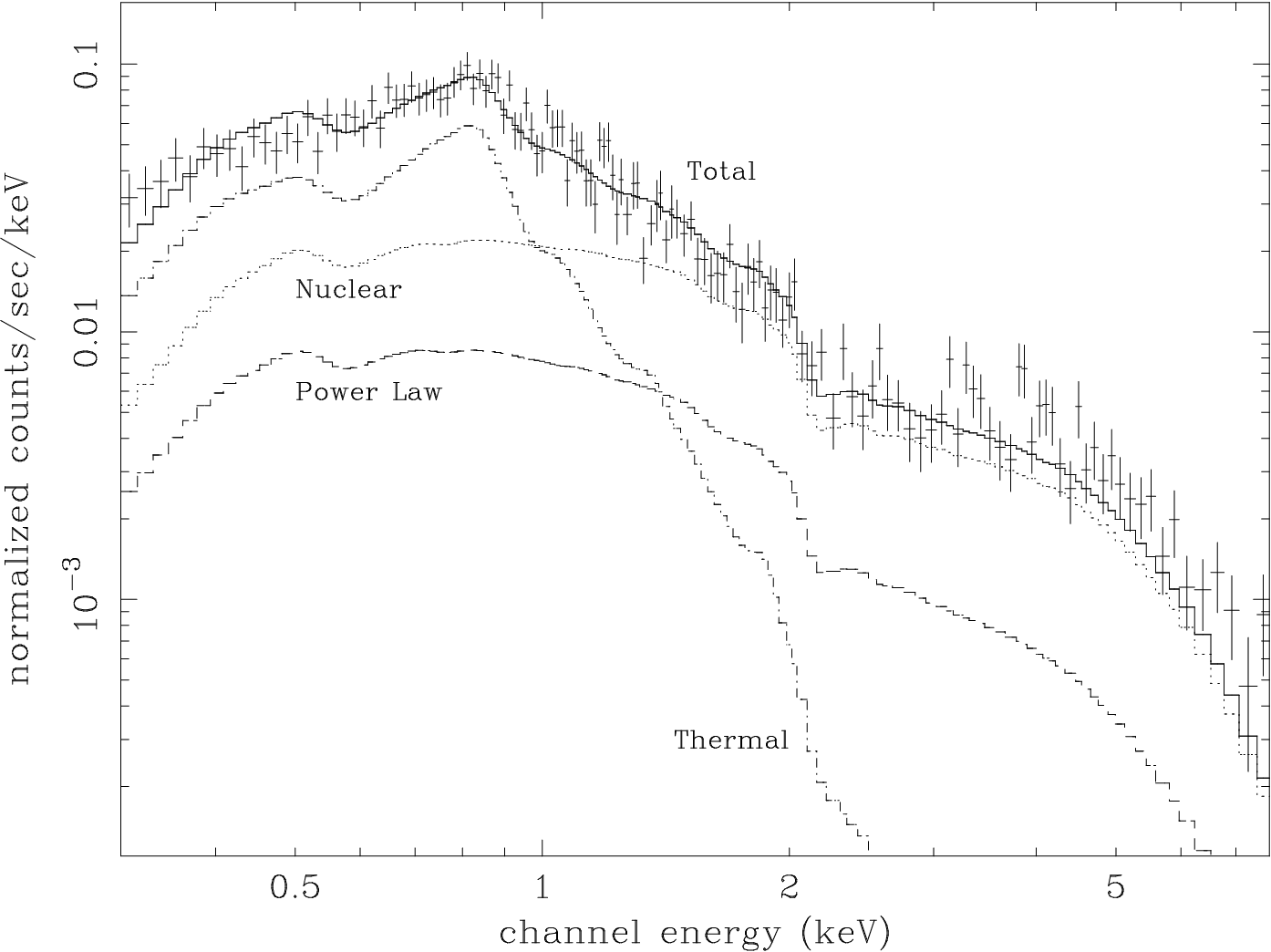}
\vspace{10pt}
\figcaption{Spectrum of the unresolved emission from an annulus spanning 10 to
 30$\arcsec$ radius from the center of M81. Shown are the contributions from the
nucleus, a power law representing unresolved point sources, and a thermal
component.
\label{f:diffuse_spec}}
\end{center}

Figure~\ref{f:diffuse_kt}
shows that the temperature profile of the thermal component decreases with
distance from the nucleus ($\left\langle kT \right\rangle =0.38\pm0.07$~keV
within 60$\arcsec$ radius and $\left\langle kT \right\rangle = 0.26\pm0.05$~keV
between 60$\arcsec$ and 120$\arcsec$, $\chi^2=7.2$ for 5 dof) and that the
absorbing column density increases away from the nucleus
(from $N_{20}=9.1\pm2.6$ to $21.2\pm5.3$ on the same ranges, respectively,
$\chi^2=13.2$ for 5 dof).
Assuming a homogeneous distribution of hot gas is responsible for the thermal
emission, its density and mass can be estimated based on the shape of the X-ray
emitting region discussed above, namely a spherical bulge in the inner
$\sim$1~kpc surrounded by a region dominated by disk emission. The spectral fit
parameters imply the number density of the gas is $n_e\sim0.01$~cm$^{-3}$ but
rises to $n_e \sim 0.07$~cm$^{-3}$ in the innermost annulus and that the total
mass of hot gas within the volume extending $\sim$2$\arcmin$ from the nucleus is
$\sim7\times10^6$~\msun. This is a crude estimate because the thickness of the
disk is unknown and was taken to be 1~kpc or roughly the thickness of the
Galactic disk. The derived mass is a small fraction of the total mass of the
bulge, $\sim3 \times 10^{10}$~\msun\ (Tenjes \etal\ 1998). The total thermal
energy in the hot gas is $\sim$$7\times 10^{54}$~erg or about  0.3\% of the
kinetic energy in bulge stellar motion (see Tenjes \etal\ 1998).

\begin{center}
\includegraphics[angle=-90,width=\columnwidth]{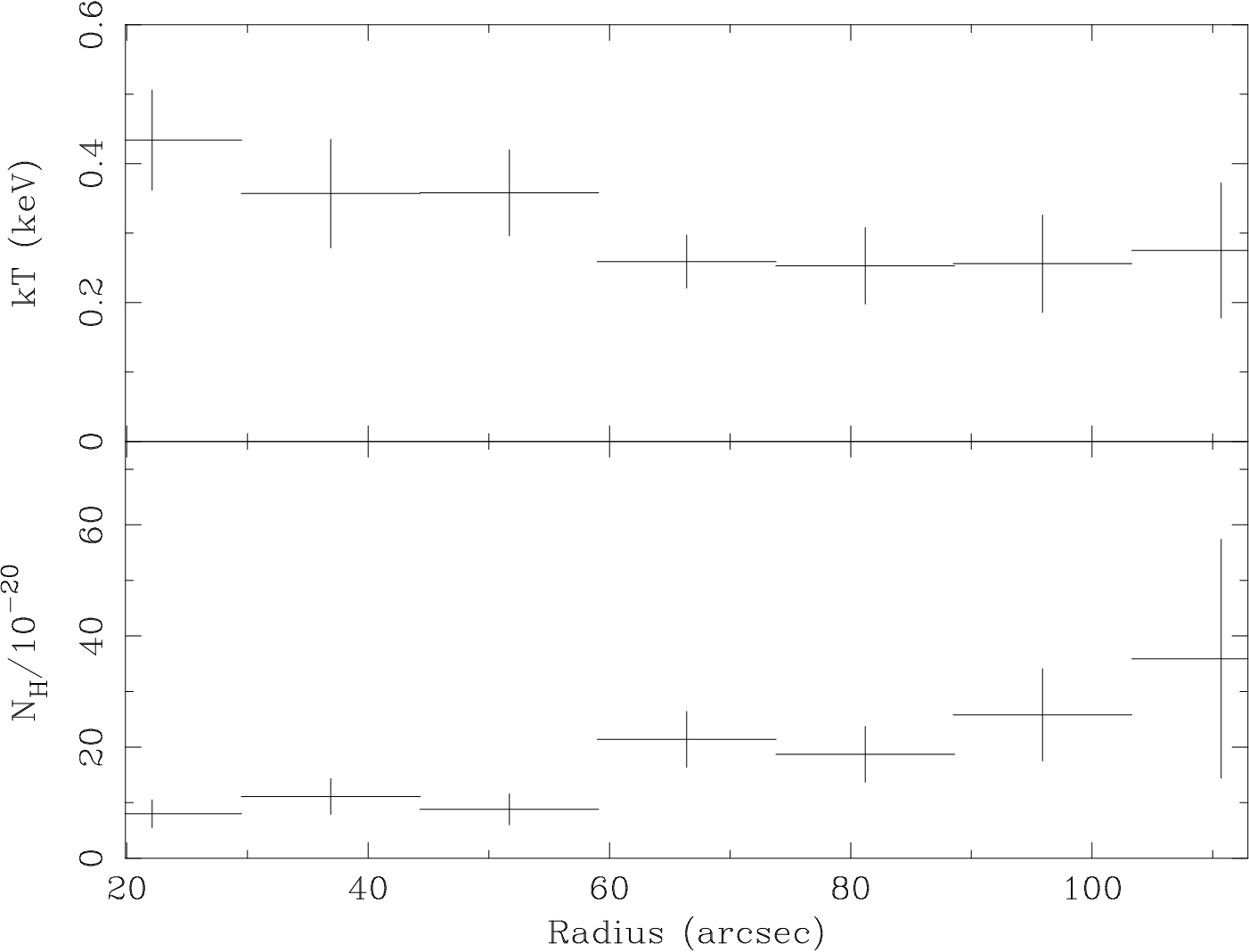}
\vspace{10pt}
\figcaption{Radial variation of the best-fit temperature ({\em top}) and
column density ({\em bottom}) for the thermal contribution to the unresolved
emission from the bulge. \label{f:diffuse_kt}} \end{center}

Page et al. (2002a,b) argued that the unresolved emission
is from collisionally excited, optically thin plasmas based
the RGS data obtained by \xmm. A model consisting of
 3 mekal components (with $kT = 0.18\pm0.04,~ 0.64\pm0.04$, and
$1.7\pm 0.2$~keV, respectively) and an absorbed power-law
(with $\Gamma \approx 1.95$, representing the nuclear contribution)
provides a
good fit to the RGS data.
From the ratio of the
forbidden line to the intercombination lines in the
O VII triplets they rule out the alternative photo-ionization
model at the 95\% confidence level. The ratio of the
resonance line to the other lines in the O VII triplet
also put an upper limit of $10^9$~cm$^{-3}$
to the electron-number density of the line emitting gas.
This result together with the luminosity of diffuse gas
inferred from the Chandra data implies that the gas is
not uniformly distributed and it fills only a small fraction
of the available volume.

Figure~\ref{f:diffuse_lx} displays the individual contributions to the X-ray
luminosity of the unresolved excess emission as a function of the distance from
the center of the galaxy. Both the thermal and power law contributions are
relatively flat compared to the nuclear contribution whose spatial-dependence is
dictated by the shape of the PSF.
Also shown in
Figure~\ref{f:diffuse_lx} is the possible contribution from thermal emission at
the center of the galaxy deduced from analysis of the nuclear readout trail
spectrum of \S\ref{s:agn}. This emission is consistent with an extrapolation of
the thermal and/or power law emission from the bulge and therefore is not
necessarily intrinsic to the active nucleus.

\begin{center}
\includegraphics[angle=-90,width=\columnwidth]{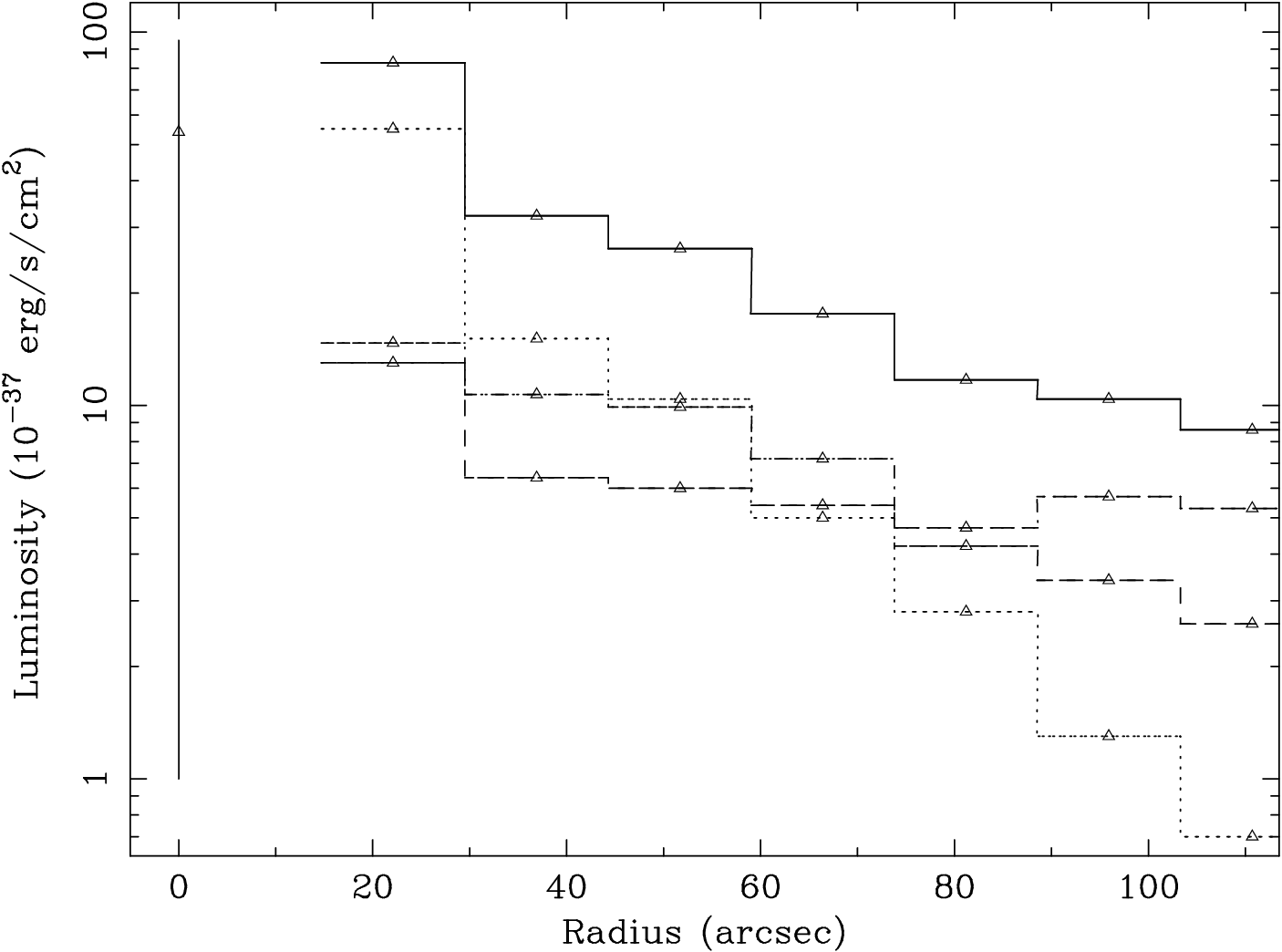}
\vspace{10pt}
\figcaption{Radial dependence of the X-ray luminosity of the unresolved
emission from the bulge. Shown are the contributions from thermal ({\em
dashed}), power law ({\em dot-dashed}), and nuclear PSF ({\em dotted}) model
components. The solid line represents the total luminosity. Also shown is the
estimated thermal emission component at the nucleus obtained by adding a thermal
model to the model of the spectrum extracted from the nuclear readout trail
(\S\ref{s:agn}) \label{f:diffuse_lx}} \end{center}

The total absorbed
luminosity in the power law component is $5.8\times 10^{38}$~\ergl\ and
$4.2\times 10^{38}$~\ergl\ in the thermal component. The total is
less than the $1.6\times10^{39}$~\ergl\ emitted by the discrete bulge sources.

The expected contribution of unresolved discrete sources to the unresolved
emission can be estimated by extrapolating the observed
luminosity function of the resolved bulge sources to lower luminosity.
The bulge luminosity function is displayed in Figure~\ref{f:bulge_3LF}. Also
shown are the luminosity functions of the subsets of bulge sources within
 1$\arcmin$ and beyond 1$\arcmin$ of the nucleus. The only systematic
 differences between the three functions is a flattening of
the distribution at low luminosities for sources near the nucleus. This is
caused by the loss of sensitivity in this region due to the
large contribution of the nucleus to the underlying background. Tennant \etal\
(2001) pointed out the break in the background-subtracted bulge luminosity
function at $\sim$$4\times 10^{37}$~\ergl\ ($\sim$200 counts). The luminosity
function shown here includes 13 more X-ray sources because of the larger
ellipse used to define the bulge and is not corrected for background sources.
Nevertheless, the break can still be seen. Another change in slope can be
envisioned at about 60 counts. This change caused by the loss of sensitivity
near the center of the bulge. The luminosity function is a power law,
$N(>C)=(194.1\pm12) C^{-0.50\pm0.02}$ between 13 and 200 counts and
$N(>C)=(135.6\pm5.0) C^{-0.37\pm0.01}$ between 13 and 60 counts. Extrapolating
these two functions to lower luminosities implies unresolved sources of the type
contributing to the resolved-source luminosity function account for only about 4
to 8\% of the excess X-ray counts or an equivalent fraction of the luminosity
(assuming the spectral shape is preserved).

\begin{center}
\includegraphics[angle=-90,width=\columnwidth]{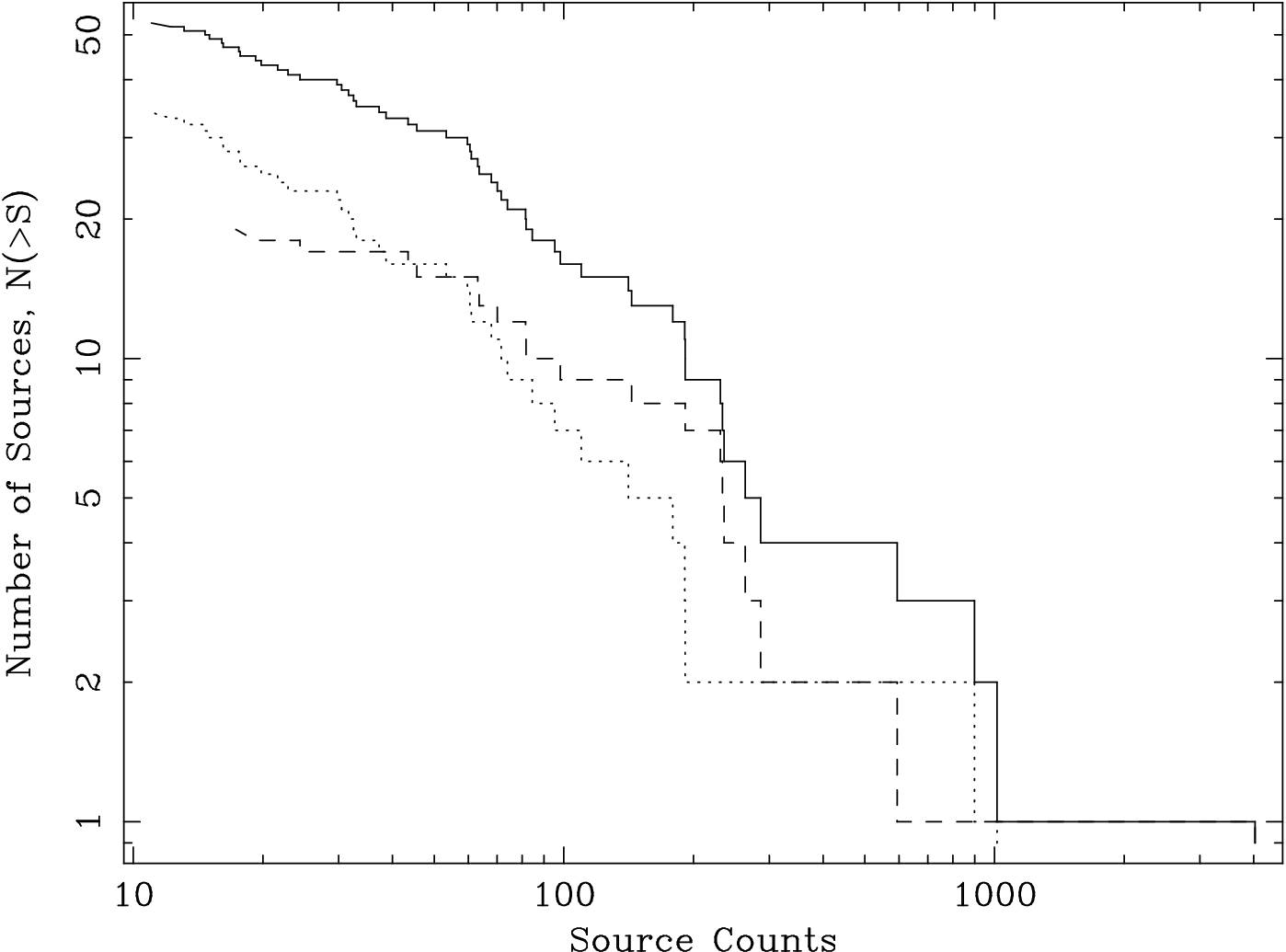}
\vspace{10pt}
\figcaption{Luminosity function for bulge sources ({\em solid}).
Also shown are the luminosity functions for sources lying within 1$\arcmin$
of the nucleus ({\em dashed}) and outside ({\em dotted}) this radius.
Note the lack of weak detected sources in the region near the
nucleus due to the high contribution from the nucleus and from the X-ray excess
to the source background in this region. The break in the luminosity function at
$\sim200$ counts corresponds to a luminosity of $\sim4\times 10^{37}$~\ergl.
\label{f:bulge_3LF}} \end{center}

In summary, the spectrum of the unresolved emission probably has contributions
from weak, unresolved sources and from diffuse hot gas. While spectral models
show roughly equal
contributions from both these components, extrapolation of the
resolved-source luminosity function predicts only a small contribution
from unresolved discrete sources. Perhaps
another, distinct, population of weaker sources is present and contributes
substantially to the unresolved emission. These sources must be quite weak (and
hence numerous) as no such population has been resolved in observations of
the nearby galaxy M31 to a limiting luminosity of $\sim6 \times 10^{35}$~\ergl\
(Shirey \etal\ 2001).  The thermal component needed to fit the observed
excess X-ray spectrum may also be from a collection of weak sources or from
truly diffuse gas at $kT\sim0.4$~keV. The mass and thermal energy content of any
X-ray-emitting gas is small compared to the total mass and energy confined in
the bulge. Unlike the distribution of far-UV light, both the unresolved emission
and the resolved bulge point-source population trace the optical profile from
old bulge stars and are confined by the same gravitational potential.
In particular, there is no evidence of enhanced activity in the core such as
heating by shocks as may explain the wispy H$\alpha$ emission (Devereux \etal\
 1995, 1997).

\section{The M81 Disk and Spiral Arms} \label{s:disk}

In contrast to the bulge,
there is no measurable excess X-ray emission in the disk imaged by \cxo. In
particular, there is no (unresolved) X-ray signature of the spiral arms which 
are seen clearly at other wavelengths and in X-rays in other galaxies such as
M83 (Soria \& Wu 2002).

 \subsection{Spatial Distribution \& Luminosity Function}

 The spiral arms are perhaps most
clearly seen in UV images 
(compare, e.g., Reichen \etal\ 1994 or Hill \etal\ 1995) 
where the population of young stars trace the principal features of
the spiral arms. The UV provides a high contrast 
by not sampling the underlying disk component
that is clearly seen in optical light. The
archival near-UV image obtained by the Ultraviolet Imaging Telescope,
with approximately 3$\arcsec$ resolution,
was used to define the locations of the
spiral arms. Distances between X-ray source positions and the nearest 
spiral arm was measured. Sources within the $7.\arcmin64$ major-axis
ellipse defining the bulge/disk interface (\S\ref{s:bulge}) were excluded as
were those sources exterior to the $D_{25}$ isophote at $26.\arcmin9$. Source
distances from spiral arms are displayed in Figure~\ref{f:disk_ctsVSdis}
against the number of detected X-ray counts. Average distances
in three luminosity ranges are also shown. Clearly, the average distance to
a spiral arm decreases as the source luminosity increases.

\begin{center}
\includegraphics[angle=-90,width=\columnwidth]{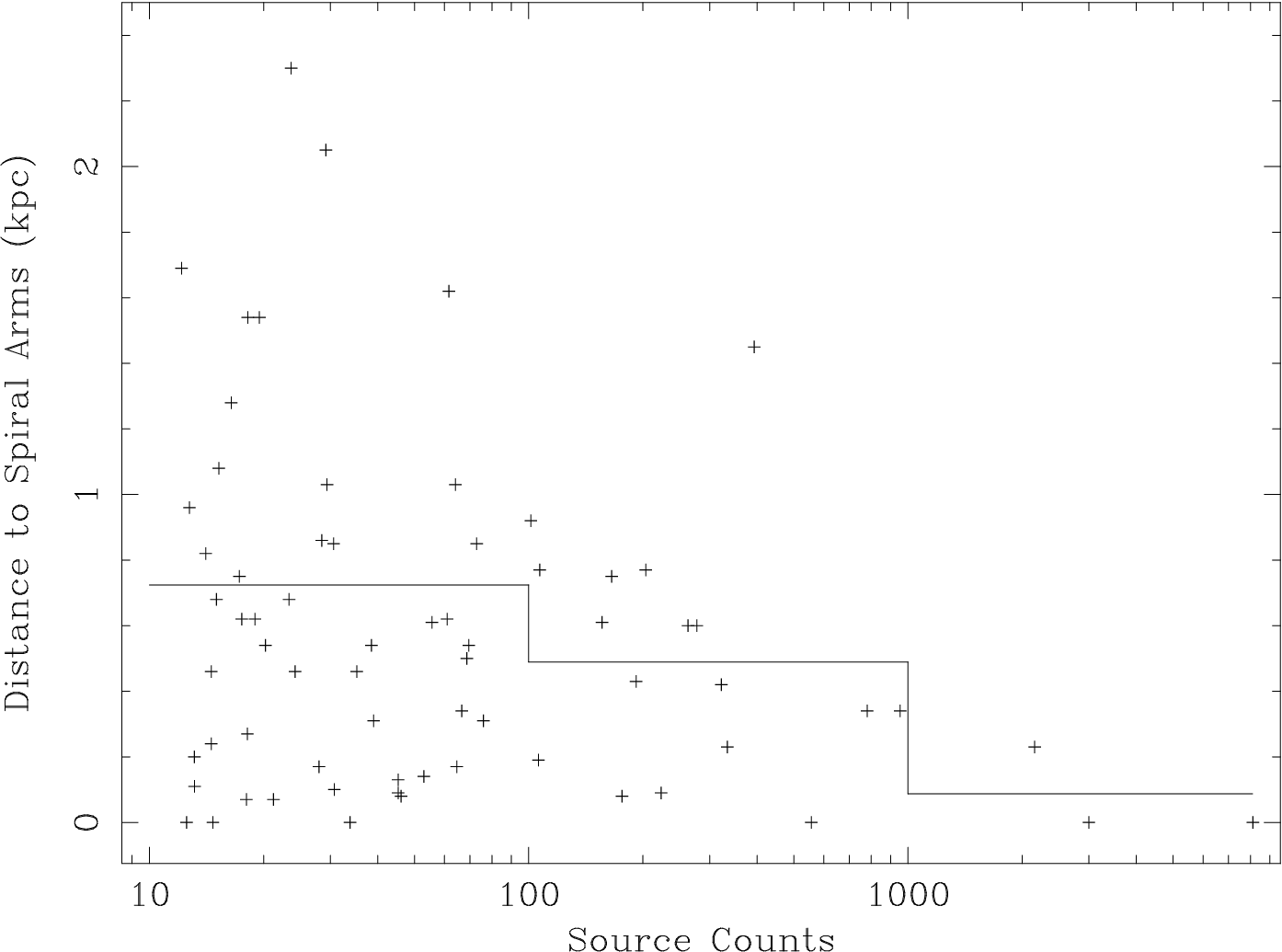}
\vspace{10pt}
\figcaption{Distribution of resolved X-ray sources in the disk. Shown is
the distance from the nearest spiral arm against the (background-corrected)
source counts. The line shows the average distance to spiral arms for sources
in the 10-100, 100-1000, and $>$1000 count ranges (100 counts corresponds to an
unabsorbed luminosity of $\sim2.4\times10^{37}$~\ergl). \label{f:disk_ctsVSdis}}
\end{center}

\begin{center}
\includegraphics[angle=-90,width=\columnwidth]{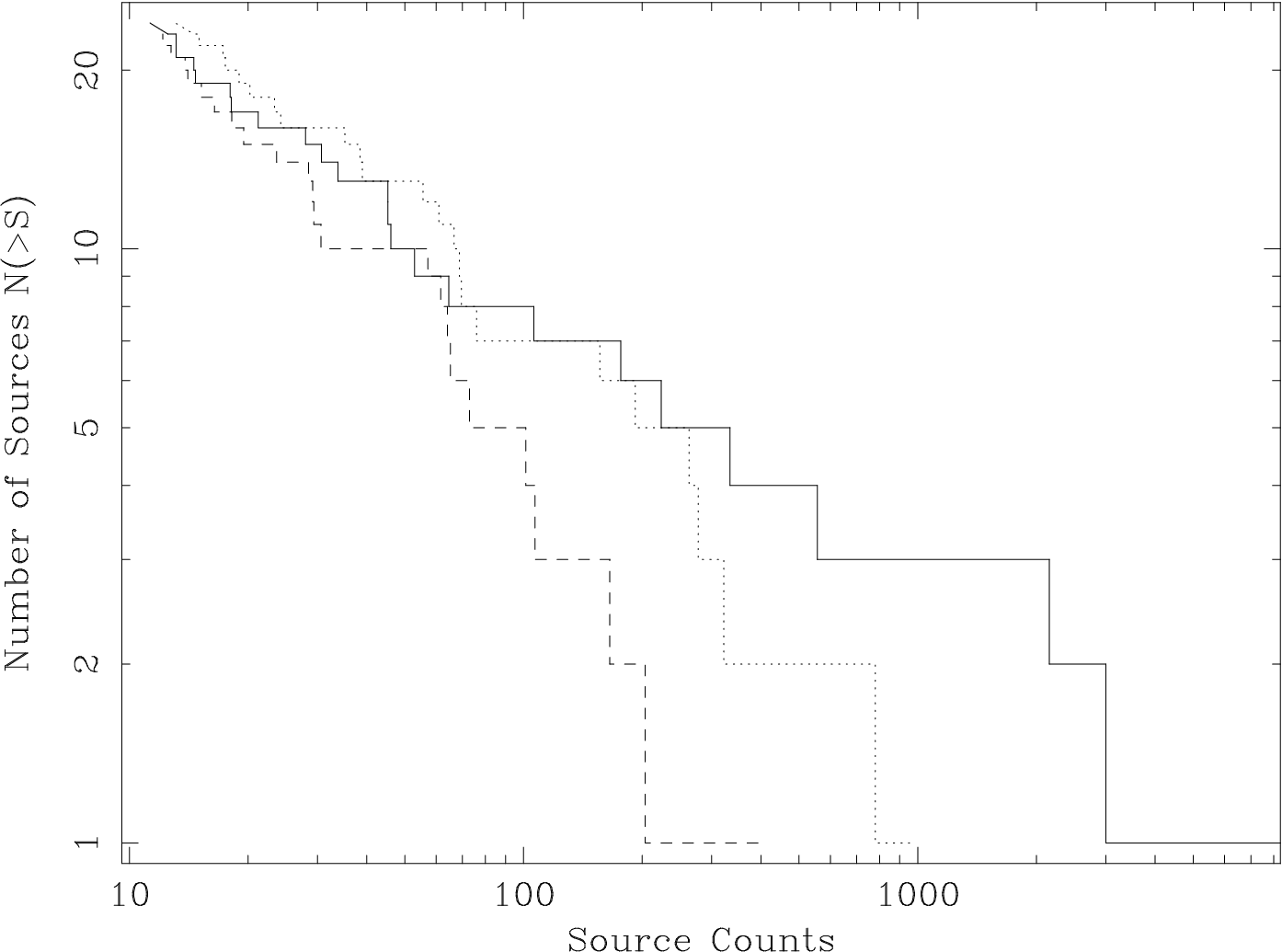}
\vspace{10pt}
\figcaption{Luminosity function for disk sources.
Shown are the luminosity functions for sources within 0.3~kpc of spiral arms
({\em solid}), sources between 0.3 and 0.75~kpc ({\em dotted}), and 
sources more than 0.75~kpc away from spiral arms ({\em dashed}).
\label{f:disk_3LF}}
\end{center}

The spatial dependence of the luminosity function for the disk sample
is shown in Figure~\ref{f:disk_3LF}.
The sample of 72 X-ray sources located interior to the  $D_{25}$ isophote
and exterior to the bulge/disk interface was binned into three groups with an
equal number of sources in each group corresponding to sources within 0.30~kpc
of a spiral arm, those 0.30~--~0.75~kpc, and those located $>$0.75~kpc from
spiral arms.
Note the strong dependence of the shape of the luminosity function
on distance from spiral arms exhibited in Figure~\ref{f:disk_3LF}.
Sources located near spiral arms are expected to be associated with the young
population of stars recently formed following passage through the density wave
while those further from the spiral arms should be related to a relatively
older population. Thus the observed X-ray luminosity functions for these 
populations suggests the younger population contains brighter objects and
exhibits a luminosity function with a constant power law index
($N(>C)=(75.7\pm5.9) C^{-0.48\pm0.03}$). Those further
from the spiral arms, in contrast, are dominated by weaker sources and their
luminosity functions steepen above $\sim$250 and $\sim$100 counts,
respectively (luminosities $\sim$$6\times 10^{37}$~\ergl\ and
$\sim$$2.4\times10^{37}$~\ergl). A power law fit to the full data range for
sources between 0.30 and 0.75~kpc gives $N(>C)=(100.3\pm10.8) C^{-0.54\pm0.03}$
and for those furthest from the spiral arms, $N(>C)=(127.1\pm16.6)
C^{-0.69\pm0.04}$.

 \subsection{X-ray Spectra}

The spectra of the 44 resolved disk sources imaged on device S3, with the
exception of SN~1993J  and source X6, were added to obtain a representative
disk source spectrum to compare to the bulge source spectrum. An absorbed
power law with the addition of a thermal component was an improvement
over an absorbed power law alone ($\Delta\chi^2=33$ for 3 additional
parameters). The resulting column density is $N_{20}=8.1$, the photon index is
$\Gamma=1.4$, and the thermal component temperature and abundance are
$kT=0.22$~keV and $Z=0.2Z_{\odot}$ ($\chi^2=193$ for 172 dof). The thermal
component accounts for $\sim$3\% of the total 0.3~--~8.0~keV flux. The power law
slope is flatter than typical of individual sources for which
reliable spectral fits could be made ($\Gamma=1.5$, \S\ref{s:discrete_spec}) and
of the composite spectrum of the resolved bulge sources ($\Gamma=1.6$). The
bulge spectrum also did not require a thermal component. The need for a thermal
component suggests some of the weaker resolved disk sources may be X-ray-bright
SNRs undetected in radio or in the survey of Matonick \& Fesen (1997).
Alternatively, a thermal contribution from black hole binaries in
high soft states cannot be discounted. On the other hand, the thermal component
contributes to the spectrum only at Fe~L and at lower energies. A large range of
column densities among the individual sources could also mimic this effect.

\section{Discussion} \label{s:discussion}
\subsection{The Discrete X-ray Source Population}

The X-ray spectra of the brightest sources in the M81 field are predominantly
moderately-absorbed power laws with photon index $\Gamma\sim1.5$. A similar
spectral shape reproduces the combined bulge resolved-source spectrum while the
combined disk source spectrum (for sources on S3) requires an additional weak
thermal component. A power law is indicative of accreting X-ray binaries (XRBs)
and a large population of bright power law sources is consistent with surveys
of our Galaxy and the Local Group where the bright X-ray source population is
dominated by low-mass XRBs ({\sl e.g.}, Grimm \etal\ 2001).

The scarcity of X-ray sources detected in the radio, or
correlated with optically-selected SNRs, or exhibiting a strong thermal X-ray
spectrum implies an insignificant number of the bright X-ray sources in M81 are 
SNRs. Supernova remnants are common in the Magellanic Clouds and in the solar
neighborhood but are relatively less luminous and short-lived compared to
typical XRBs. Perhaps some of the weaker disk sources are SNRs and account for
the thermal emission present in the combined disk source spectrum.
Multi-wavelength observations of optically-identified extra-galactic SNRs have
revealed that these sources are typically very weak X-ray and radio emitters
(e.g. Pannuti \etal\ 2000, Lacey \& Duric 2001, Pannuti \etal\ 2002).

Four X-ray sources are coincident with known M81 globular clusters 3 other
X-ray sources are coincident with optically-bright objects with colors
consistent with globular clusters and two other X-ray sources have candidate
optical counterparts that appear extended in \hst\ images. Approximately 10\% of
Galactic globular clusters contain X-ray sources. In the case of M81, two
surveys of the globular cluster population report a total of 139 globular
clusters including 98 within the \cxo\ field of view. Thus, 4\% to 7\% are
coincident with X-ray sources. DiStefano \etal\ (2002) find 25\% of confirmed
clusters in their M31 field of view contain X-ray sources and 10\% of all
globular cluster candidates have X-ray sources. They also report that most of
the luminous M31 X-ray sources are in globular clusters. In contrast, only one
of the 11 brightest M81 X-ray sources, \ein\ source X7, is coincident with a
globular cluster.

In general, there are remarkably few
counterparts to the resolved X-ray sources identified in our assessment of the
extensive literature and available archival images of M81. If the majority of
the resolved sources are XRBs, then they have companion stars and accretion
disks that may be detectable in optical light. The \hst\ images of M81
approach a limiting magnitude of $V\sim27$~mag or $M_V\sim-0.5$~mag. Thus only
O and B main sequence or giant companions or highly-luminous
accretion disks from XRBs in active states will be detectable by \hst.
Later-type companion stars will not appear in the optical data.

Low-mass XRBs
are common in galaxies because they are long-lived and slowly-evolving.
The last encounter of M81 and its companion galaxy M82 occurred some 500 Myr ago
  based on the study of the ages of young star clusters in M82
  (de~Grijs, O'Connell, \& Gallagher 2001).
If the onset of the last major star formation episode in the bulge of M81
  was triggered by this encounter, then
  the most massive members of the current population of main-sequence stars
  have masses \LA2.5~\msun\ (Maeder \& Meynet 1988).
If these constitute the population of companion stars in the currently-active
XRBs, then the resolved bulge sources are mostly low-mass XRBs.
They will not have detectable optical counterparts because they do not have O,
B, or giant star companions (and because of the bright, amorphous optical
background of the bulge). In contrast, on-going star formation along the spiral
arms should produce a population of high-mass XRBs with massive O and B star
companions. This environment is, however, also the location of obscuring atomic
and molecular gas. The few correlations with HII regions suggests some of the
X-ray sources are located in star forming regions that may be populated by
massive stars.
The high percentage of \hst\ potential counterparts identified in the disk
relative to the bulge also suggests an abundance of early-type stars in the
vicinity of the disk sources and, potentially, a preference for high-mass XRB
systems in this environment. This is consistent with the distribution of XRBs in
our Galaxy where high-mass XRBs are concentrated towards the Galactic plane and
along spiral arms while low-mass XRBs show a concentration towards the Galactic
center (Grimm \etal\ 2001).

\subsection{The Brightest M81 Sources}

In-depth analysis of three of the 4 brightest sources in the M81 field was
presented in \S\ref{s:bright_src} and of the third-brightest source in Swartz
\etal\ (2002). Interestingly, all three of the brightest non-nuclear source are
far from typical XRBs as seen in our Galaxy. SN~1993J is a supernova, \ein\ 
source X6 is a rare ultra-luminous X-ray source with possible optical and radio
counterparts, and source number 132 is an exceptionally-bright and hot supersoft
source candidate (Swartz \etal\ 2002).

\subsubsection{SN 1993J}

SN~1993J appears to be evolving as expected
based on the standard CSM interaction model of Chevalier (1982; Fransson \etal\
 1996) though a complete picture incorporating models of the exploding
star and its pre-supernova environment awaits detailed numerical calculation.
The X-ray properties of SN~1993J reported here provide an important
constraint on any future theoretical investigations because the X-ray light
curve is declining steadily, even at $\sim$7~yr, whereas the most-detailed
numerical simulations to date (Suzuki \& Nomoto 1995) predicted the light curve
would drop precipitously long ago unless the CSM were clumpy. A clumpy CSM would
produce a varying light curve with episodes of high X-ray flux occuring whenever
clumps are overtaken by the outgoing shock wave (Chugai 1993). In this scenario,
the CSM consists of a rarified wind embedded with relatively dense clouds and
the X-rays emanate from the shocked gas of the clouds with little or no
reverse shock emission. This is not what is observed spectroscopically. The
spectrum of SN~1993J is best modeled with a combination of thermal emission from
a reverse shock and a hard component from a forward shock.

\subsubsection{Einstein Source X6}

The multi-wavelength properties of \ein\ source X6 are intriguing. The X-ray
spectrum of the source is best-fit with a disk blackbody model. In this model,
the X-rays come from the inner portions of an accretion disk surrounding a
compact object. The inferred mass of the central object is $\sim$18~\msun\
assuming the innermost disk radius derived from the model corresponds to the
last stable Keplerian orbit of a non-spinning black hole. The
X-ray-model-derived bolometric luminosity is near the Eddington limit for an
object of this mass. The X-ray flux from X6 has been persistent throughout the
$>$20~yr of observation.

X6 is
located within a 5$\arcsec$-diameter ($\sim$90~pc) SNR candidate according to
Matonick \& Fesen (1997) based on a high [SII]/H$\alpha$ ratio indicative of
collisional excitation in the cooling region behind a SNR shock. There are no
emission lines present in the X-ray spectrum of X6 and an optically-thin thermal
plasma model is a notably poorer fit to the X-ray data. The X-ray morphology of
X6 is that of a point source with no evidence for extension. Thus, no X-ray
evidence, besides a steady flux, supports the conjecture that X6 is a SNR.

A weak radio source is  present at the
location of X6. Synchrotron emission is observed at radio wavelengths from
relativistic electrons accelerated in SNR shocks and in jets emanating from some
(Galactic) XRBs. A radio (or optical) light curve of the source at the location
of X6 has not yet been constructed. The radio source was present at 3.6~cm in
late 1994 but not seen in a 6~cm image taken in late 1999. Analysis of other
radio images is in progress. If 
the radio source proves to be variable, then it is not from a SNR.
The observed radio flux density of
$\sim$95~$\mu$Jy is typical of, for example, Magellanic Cloud SNRs (Filipovic
\etal\ 1998) after accounting for the disparate distances. In comparison, radio
jets associated with Galactic XRBs are weaker except during extreme outbursts.

There is also an optical point source coincident with X6. If associated with X6,
the optical emission may either be from a moderately-massive, O9~--~B1,
companion, which may be a Be star, or from the accretion disk itself but does
not come from an extended source at \hst\ resolution.

X6 can be compared to well-studied nearby XRBs. An example of a high-mass system
with a massive compact accretor is Cyg~X-1 (e.g., van~Paradijs 1995). The
optical counterpart to Cyg~X-1 is a O9.7 supergiant with colors similar to those
of the X6 counterpart but with the higher optical luminosity of a supergiant
compared to a main-sequence star. The putative black hole in Cyg~X-1 exceeds
$7$~\msun\ and is most probably $\sim$16~\msun. Cyg~X-1 is a persistent X-ray
source as is X6. It displays the charactersitic high-soft and low-hard states
typical of black hole XRBs (Tanaka \& Lewin 1995) and does not exceed an X-ray
luminosity of $\sim2\times10^{38}$~\ergl. Cyg~X-1 is radio-bright during its
low-hard state with a flux of $\sim$15~mJy or $0.007$~$\mu$Jy if it were placed
at the distance of M81. Scaling this value upward by the ratio of the X-ray
luminosities of X6 to Cyg~X-1 in its low-hard state ($\sim$700) results in a
radio flux density of only $6$~$\mu$Jy which would not be detectable.

An example of a low-mass system with a massive compact object and strong radio
emission is the microquasar GRS~1915+105 (e.g., Mirabel \& Rodr\`{i}guez 1999).
This system is a rapidly variable X-ray and radio transient reaching a peak
X-ray luminosity of $\sim1.5\times10^{39}$~\ergl\ in its high-soft state,
comparable to X6, and an average luminosity of $\sim3.7\times10^{38}$~\ergl.
High extinction along the line of sight to GRS~1915+105 obscures the optical
counterpart and accretion disk. Near-infrared spectroscopy (Greiner \etal\
 2001), however, shows the companion to be a K~--~M main-sequence star and, 
along with the orbital period, constrained the compact object mass to be
$14\pm4$~\msun. GRS~1915+105 is a strong radio emitter. Scaling to the distance
of M81 and to the X-ray luminosity of X6 (a factor of $\sim$13 when GRS~1915+105
is in its hard state) results in a radio flux density of about one-half the X6
value.

Thus, while monitoring at many wavelengths is required before any
definitive statment can be made, it is intriguing to consider X6
may be an X-ray- and radio-bright member of the class of microquasars
(see Mirabel \& Rodr\`{i}guez 1999 for a review) consisting of an accreting
black hole with a radio-bright jet but with unusually-steady X-ray flux.

\subsubsection{The M81 Nucleus}

The X-ray properties of the nucleus of M81 are difficult to deduce from the
present dataset because of severe pileup. A relatively weak spectrum
extracted from the readout trail was analyzed and found to be a
power law of photon index $\Gamma = 1.98\pm0.08$, consistent with numerous
previous X-ray studies. The presence of Fe~K$\alpha$ emission
could not be confirmed because of the lack of counts above $\sim$5~keV.
Variability of the source also could not be assessed. However, it was shown,
with the aid of the high angular resolution of the \cxo\ image, that the
contribution to the nuclear spectrum from thermal emission is small or
non-existent. Any thermal X-ray component present in the region is consistent
with an extrapolation of the unresolved bulge emission observed surrounding the
nucleus and extending over an $\sim$4~kpc diameter region.

\subsection{The M81 Bulge}

In addition to 53 X-ray sources resolved in the \cxo\ image, the bulge of M81
emits $\sim10^{39}$~\ergl\ in unresolved emission. This is $\sim$12\% of the
total non-nuclear emission from the entire galaxy and is distributed over an
$\sim$2$\arcmin$-radius region centered on the nucleus. Both the resolved
sources and unresolved emission trace the optical light from the old population
of bulge stars.

If the unresolved
emission is also produced by stellar systems, then they are systems distinct
from the resolved sources because extrapolation of the luminosity function of
the resolved sources contributes $<$10\% of the unresolved emission.
The possible X-ray-luminous stellar systems below the detection limit are
  massive OB stars, Be XRBs, CVs, RS CVn stars, and, at a lower luminosity,
late-type stars. However, individual late-type stars have X-ray luminosities
only of order a few $10^{27}$ to a few $10^{28}$~\ergl, requiring some
$10^{12}$ stars to produce the unresolved emission.

Massive OB stars with colliding winds can be strong X-ray emitters
  but are rare.
None are found in the bulge of M81 (Devereux, Ford, \& Jacoby 1997).
Be XRBs are young high-mass systems
  and the Be companion star is optically bright.
They are therefore also  unlikely to be abundant in the galactic bulge.

CVs are short-period (typically $<$1 day) binaries
  consisting of a white dwarf and a late-type low-mass companion (Warner 1995).
They are numerous and are long-lived.
The magnetic CVs, with a magnetic white dwarf, are known to
  have X-ray luminosities as high as $\sim10^{32}$~\ergl.
The space density of magnetic CVs in the solar neighborhood
  is $\sim 10^{-6}$~pc$^{-3}$ (Warner 1995) while
the stellar density is about 0.7~\msun~pc$^{-3}$ (Allen 1973). This implies
  the density of magnetic CVs is $\sim 10^{-5}$~\msun$^{-1}$.
If M81 has a similar space density of magnetic CVs, then
  there will be roughly $10^5$ CV systems in the bulge of M81.
If about 10\% are active
  (a rough estimate based on the properties of the local systems), then
  only $\sim10^{36}$~\ergl\ of the unresolved bulge X-ray emission can come from
CVs.

RS~CVn systems,
composed of chromospherically active G or K stars with late-type main
sequence or subgiant companions, also have high X-ray luminosities. Typical
X-ray luminosities of RS~CVn systems range from $\sim10^{29}$~\ergl\ to
$\sim3\times10^{31}$~\ergl\ (Rosner, Golub, \& Vaiana 1985).
Thus some $10^7$ to $10^{10}$ RS~CVn systems
are required to produce the unresolved bulge emission. If all stars in the bulge
are $\sim$1~\msun\ and half are in binary systems, then there are
$\sim$$10^{10}$ binary systems in the bulge. Assuming about 20\% of these
systems become RS~CVns and that G-K stars spend only a few percent of their
lifetimes in their giant stage, an uncomfortably large fraction must currently
be in an RS~CVn phase.

Individually, therefore, none of the these stellar systems can readily
account for the observed unresolved X-ray emission from the M81 bulge.
If, instead, some portion of the unresolved bulge X-ray emission is from hot
 diffuse gas, as suggested by its spectral distinction from the simple power law
shape of the resolved sources, then only a small fraction ($\sim$0.02\%) of the
total bulge mass is needed to account for the observed emission. The X-ray
emission, however, does not appear filamentary like the H$\alpha$ emission does
(Devereux \etal\ 1995). A filamentary morphology would be expected if the
emission is from ionization by shocks. A source for producing shocks is also not
obvious. Devereux \etal\ (1997) suggest shocks originating from nuclear activity
can account for the wispy ``nuclear spiral'' of H$\alpha$ emission confined to
the central $\sim$1$\arcmin$  but the unresolved X-ray emission is rather
smoothly extended over a region of 2$\arcmin$ radius. While ionizing radiation
from hot evolved post-AGB stars may produce the observed UV excess in the core
of M81 (O'Connell \etal\ 1992, Devereux \etal\ 1995), these stars do not produce
adequate ionizing radition in the \cxo\ energy band to account for the X-ray
emission (Binette \etal\ 1994).

\subsection{The M81 Disk and Spiral Arms}

One of the most spectacular features of M81 is its grand design spiral arm
structure.
The spiral arms
trace the location of recent star-forming activity induced by the passage
of spiral density waves. Applications of classical density wave models to M81
(e.g., Visser 1980) predict that
material travels faster than the spiral pattern, entering an arm on the
inside ``upstream'' edge.
Stars forming at the spiral shock front travel
at the local circular velocity of galactic rotation so that the youngest stars
would be immediately downstream of the shock
or toward the outside edge of the arm. The most massive stars are the quickest 
to evolve. They end their lives in core-collapse SN explosions leaving behind a
neutron star or, perhaps, a black hole remant. Supernova explosions produce
X-ray emitting SNRs and compact stars in binaries may become XRBs. Thus, the
spiral arms are not only the site of star formation but also a stellar graveyard
and the birthplace of X-ray sources.

The brightest X-ray sources in the disk of M81 correlate spatially with the
spiral arms. Accepting that the majority of the resolved sources are XRBs and
 that the X-ray flux is generally proportional to the mass accretion rate, then
the brightest XRBs are young high-mass XRBs with high mass-transfer rates.
The onset of
mass-transfer in these systems, and hence of the X-ray-bright phase, can begin
immediately following the formation of the compact object because of the strong
stellar wind from the massive companion.
This is in contrast to the low-mass XRBs
  in which mass transfer begins only
  after the companion star evolves to a (sub)giant stage
  or when the binary orbit has decayed sufficiently
  so that Roche lobe overflow can begin.
Thus, the young high-mass systems
  become X-ray emitters
  while still within the spiral arm region of their origin.
For this reason the luminosity distribution of the young XRBs in the spiral arms
  are expected to differ from the distribution of the older XRBs
  elsewhere in the galaxy.
In particular, it will not show the characteristic luminosity break
  induced by aging of the XRB population as predicted by Wu (2001) and Wu \etal\
(2002a,b).

Core-collapse supernova only come from stars
  more massive than $\sim$8--10~\msun\ and are X-ray bright SNRs
  only for a relatively short time.
They, too, should be found preferentially near their place of origin, the spiral arms.

\section{Contributions to the M81 X-ray Luminosity Function} \label{s:summary}

The observed M81 X-ray luminosity functions reported by Tennant \etal\ (2001)
prompted Wu (2001) and Wu \etal\ (2002a,b) to consider the physical 
underpinnings that give rise to a cutoff in the luminosity function for the
bulge sources and to the absence of this feature in the disk population. Wu
(2001) showed that the shape of the luminosity function is governed to first
order simply by the birth rates and (X-ray active) lifespans of the XRBs that
dominate the luminosity function and hence is a measure of the star formation
history of the local environment and of galaxy evolution in the broader context.
Further investigation (Wu \etal\ 2002a,b) reveals that several complicating 
issues must be considered before this hypothesis can be rigorously applied.
Some of these issues have been addressed in the present work.

The first issue is the presence of a population of SNRs. The onset and duration
of the X-ray active phase of XRBs depends mainly on the donor star mass and
its consequent evolutionary path while the X-ray luminosity depends on the
accretion rate. This is fundamentally different than the X-ray evolution of
SNRs. Here we have shown, however, that SNRs are not important contributors to
the total X-ray source population in M81 with the exception of SN~1993J, the
fourth-brightest source in the M81 field at the time of observation.

Another issue is
the occurrence of XRBs in globular clusters. If capture processes govern the
formation of XRBs in globular clusters, as seems likely to account for the
excess of XRBs in these environments, then the characteristic lifetime of the
XRB is not just correlated with the nuclear or orbital evolution
timescales of the system but is also a function of the encounter frequency.
However, only a few percent of the X-ray sources in M81 appear to be in globular
clusters. As with Galactic globular cluster XRBs (Verbunt \& van~den~Heuvel
 1995), the impact on the luminosity function is further minimized by the fact
that the globular cluster XRBs in M81 are not among the brightest X-ray sources.
Again, the exception is \ein\ source X7, the fifth-brightest source in the
field.

A third factor with potential impact to the basic hypothesis of Wu
(2001) is the presence of XRBs with a nuclear-burning white dwarf accretor,
i.e., members of the class of supersoft sources (e.g., Kahabka \& van~den~Heuvel
 1997). While the lifespans of these objects depend on the companion mass and
mass-transfer rates as with other XRBs, only a narrow range of mass-transfer
rates result in {\sl steady} nuclear burning. Wu \etal\ (2002a,b) argue,
therefore, a narrow X-ray luminosity range for this source population and a
sharp decline in the number of sources with luminosities above the Eddington
limit for a Chandrasekar-mass accretor. Swartz \etal\ (2002) found 9 supersoft
source candidates in the M81 field. Six of these are relatively weak sources
with luminosities in a narrow range around $\sim10^{37}$~\ergl. The three
brightest candidates, however, radiate at or above the Eddington limit depending
on the adopted spectral model with one source (the third-brightest in the
entire field) approaching $\sim10^{39}$~\ergl. Thus the most important
contribution to the luminosity function from supersoft sources is at the high
luminosity end and is dominated by one bright source. The effect of the weaker
supersoft sources is obscured by the large number of other sources contributing
at low luminosities.

\begin{center}
\includegraphics[angle=270,width=\columnwidth]{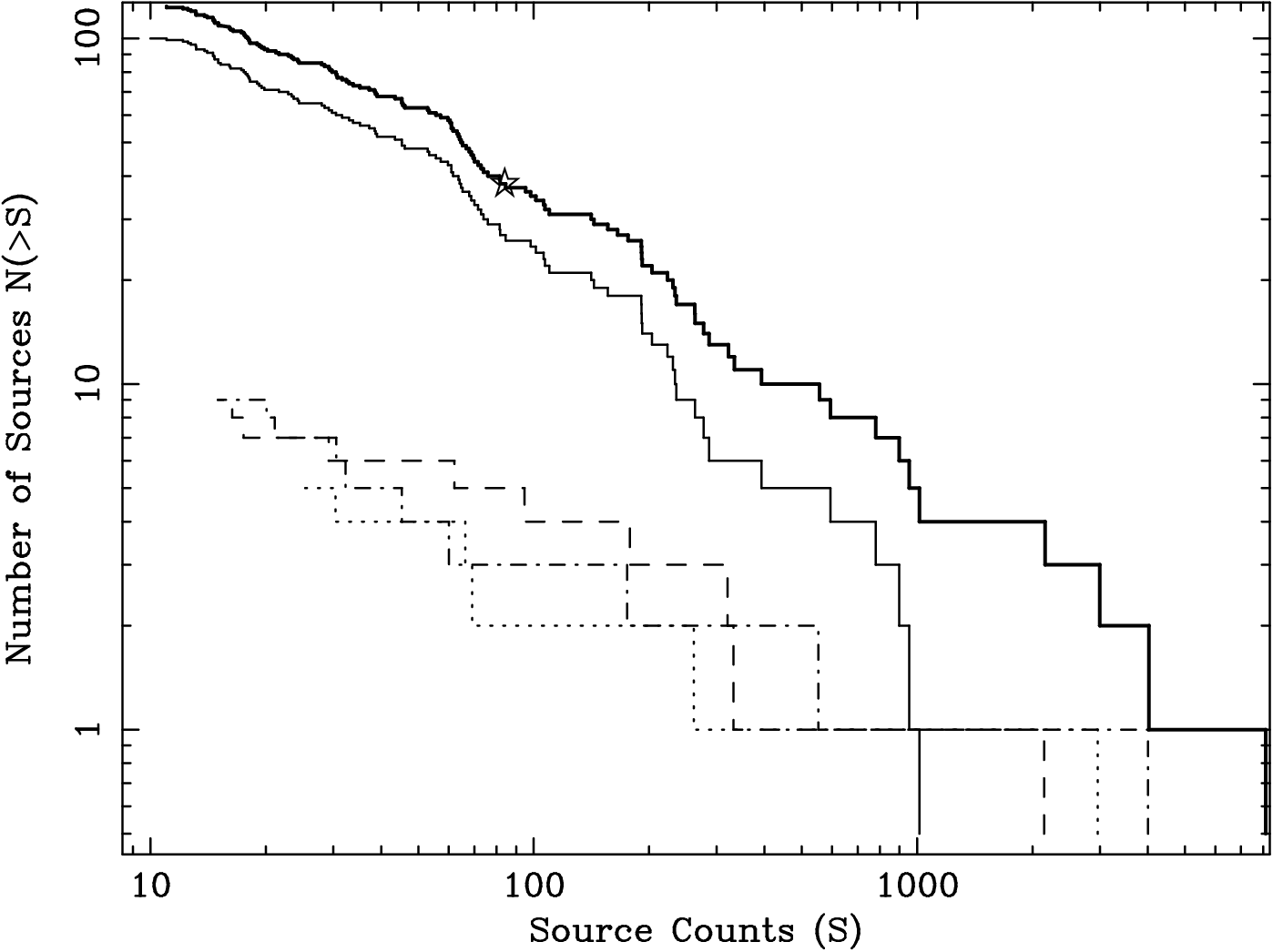}
\vspace{10pt}
\figcaption{Observed X-ray luminosity function including all 123
non-nuclear X-ray sources detected within the $D_{25}$ isophote of M81 ({\em
heavy solid line}). Also shown are the luminosity functions
of the SNRs ({\em dotted}), of X-rays sources spatially-coincident with globular
clusters {\em dashed}), and of the supersoft sources ({\em dot-dashed}). The
luminosity function with these three populations and the Ultra-Luminous
X-ray source X6 omitted is shown as a thin solid line. The symbol ($\star$)
marks the division between bright sources for which spectral analysis has been
reported in this work and of weak sources for which no spectral fits were made.
\label{f:f17}}
\end{center}

The luminosity function for all non-nuclear sources detected within the $D_{25}$
isophote of M81 is shown in Figure~\ref{f:f17}. Also shown are the luminosity
functions of the supersoft sources, of X-ray sources spatially-coincident with
globular clusters, and with SNRs.
   As shown in Figure~\ref{f:f17},
     these three populations all
     have relatively flat power-law luminosity functions
     and they affect only the bright end of the overall luminosity distribution.
  XRBs, the dominant population of X-ray sources in M81,
     however, have a steep luminosity function
     and hence determine the overall shape of the luminosity functions,
     especially at the faint ends.
  The break at the luminosity of $\sim 4 \times 10^{37}$~erg~s$^{-1}$
     that we have found (see also Tennant \etal\ 2001)
     is therefore a characteristic imprint of the XRBs.
  We have argued that the formation of such a break
     is due to the age of a population of XRBs
     which were born at a star-burst episode in the recent past
  (Wu 2001, Wu \etal\ 2002a,b).
  This break is distinguishable from another possible break,
     expected to occur at $\sim 2 \times 10^{38}$~erg~s${-1}$,
     the Eddington luminosity of a 1.5~\msun\ accreting object.
  The latter is attributed (Sarazin, Irwin, \& Bregman 2001) to the presence
     of a population of neutron stars which accrete at rates
     close to the Eddington limit.
  Whether or not this break is visible in a given population
     depends on the relative proportion of neutron-star XRBs and black-hole XRBs.
  It also requires that the host galaxies (e.g. giant ellipticals)
     have a sufficiently large X-ray source population
     that the break becomes statistically significant.
  Nevertheless, we do see hints of this break
     in the luminosity function of the X-ray sources in M81,
     when we remove the SNRs, globular clusters XRBs and the supersoft sources.

In summary, most of the factors complicating the simple birth-death model are
unimportant for M81. Nevertheless, careful examination of the brightest sources
is warranted because they have the largest influence on the luminosity function
and yet are certainly not typical of the dominant class of X-ray sources, the
XRBs.

\vspace{0.125in} 
We thank L. Townsley for applying her CTI-corrector algorithm to the \cha\ data
and for providing response matrices. We thank J. Davis for discussions and
his independent analysis of the spectrum of the nucleus. We are grateful to N.
Bartel and M. Beitenholtz for sharing their radio data. T.~G.~P. acknowledges a
travel grant from the NRAO to reduce the radio data and he is very
grateful to M. Rupen for guidance during the reduction
process. K.~W. thanks M. Weisskopf for  providing support for visits to
MSFC. Support for this research was provided in part by
NASA/\cha\ grants GO0-1058X and AR2-3008X to D.~A.~S.


\clearpage
\begin{center}
\small{
\begin{tabular}{rrrrrrccrl}
\multicolumn{10}{c}{{\sc Table 2}} \\
\multicolumn{10}{c}{M81 Discrete X-Ray Sources} \\
\hline \hline
 &  \multicolumn{1}{c}{R.A.} & \multicolumn{1}{c}{Dec.} & Count Rate & S/N & $m_V$ & CCD & region & \multicolumn{1}{c}{$L_X$} & Comment \\
 &  \multicolumn{1}{c}{(J2000)} & \multicolumn{1}{c}{(J2000)} & ($10^{-4}$~s$^{-1}$) & & & & & \multicolumn{1}{c}{($10^{37}$~erg~s$^{-1}$)} & \\ 
\hline
  1 &  9 52 38.50 &  68 56 37.5 & $  24.05\pm  2.19$ &  7.45 & $ $  ---   & i2 & $D_{25}$ & $   4.40\pm  0.40$ &  \\
  2 &  9 52 39.86 &  69 03 59.8 & $  12.73\pm  1.59$ &  4.82 & $>$16.52   & i3 & $D_{25}$ & $   2.33\pm  0.29$ & P1 \\
  3 &  9 52 41.52 &  68 55 26.6 & $  18.07\pm  1.90$ &  6.10 & $ $  ---   & i2 & $D_{25}$ & $   3.31\pm  0.35$ &  \\
  4 &  9 52 49.86 &  68 59 25.3 & $  12.75\pm  1.59$ &  5.18 & $ $  ---   & i2 & $D_{25}$ & $   2.33\pm  0.29$ & P4 \\
  5 &  9 52 59.76 &  69 07 43.1 & $  15.90\pm  1.78$ &  6.49 & $ $  ---   & i3 & $D_{25}$ & $   2.91\pm  0.33$ & P5 \\
  6 &  9 53 02.06 &  69 09 38.8 & $  10.82\pm  1.47$ &  4.53 & $>$21.02   & i3 & $D_{25}$ & $   1.98\pm  0.27$ &  \\
  7 &  9 53 04.08 &  69 01 43.7 & $  10.00\pm  1.41$ &  4.12 & $ $  ---   & i2 & $D_{25}$ & $   1.83\pm  0.26$ &  \\
  8 &  9 53 10.54 &  69 00 02.8 & $   7.44\pm  1.22$ &  4.13 & $>$16.72   & i2 & $D_{25}$ & $   1.36\pm  0.22$ & H2, P6, Gal? \\
  9 &  9 53 16.05 &  69 00 03.4 & $   6.69\pm  1.15$ &  3.73 & $ $  ---   & i2 & $D_{25}$ & $   1.22\pm  0.21$ &  \\
 10 &  9 53 17.96 &  69 06 43.2 & $  14.36\pm  1.69$ &  5.89 & $>$20.83   & i3 & $D_{25}$ & $   2.63\pm  0.31$ & H3, P7 \\
 11 &  9 53 18.71 &  69 02 19.1 & $   6.19\pm  1.11$ &  3.81 & $ $  ---   & i2 & $D_{25}$ & $   1.13\pm  0.20$ &  \\
 12 &  9 53 19.37 &  69 10 45.1 & $   7.27\pm  1.20$ &  3.28 & $ $  ---   & i3 & $D_{25}$ & $   1.33\pm  0.22$ &  \\
 13 &  9 53 27.53 &  69 04 19.3 & $  16.56\pm  1.81$ &  6.95 & $ $  ---   & i3 & $D_{25}$ & $   3.03\pm  0.33$ & P8 \\
 14 &  9 53 32.65 &  69 02 20.3 & $   5.29\pm  1.03$ &  3.53 & $ $  ---   & i2 & $D_{25}$ & $   0.97\pm  0.19$ &  \\
 15 &  9 53 33.32 &  69 03 42.0 & $   7.56\pm  1.23$ &  4.33 & $ $  ---   & i3 & $D_{25}$ & $   1.38\pm  0.22$ &  \\
 16 &  9 53 33.87 &  68 58 20.7 & $  12.42\pm  1.57$ &  6.35 & $ $\hspace{5pt}18.65 & i2 & $D_{25}$ & $   2.27\pm  0.29$ & P9, Extended? \\
 17 &  9 53 36.05 &  69 05 45.4 & $   5.71\pm  1.07$ &  3.61 & $ $  ---   & i3 & $D_{25}$ & $   1.05\pm  0.20$ &  \\
 18 &  9 53 37.69 &  68 59 19.1 & $  16.57\pm  1.82$ &  7.24 & $ $  ---   & i2 & $D_{25}$ & $   3.03\pm  0.33$ &  \\
 19 &  9 53 42.03 &  68 59 17.9 & $  10.33\pm  1.43$ &  5.61 & $>$21.11   & i2 & $D_{25}$ & $   1.89\pm  0.26$ & P11 \\
 20 &  9 53 44.48 &  69 05 26.5 & $   2.98\pm  0.77$ &  2.98 & $ $  ---   & i3 & $D_{25}$ & $   0.55\pm  0.14$ &  \\
 21 &  9 53 48.61 &  69 00 21.0 & $   2.85\pm  0.75$ &  2.93 & $ $  ---   & i2 & $D_{25}$ & $   0.52\pm  0.14$ &  \\
 22 &  9 53 50.85 &  69 05 26.5 & $   4.07\pm  0.90$ &  3.09 & $ $  ---   & i3 & $D_{25}$ & $   0.75\pm  0.16$ &  \\
 23 &  9 53 51.59 &  68 55 37.4 & $   4.82\pm  0.98$ &  3.47 & $ $  ---   & i2 & $D_{25}$ & $   0.88\pm  0.18$ &  \\
 24 &  9 53 51.88 &  69 02 49.9 & $  19.14\pm  1.95$ &  8.06 & $ $  ---   & i3 & $D_{25}$ & $   3.50\pm  0.36$ & H5, P14 \\
 25 &  9 53 53.41 &  69 03 59.4 & $   4.48\pm  0.94$ &  3.36 & $ $  ---   & i3 & $D_{25}$ & $   0.82\pm  0.17$ &  \\
 26 &  9 53 53.67 &  69 03 18.0 & $   3.24\pm  0.80$ &  2.87 & $ $  ---   & i3 & $D_{25}$ & $   0.59\pm  0.15$ &  \\
 27 &  9 53 57.47 &  69 03 53.8 & $  16.92\pm  1.83$ &  7.49 & $ $  ---   & i3 & $D_{25}$ & $   3.10\pm  0.34$ & H6, P15 \\
 28 &  9 54 02.15 &  69 01 26.7 & $   3.47\pm  0.83$ &  3.02 & $ $  ---   & i2 & $D_{25}$ & $   0.64\pm  0.15$ &  \\
 29 &  9 54 06.73 &  69 08 41.7 & $  11.37\pm  1.50$ &  5.95 & $ $  ---   & i3 &     d    & $   2.08\pm  0.28$ &  \\
 30 &  9 54 14.00 &  69 05 37.9 & $  12.96\pm  1.61$ &  6.71 & $ $  ---   & i3 &     d    & $   2.37\pm  0.29$ &  \\
 31 &  9 54 14.92 &  69 06 12.1 & $   2.75\pm  0.74$ &  2.84 & $ $  ---   & i3 &     d    & $   0.50\pm  0.14$ &  \\
 32 &  9 54 21.30 &  68 44 41.6 & $  17.03\pm  1.84$ &  4.19 & $ $  ---   & s1 & $D_{25}$ & $   2.32\pm  0.25$ &  \\
 33 &  9 54 25.49 &  68 46 51.7 & $  29.01\pm  2.40$ &  6.75 & $ $  ---   & s1 & $D_{25}$ & $   3.94\pm  0.33$ &  \\
 34 &  9 54 26.42 &  68 43 43.5 & $  18.67\pm  1.93$ &  3.14 & $ $  ---   & s1 & $D_{25}$ & $   2.54\pm  0.26$ &  \\
 35 &  9 54 27.84 &  68 53 11.3 & $   3.44\pm  0.83$ &  3.07 & $ $  ---   & s2 & $D_{25}$ & $   0.63\pm  0.15$ &  \\
 36 &  9 54 32.66 &  68 47 44.1 & $  26.95\pm  2.31$ &  5.77 & $ $  ---   & s1 & $D_{25}$ & $   3.66\pm  0.31$ &  \\
 37 &  9 54 33.16 &  68 52 29.0 & $  27.96\pm  2.36$ &  9.51 & $ $  ---   & s2 & $D_{25}$ & $   6.30\pm  1.10$ & P18 \\
 38 &  9 54 38.67 &  68 52 42.9 & $   8.18\pm  1.28$ &  4.90 & $ $  ---   & s2 & $D_{25}$ & $   1.50\pm  0.23$ &  \\
 39 &  9 54 39.23 &  68 45 49.4 & $  73.05\pm  3.81$ & 12.55 & $ $  ---   & s1 & $D_{25}$ & $   9.93\pm  0.52$ &  \\
 40 &  9 54 41.82 &  68 56 47.6 & $  11.91\pm  1.54$ &  6.51 & $ $  ---   & s2 & $D_{25}$ & $   2.18\pm  0.28$ &  \\
 41 &  9 54 41.99 &  69 02 43.7 & $  11.05\pm  1.48$ &  6.38 & $>$HST     & s3 &     d    & $   1.50\pm  0.20$ & P21 \\
 42 &  9 54 44.34 &  68 56 11.0 & $   5.61\pm  1.06$ &  4.44 & $>$19.81   & s2 & $D_{25}$ & $   1.03\pm  0.19$ &  \\
 43 &  9 54 45.30 &  68 56 58.6 & $  58.15\pm  3.40$ & 14.30 & $>$13.81   & s2 & $D_{25}$ & $  10.64\pm  0.62$ & H8, P22, $\star$ \\
 44 &  9 54 46.79 &  69 05 12.6 & $   3.58\pm  0.84$ &  3.07 & $ $\hspace{5pt}23.01 & s3 &     d    & $   0.49\pm  0.11$ &  \\
 45 &  9 54 47.18 &  69 01 01.4 & $   4.64\pm  0.96$ &  3.77 & $ $  ---   & s3 &     d    & $   0.63\pm  0.13$ &  \\
 46 &  9 54 51.49 &  68 51 43.5 & $   9.50\pm  1.37$ &  5.31 & $ $  ---   & s2 & $D_{25}$ & $   1.74\pm  0.25$ &  \\
 47 &  9 54 53.96 &  68 54 55.0 & $   6.40\pm  1.13$ &  4.62 & $ $  ---   & s2 & $D_{25}$ & $   1.17\pm  0.21$ &  \\
 48 &  9 54 55.15 &  69 04 20.3 & $   2.99\pm  0.77$ &  3.40 & $ $  ---   & s3 &     d    & $   0.41\pm  0.10$ &  \\
 49 &  9 54 55.60 &  68 51 59.6 & $   4.15\pm  0.91$ &  2.85 & $ $  ---   & s2 & $D_{25}$ & $   0.76\pm  0.17$ &  \\
 50 &  9 54 56.05 &  69 05 17.4 & $   3.43\pm  0.83$ &  3.18 & $ $\hspace{5pt}23.79 & s3 &     d    & $   0.47\pm  0.11$ &  \\
\hline 
\end{tabular}
} 
\end{center}

\clearpage
\begin{center}
\small{
\begin{tabular}{rrrrrrccrl}
\multicolumn{10}{c}{{\sc Table} 2 --- Continued} \\
\multicolumn{10}{c}{M81 Discrete X-Ray Sources} \\
\hline \hline
 &  \multicolumn{1}{c}{R.A.} & \multicolumn{1}{c}{Dec.} & Count Rate & S/N & $m_V$ & CCD & region & \multicolumn{1}{c}{$L_X$} & Comment \\
 &  \multicolumn{1}{c}{(J2000)} & \multicolumn{1}{c}{(J2000)} & ($10^{-4}$~s$^{-1}$) & & & & & \multicolumn{1}{c}{($10^{37}$~erg~s$^{-1}$)} & \\ 
\hline
 51 &  9 54 57.59 &  69 02 41.1 & $  32.91\pm  2.56$ & 10.82 & $ $  ---   & s3 &     d    & $   5.00\pm  0.60$ &  \\
 52 &  9 55 00.11 &  69 07 45.2 & $ 189.55\pm  6.14$ & 25.64 & $ $22.7(I) & s3 &     d    & $  27.00\pm  0.60$ & X10, H10, P25 \\
 53 &  9 55 00.28 &  69 04 36.9 & $   2.61\pm  0.72$ &  2.86 & $ $  ---   & s3 &     d    & $   0.36\pm  0.10$ &  \\
 54 &  9 55 00.36 &  69 01 48.9 & $   2.80\pm  0.75$ &  2.90 & $ $  ---   & s3 &     d    & $   0.38\pm  0.10$ &  \\
 55 &  9 55 00.48 &  68 56 32.8 & $   2.42\pm  0.69$ &  2.91 & $ $  ---   & s2 &     d    & $   0.44\pm  0.13$ &  \\
 56 &  9 55 01.00 &  68 56 22.1 & $  10.48\pm  1.44$ &  6.11 & $ $  ---   & s2 & $D_{25}$ & $   1.92\pm  0.26$ & $\star$ \\
 57 &  9 55 01.05 &  69 07 27.1 & $  55.15\pm  3.31$ & 13.88 & $ $  ---   & s3 &     d    & $  22.00\pm  2.80$ &  \\
 58 &  9 55 01.40 &  68 53 29.7 & $  13.66\pm  1.65$ &  6.50 & $ $  ---   & s2 & $D_{25}$ & $   2.50\pm  0.30$ &  \\
 59 &  9 55 01.65 &  69 10 42.3 & $   9.00\pm  1.34$ &  5.03 & $ $  ---   & s4 &     d    & $   1.65\pm  0.24$ & P26 \\
 60 &  9 55 02.57 &  68 56 21.2 & $  12.73\pm  1.59$ &  6.78 & $ $  ---   & s2 & $D_{25}$ & $   2.33\pm  0.29$ & H11, P27, $\star$ \\
 61 &  9 55 05.43 &  68 44 22.8 & $  85.31\pm  4.12$ & 12.94 & $ $  ---   & s1 & $D_{25}$ & $  11.60\pm  0.56$ &  \\
 62 &  9 55 05.62 &  68 58 52.1 & $   3.60\pm  0.85$ &  3.11 & $ $  ---   & s2 &     d    & $   0.66\pm  0.15$ & P28, HII \\
 63 &  9 55 06.34 &  69 04 05.7 & $  10.52\pm  1.45$ &  5.79 & $ $  ---   & s3 &     d    & $   1.43\pm  0.20$ &  \\
 64 &  9 55 08.91 &  68 57 22.9 & $   2.89\pm  0.76$ &  3.03 & $>$19.69   & s2 &     d    & $   0.53\pm  0.14$ &  \\
 65 &  9 55 09.28 &  68 53 35.6 & $  10.34\pm  1.43$ &  5.94 & $ $  ---   & s2 & $D_{25}$ & $   1.89\pm  0.26$ &  \\
 66 &  9 55 09.66 &  69 07 43.4 & $  21.26\pm  2.06$ &  8.43 & $ $\hspace{5pt}23.19 & s3 &     d    & $   2.89\pm  0.28$ & HII \\
 67 &  9 55 09.77 &  69 04 07.8 & $  18.93\pm  1.94$ &  8.29 & $>$17.29   & s3 &     b    & $   3.50\pm  0.70$ & X2, H13, P29, GC? \\
 68 &  9 55 09.80 &  69 08 35.4 & $   5.57\pm  1.05$ &  2.97 & $ $\hspace{5pt}21.45 & s4 &     d    & $   1.02\pm  0.19$ &  \\
 69 &  9 55 10.29 &  69 05 02.4 & $ 201.59\pm  6.33$ & 27.36 & $ $\hspace{5pt}22.79 & s3 &     b    & $  26.70\pm  1.00$ & X3, H14, P31 \\
 70 &  9 55 10.71 &  69 08 43.7 & $  13.25\pm  1.62$ &  6.23 & $ $\hspace{5pt}22.86 & s4 &     d    & $   2.42\pm  0.30$ & P30, SNR \\
 71 &  9 55 11.81 &  68 57 47.9 & $   2.92\pm  0.76$ &  2.91 & $ $  ---   & s2 &     d    & $   0.53\pm  0.14$ &  \\
 72 &  9 55 12.44 &  69 01 21.5 & $   2.53\pm  0.71$ &  2.83 & $>$HST     & s3 &     d    & $   0.34\pm  0.10$ &  \\
 73 &  9 55 14.12 &  69 12 36.1 & $  12.74\pm  1.59$ &  5.56 & $ $  ---   & s4 &     d    & $   2.33\pm  0.29$ & P32 \\
 74 &  9 55 14.61 &  69 06 40.4 & $   3.94\pm  0.89$ &  3.37 & $ $\hspace{5pt}24.03 & s3 &     b    & $   0.54\pm  0.12$ &  \\
 75 &  9 55 15.22 &  69 05 38.0 & $   7.40\pm  1.21$ &  4.83 & $ $\hspace{5pt}24.36 & s3 &     b    & $   1.01\pm  0.16$ &  \\
 76 &  9 55 15.56 &  68 54 27.5 & $   4.70\pm  0.97$ &  4.18 & $ $  ---   & s2 &     d    & $   0.86\pm  0.18$ &  \\
 77 &  9 55 15.99 &  68 51 59.6 & $   9.70\pm  1.39$ &  5.38 & $ $  ---   & s2 & $D_{25}$ & $   1.77\pm  0.25$ &  \\
 78 &  9 55 19.76 &  69 07 33.7 & $   6.08\pm  1.10$ &  3.90 & $ $\hspace{5pt}22.05 & s3 &     d    & $   0.83\pm  0.15$ & HII, SNR \\
 79 &  9 55 19.95 &  69 03 52.0 & $   2.43\pm  0.70$ &  2.99 & $>$HST     & s3 &     b    & $   0.33\pm  0.09$ & Radio \\
 80 &  9 55 21.85 &  69 03 44.9 & $   3.52\pm  0.84$ &  3.34 & $>$HST     & s3 &     b    & $   0.48\pm  0.11$ &  \\
 81 &  9 55 21.87 &  69 05 22.3 & $  37.95\pm  2.75$ & 11.53 & $>$HST     & s3 &     b    & $   4.80\pm  0.50$ &  \\
 82 &  9 55 21.99 &  69 06 37.6 & $  35.61\pm  2.66$ & 11.08 & $ $\hspace{5pt}17.14 & s3 &     b    & $   5.10\pm  1.10$ & H16, P33, GC? \\
 83 &  9 55 22.16 &  69 05 10.6 & $ 178.55\pm  5.96$ & 25.84 & $>$HST     & s3 &     b    & $  22.00\pm  0.90$ & H15, P34 \\
 84 &  9 55 23.71 &  68 58 48.9 & $   3.61\pm  0.85$ &  3.85 & $ $  ---   & s2 &     d    & $   0.66\pm  0.16$ &  \\
 85 &  9 55 24.30 &  69 04 39.3 & $   6.06\pm  1.10$ &  4.46 & $>$HST     & s3 &     b    & $   0.82\pm  0.15$ &  \\
 86 &  9 55 24.36 &  69 09 57.9 & $ 155.20\pm  5.55$ & 23.19 & $ $  ---   & s4 &     d    & $  35.00\pm  3.00$ & X4, H17, P35 \\
 87 &  9 55 24.41 &  69 14 50.7 & $  12.24\pm  1.56$ &  4.80 & $>$15.82   & s4 &     d    & $   2.24\pm  0.29$ &  \\
 88 &  9 55 24.77 &  69 01 13.4 & $ 596.32\pm 10.89$ & 46.98 & $>$20.22   & s3 &     d    & $  48.00\pm  2.00$ & SN 1993J, H18, P36 \\
 89 &  9 55 26.31 &  69 04 37.3 & $  12.55\pm  1.58$ &  6.52 & $>$HST     & s3 &     b    & $   1.71\pm  0.21$ &  \\
 90 &  9 55 26.57 &  69 04 00.4 & $   3.82\pm  0.87$ &  2.83 & $>$HST     & s3 &     b    & $   0.52\pm  0.12$ &  \\
 91 &  9 55 26.93 &  69 05 42.4 & $   6.29\pm  1.12$ &  4.18 & $ $  ---   & s3 &     b    & $   0.86\pm  0.15$ &  \\
 92 &  9 55 27.01 &  69 04 15.3 & $  16.24\pm  1.80$ &  7.45 & $>$HST     & s3 &     b    & $   2.21\pm  0.24$ &  \\
 93 &  9 55 27.28 &  69 02 48.0 & $  28.07\pm  2.36$ &  9.77 & $>$27.76   & s3 &     b    & $   3.70\pm  0.50$ &  \\
 94 &  9 55 27.85 &  68 49 52.9 & $  10.25\pm  1.43$ &  4.05 & $ $  ---   & s1 & $D_{25}$ & $   1.39\pm  0.19$ &  \\
 95 &  9 55 28.03 &  69 04 07.9 & $  38.02\pm  2.75$ & 11.48 & $>$HST     & s3 &     b    & $   5.70\pm  0.60$ &  \\
 96 &  9 55 28.44 &  69 02 44.5 & $  12.02\pm  1.55$ &  6.04 & $>$HST     & s3 &     b    & $   1.64\pm  0.21$ & SSS \\
 97 &  9 55 28.82 &  69 06 12.9 & $  11.87\pm  1.54$ &  6.17 & $ $  ---   & s3 &     b    & $   1.61\pm  0.21$ & H20 \\
 98 &  9 55 29.20 &  69 03 21.1 & $   4.85\pm  0.98$ &  4.11 & $>$HST     & s3 &     b    & $   0.66\pm  0.13$ &  \\
 99 &  9 55 29.28 &  69 15 57.2 & $   9.92\pm  1.40$ &  3.21 & $ $  ---   & s4 & $D_{25}$ & $   1.82\pm  0.26$ & H19 \\
100 &  9 55 30.21 &  69 03 18.4 & $  12.64\pm  1.59$ &  6.47 & $>$HST     & s3 &     b    & $   1.72\pm  0.22$ &  \\
\hline 
\end{tabular}
} 
\end{center}

\clearpage
\begin{center}
\small{
\begin{tabular}{rrrrrrccrl}
\multicolumn{10}{c}{{\sc Table} 2 --- Continued} \\
\multicolumn{10}{c}{M81 Discrete X-Ray Sources} \\
\hline \hline
 &  \multicolumn{1}{c}{R.A.} & \multicolumn{1}{c}{Dec.} & Count Rate & S/N & $m_V$ & CCD & region & \multicolumn{1}{c}{$L_X$} & Comment \\
 &  \multicolumn{1}{c}{(J2000)} & \multicolumn{1}{c}{(J2000)} & ($10^{-4}$~s$^{-1}$) & & & & & \multicolumn{1}{c}{($10^{37}$~erg~s$^{-1}$)} & \\ 
\hline
101 &  9 55 30.25 &  69 02 46.8 & $   6.55\pm  1.14$ &  4.48 & $>$HST     & s3 &     b    & $   0.89\pm  0.16$ &  \\
102 &  9 55 31.38 &  69 04 19.5 & $  46.44\pm  3.04$ & 12.62 & $>$HST     & s3 &     b    & $   6.80\pm  0.60$ &  \\
103 &  9 55 32.61 &  69 05 13.0 & $   2.99\pm  0.77$ &  2.85 & $>$HST     & s3 &     b    & $   0.41\pm  0.10$ &  \\
104 &  9 55 32.66 &  69 02 31.4 & $   3.22\pm  0.80$ &  2.95 & $ $\hspace{5pt}22.87 & s3 &     b    & $   0.44\pm  0.11$ &  \\
105 &  9 55 32.99 &  69 00 33.3 & $   1614\pm    18$ & 73.39 & $ $\hspace{5pt}24.13 & s3 &     d    & $    270\pm    10$ & X6, H21, P37 \\   
106 &  9 55 33.17 &  69 03 55.1 & $   5964\pm    34$ & 140.0 & $ $\hspace{5pt}15.42 & s3 &     b    & $   3400\pm   233$ & X5, H22, P38 \\
107 &  9 55 33.92 &  69 03 43.3 & $  13.92\pm  1.66$ &  6.21 & $>$HST     & s3 &     b    & $   1.89\pm  0.23$ &  \\
108 &  9 55 34.12 &  69 07 13.1 & $   7.75\pm  1.24$ &  4.73 & $ $  ---   & s3 &     d    & $   1.05\pm  0.17$ &  \\
109 &  9 55 34.32 &  69 03 50.9 & $  45.90\pm  3.02$ &  9.31 & $>$HST     & s3 &     b    & $   4.20\pm  0.40$ &  \\
110 &  9 55 34.56 &  69 03 39.0 & $   8.64\pm  1.31$ &  4.35 & $ $\hspace{5pt}19.52 & s3 &     b    & $   1.18\pm  0.18$ &  \\
111 &  9 55 34.62 &  69 02 50.0 & $  16.79\pm  1.83$ &  7.58 & $>$HST     & s3 &     b    & $   2.28\pm  0.25$ &  \\
112 &  9 55 34.65 &  69 03 51.4 & $  52.38\pm  3.23$ &  8.47 & $ $\hspace{5pt}22.73 & s3 &     b    & $   5.80\pm  0.60$ &  \\
113 &  9 55 34.71 &  69 04 53.9 & $  38.03\pm  2.75$ & 11.84 & $>$HST     & s3 &     b    & $   4.30\pm  0.50$ &  \\
114 &  9 55 34.81 &  69 03 13.5 & $  16.21\pm  1.79$ &  7.23 & $ $\hspace{5pt}20.78 & s3 &     b    & $   2.20\pm  0.24$ &  \\
115 &  9 55 34.90 &  69 04 07.9 & $  19.51\pm  1.97$ &  7.17 & $>$HST     & s3 &     b    & $   3.10\pm  0.50$ &  \\
116 &  9 55 34.98 &  69 03 42.3 & $ 118.14\pm  4.85$ & 20.38 & $>$HST     & s3 &     b    & $  19.50\pm  1.00$ &  \\
117 &  9 55 35.28 &  68 55 10.6 & $  31.02\pm  2.48$ & 10.59 & $ $  ---   & s2 &     d    & $   6.00\pm  1.00$ & H24 \\
118 &  9 55 35.29 &  69 03 15.9 & $  56.97\pm  3.37$ & 13.90 & $>$HST     & s3 &     b    & $   7.40\pm  0.50$ & H23 \\
119 &  9 55 35.40 &  69 05 57.7 & $   4.55\pm  0.95$ &  3.64 & $ $  ---   & s3 &     b    & $   0.62\pm  0.13$ &  \\
120 &  9 55 35.56 &  69 03 54.3 & $  46.87\pm  3.05$ & 10.55 & $>$HST     & s3 &     b    & $   3.60\pm  0.60$ &  \\
121 &  9 55 35.72 &  69 06 37.7 & $   7.65\pm  1.23$ &  3.86 & $>$20.66   & s3 &     d    & $   1.04\pm  0.17$ &  \\
122 &  9 55 36.29 &  69 02 44.8 & $  13.49\pm  1.64$ &  6.76 & $>$HST     & s3 &     b    & $   1.83\pm  0.22$ &  \\
123 &  9 55 36.45 &  69 02 40.5 & $   7.69\pm  1.24$ &  4.67 & $>$HST     & s3 &     b    & $   1.05\pm  0.17$ &  \\
124 &  9 55 36.75 &  69 06 33.2 & $  13.82\pm  1.66$ &  6.90 & $>$19.83   & s3 &     d    & $   1.88\pm  0.23$ & HII, SNR \\
125 &  9 55 36.87 &  68 56 56.2 & $   3.03\pm  0.78$ &  3.40 & $ $  ---   & s2 &     d    & $   0.55\pm  0.14$ &  \\
126 &  9 55 37.05 &  69 04 33.3 & $  28.54\pm  2.38$ & 10.34 & $>$HST     & s3 &     b    & $   6.40\pm  1.00$ &  \\
127 &  9 55 37.28 &  69 02 07.3 & $   3.50\pm  0.83$ &  3.11 & $ $\hspace{5pt}18.41 & s3 &     b    & $   0.48\pm  0.11$ &  GC? \\
128 &  9 55 37.60 &  69 04 57.7 & $  10.60\pm  1.45$ &  6.08 & $ $\hspace{5pt}23.10 & s3 &     b    & $   1.44\pm  0.20$ &  \\
129 &  9 55 37.66 &  69 03 16.2 & $   9.06\pm  1.34$ &  5.43 & $>$HST     & s3 &     b    & $   1.23\pm  0.18$ &  SSS \\
130 &  9 55 38.62 &  68 49 22.9 & $  10.98\pm  1.48$ &  4.09 & $ $  ---   & s1 & $D_{25}$ & $   1.49\pm  0.20$ &  \\
131 &  9 55 40.69 &  69 01 05.0 & $   3.20\pm  0.80$ &  3.21 & $>$HST     & s3 &     b    & $   0.43\pm  0.11$ &  \\
132 &  9 55 42.21 &  69 03 36.3 & $ 800.98\pm 12.62$ & 53.88 & $ $\hspace{5pt}21.67 & s3 &     b    & $  30.00\pm  1.00$ & SSS, H25 \\
133 &  9 55 42.86 &  69 03 07.6 & $  21.84\pm  2.08$ &  8.82 & $ $  ---   & s3 &     b    & $   3.60\pm  1.70$ &  \\
134 &  9 55 43.17 &  69 04 45.0 & $   5.91\pm  1.08$ &  4.21 & $>$HST     & s3 &     b    & $   0.80\pm  0.15$ &  \\
135 &  9 55 43.34 &  69 04 23.2 & $   2.92\pm  0.76$ &  3.33 & $>$HST     & s3 &     b    & $   0.40\pm  0.10$ &  \\
136 &  9 55 43.76 &  68 59 04.8 & $  38.18\pm  2.75$ & 11.44 & $ $  ---   & s3 &     d    & $   4.00\pm  0.40$ & H27 \\
137 &  9 55 44.63 &  69 10 05.2 & $  14.50\pm  1.70$ &  6.80 & $>$20.77   & s4 &     d    & $   2.65\pm  0.31$ &  \\
138 &  9 55 44.71 &  69 05 34.5 & $  12.12\pm  1.55$ &  6.17 & $ $\hspace{5pt}22.77 & s3 &     d    & $   1.65\pm  0.21$ &  \\
139 &  9 55 45.91 &  69 03 00.4 & $  14.70\pm  1.71$ &  7.13 & $>$17.60   & s3 &     b    & $   2.00\pm  0.23$ &  \\
140 &  9 55 46.16 &  68 53 40.7 & $   3.87\pm  0.88$ &  3.27 & $ $  ---   & s2 &     d    & $   0.71\pm  0.16$ &  \\
141 &  9 55 47.05 &  69 05 51.1 & $  66.39\pm  3.63$ & 15.28 & $ $\hspace{5pt}18.63 & s3 &     d    & $  11.80\pm  1.30$ & H28, HII, GC \\
142 &  9 55 47.96 &  68 59 28.2 & $   6.11\pm  1.10$ &  4.52 & $ $  ---   & s3 &     d    & $   0.83\pm  0.15$ & SSS, HII \\
143 &  9 55 48.19 &  68 59 15.1 & $   4.22\pm  0.92$ &  3.88 & $>$21.00   & s3 &     d    & $   0.57\pm  0.12$ & SSS \\
144 &  9 55 49.41 &  68 58 36.3 & $  52.30\pm  3.22$ & 13.31 & $ $  ---   & s3 &     d    & $  14.50\pm  1.50$ & H29, P40, SNR \\
145 &  9 55 49.52 &  69 08 12.0 & $  78.12\pm  3.94$ & 16.74 & $ $  ---   & s4 &     d    & $  14.80\pm  1.60$ & H30, P42 \\
146 &  9 55 49.87 &  69 05 32.0 & $ 428.91\pm  9.23$ & 39.54 & $ $\hspace{5pt}20.77 & s3 &     d    & $  59.80\pm  1.60$ & X7, H31, P41, GC \\
147 &  9 55 51.54 &  69 04 10.5 & $   2.61\pm  0.72$ &  2.86 & $ $  ---   & s3 &     b    & $   0.36\pm  0.10$ &  \\
148 &  9 55 51.58 &  69 07 43.1 & $   3.27\pm  0.81$ &  3.26 & $ $\hspace{5pt}18.81 & s4 &     d    & $   0.60\pm  0.15$ & HII, GC \\
149 &  9 55 53.13 &  69 05 20.1 & $  35.05\pm  2.64$ & 11.05 & $ $\hspace{5pt}22.20 & s3 &     d    & $   1.00\pm  0.00$ & SSS \\
150 &  9 55 53.31 &  69 02 06.5 & $  14.22\pm  1.68$ &  7.02 & $ $  ---   & s3 &     b    & $   1.93\pm  0.23$ &  \\
\hline 
\end{tabular}
} 
\end{center}

\clearpage
\begin{center}
\small{
\begin{tabular}{rrrrrrccrl}
\multicolumn{10}{c}{{\sc Table} 2 --- Continued} \\
\multicolumn{10}{c}{M81 Discrete X-Ray Sources} \\
\hline \hline
 &  \multicolumn{1}{c}{R.A.} & \multicolumn{1}{c}{Dec.} & Count Rate & S/N & $m_V$ & CCD & region & \multicolumn{1}{c}{$L_X$} & Comment \\
 &  \multicolumn{1}{c}{(J2000)} & \multicolumn{1}{c}{(J2000)} & ($10^{-4}$~s$^{-1}$) & & & & & \multicolumn{1}{c}{($10^{37}$~erg~s$^{-1}$)} & \\ 
\hline
151 &  9 55 53.68 &  69 04 34.8 & $   7.00\pm  1.18$ &  4.60 & $ $\hspace{5pt}24.28 & s3 &     d    & $   0.95\pm  0.16$ &  \\
152 &  9 55 55.01 &  69 02 38.9 & $   4.30\pm  0.93$ &  3.77 & $ $  ---   & s3 &     b    & $   0.59\pm  0.13$ &  \\
153 &  9 55 55.37 &  68 58 58.5 & $   2.49\pm  0.70$ &  2.86 & $ $  ---   & s3 &     d    & $   0.34\pm  0.10$ &  \\
154 &  9 55 55.79 &  69 10 08.5 & $   9.00\pm  1.34$ &  5.31 & $ $  ---   & s4 &     d    & $   1.65\pm  0.24$ &  \\
155 &  9 55 56.11 &  69 03 12.2 & $   6.46\pm  1.13$ &  4.87 & $ $  ---   & s3 &     b    & $   0.88\pm  0.15$ & SSS, Radio \\
156 &  9 55 56.21 &  69 05 14.7 & $   9.16\pm  1.35$ &  5.56 & $ $\hspace{5pt}23.56 & s3 &     d    & $   1.25\pm  0.18$ &  \\
157 &  9 55 56.74 &  69 08 02.6 & $   5.84\pm  1.08$ &  4.25 & $ $20.5(I) & s4 &     d    & $   1.07\pm  0.20$ & Extended? \\
158 &  9 55 58.61 &  69 05 26.2 & $  64.07\pm  3.57$ & 15.11 & $ $\hspace{5pt}26.37 & s3 &     d    & $  16.00\pm  1.70$ & H32, GC \\
159 &  9 55 59.15 &  69 06 17.4 & $  20.14\pm  2.00$ &  8.19 & $ $\hspace{5pt}25.28 & s3 &     d    & $   2.10\pm  0.50$ &  \\
160 &  9 56 01.97 &  68 58 59.3 & $  44.41\pm  2.97$ & 12.81 & $ $  ---   & s3 &     d    & $   2.00\pm  0.30$ & H33, P43 \\
161 &  9 56 02.69 &  68 59 35.2 & $  40.46\pm  2.84$ & 12.24 & $ $  ---   & s3 &     d    & $   3.40\pm  1.10$ & H34 \\
162 &  9 56 02.78 &  68 58 44.0 & $  12.85\pm  1.60$ &  6.88 & $ $  ---   & s3 &     d    & $   1.75\pm  0.22$ &  \\
163 &  9 56 03.15 &  69 02 16.8 & $   4.81\pm  0.98$ &  4.06 & $ $  ---   & s3 &     d    & $   0.65\pm  0.13$ &  \\
164 &  9 56 03.29 &  69 01 07.3 & $   3.77\pm  0.87$ &  3.64 & $ $  ---   & s3 &     d    & $   0.51\pm  0.12$ &  \\
165 &  9 56 04.36 &  69 11 59.7 & $   5.81\pm  1.07$ &  3.43 & $ $  ---   & s4 &     d    & $   1.06\pm  0.20$ &  \\
166 &  9 56 04.69 &  68 58 39.2 & $   2.90\pm  0.76$ &  3.24 & $ $  ---   & s3 &     d    & $   0.39\pm  0.10$ &  \\
167 &  9 56 04.93 &  69 03 43.7 & $   2.61\pm  0.72$ &  2.90 & $>$19.09   & s3 &     d    & $   0.36\pm  0.10$ & HII \\
168 &  9 56 06.07 &  68 59 40.7 & $   5.66\pm  1.06$ &  4.54 & $ $  ---   & s3 &     d    & $   0.77\pm  0.14$ & HII \\
169 &  9 56 06.09 &  69 08 33.5 & $   6.72\pm  1.16$ &  4.67 & $ $  ---   & s4 &     d    & $   1.23\pm  0.21$ &  \\
170 &  9 56 07.84 &  69 03 25.2 & $  15.10\pm  1.73$ &  7.08 & $ $  ---   & s3 &     d    & $   2.05\pm  0.24$ &  \\
171 &  9 56 09.05 &  69 01 06.4 & $ 110.56\pm  4.69$ & 20.30 & $ $  ---   & s3 &     d    & $   4.00\pm  0.50$ & SSS, H36, P44 \\
172 &  9 56 09.48 &  69 12 49.5 & $  22.46\pm  2.11$ &  8.32 & $ $  ---   & s4 & $D_{25}$ & $   5.10\pm  0.80$ & H35, P45 \\
173 &  9 56 13.74 &  69 06 30.6 & $  21.10\pm  2.05$ &  8.07 & $ $\hspace{5pt}24.75 & s3 &     d    & $   2.87\pm  0.28$ & H37, P46 \\
174 &  9 56 14.21 &  69 02 24.3 & $   4.02\pm  0.89$ &  3.58 & $>$19.05   & s3 &     d    & $   0.55\pm  0.12$ & SSS, HII \\
175 &  9 56 14.42 &  69 02 47.9 & $  13.66\pm  1.65$ &  6.81 & $ $  ---   & s3 &     d    & $   1.86\pm  0.22$ &  \\
176 &  9 56 14.85 &  69 03 37.7 & $   3.48\pm  0.83$ &  3.36 & $ $  ---   & s3 &     d    & $   0.47\pm  0.11$ & HII \\
177 &  9 56 27.46 &  69 10 12.1 & $  22.57\pm  2.12$ &  8.46 & $ $  ---   & s4 & $D_{25}$ & $   3.10\pm  0.50$ &  \\
\hline 
\end{tabular}
} 
\end{center}

\clearpage


\clearpage

\begin{center}
\small{
\begin{tabular}{rrrrrr}
\multicolumn{6}{c}{{\sc TABLE 3}} \\
\multicolumn{6}{c}{M81 Bright X-Ray Sources} \\
\hline \hline
 & \multicolumn{1}{c}{$N_H$} &\multicolumn{1}{c}{Spectral} &\multicolumn{1}{c}{$\chi^2$/dof} &\multicolumn{1}{c}{$L_X$} &\multicolumn{1}{c}{$P_{KS}$} \\
 & ($10^{20}$~cm$^{-2}$) &\multicolumn{1}{c}{Parameter$^a$} & & \multicolumn{1}{c}{($10^{37}$~erg~s$^{-1}$)} & \\ 
\hline
 37 & $  27.2\pm 19.0$ & $   1.63\pm  0.60$ & $   2.03/  5$ & $   6.3\pm  1.1$ &   74.50 \\
 51 & $  14.9\pm 10.0$ & $   1.76\pm  0.42$ & $   5.36/  8$ & $   5.0\pm  0.6$ &   81.80 \\
 52 & $  13.6\pm  4.0$ & $   1.59\pm  0.14$ & $  38.87/ 38$ & $  27.0\pm  0.6$ &   65.10 \\
 57 & $  99.1\pm 40.0$ & $   1.90\pm  0.51$ & $   8.00/ 13$ & $  22.0\pm  2.8$ &   12.10 \\
 67 & $   4.0\pm  0.0$ & $   1.31\pm  0.28$ & $   4.42/  4$ & $   3.5\pm  0.7$ &    1.89 \\
 69 & $  11.8\pm  3.0$ & $   1.47\pm  0.11$ & $  58.11/ 43$ & $  26.7\pm  1.0$ &   58.50 \\
 81 & $  13.5\pm  7.0$ & $   2.32\pm  0.35$ & $  14.01/  7$ & $   4.8\pm  0.5$ &   45.90 \\
 82 & $  10.0\pm 10.0$ & $   1.43\pm  0.55$ & $   3.65/  8$ & $   5.1\pm  1.1$ &   53.40 \\
 83 & $  14.4\pm  4.0$ & $   1.79\pm  0.13$ & $  60.26/ 35$ & $  22.0\pm  0.9$ &   92.60 \\
 86 & $   4.0\pm  0.0$ & $   1.08\pm  0.13$ & $  42.43/ 35$ & $  35.0\pm  3.0$ &    4.03 \\
 93 & $  15.6\pm  7.0$ & $   2.03\pm  0.45$ & $   1.98/  4$ & $   3.7\pm  0.5$ &   59.80 \\
 95 & $   4.0\pm  0.0$ & $   1.61\pm  0.17$ & $  14.45/ 11$ & $   5.7\pm  0.6$ &   17.60 \\
102 & $   4.0\pm  0.0$ & $   1.80\pm  0.24$ & $  12.14/ 11$ & $   6.8\pm  0.6$ &   82.00 \\
109 & $   9.7\pm  5.0$ & $   2.20\pm  0.33$ & $   8.81/  7$ & $   4.2\pm  0.4$ &   70.30 \\
112 & $   4.0\pm  0.0$ & $   1.17\pm  0.16$ & $  13.81/  8$ & $   5.8\pm  0.6$ &   70.30 \\
113 & $  14.2\pm 10.0$ & $   2.24\pm  0.50$ & $   6.54/  8$ & $   4.3\pm  0.5$ &    9.76 \\
115 & $   7.7\pm  7.0$ & $   1.26\pm  0.85$ & $   2.22/  4$ & $   3.1\pm  0.5$ &   25.40 \\
116 & $   8.7\pm  3.0$ & $   1.44\pm  0.12$ & $  32.71/ 30$ & $  19.5\pm  1.0$ &   68.70 \\
117 & $  16.0\pm 13.0$ & $   1.92\pm  0.43$ & $   6.49/  6$ & $   6.0\pm  1.0$ &   58.20 \\
118 & $   5.2\pm  3.0$ & $   1.53\pm  0.25$ & $  13.28/ 12$ & $   7.4\pm  0.5$ &   99.10 \\
120 & $  11.5\pm  5.0$ & $   2.57\pm  0.50$ & $   8.00/  5$ & $   3.6\pm  0.6$ &   98.40 \\
126 & $   4.0\pm  0.0$ & $   1.71\pm  0.15$ & $  30.69/ 14$ & $   6.4\pm  1.0$ &   56.30 \\
133 & $  24.5\pm 20.0$ & $   2.15\pm  0.80$ & $   1.16/  6$ & $   3.6\pm  1.7$ &   21.00 \\
136 & $  10.3\pm  5.0$ & $   1.95\pm  0.47$ & $   7.68/  8$ & $   4.0\pm  0.4$ &   91.00 \\
141 & $  12.3\pm  7.0$ & $   1.22\pm  0.20$ & $   8.84/ 15$ & $  11.8\pm  1.3$ &   22.90 \\
144 & $  13.6\pm  5.0$ & $   2.61\pm  0.39$ & $  12.15/  8$ & $  14.5\pm  1.5$ &    0.61 \\
145 & $   4.0\pm  0.0$ & $   1.41\pm  0.18$ & $  18.82/ 22$ & $  14.8\pm  1.6$ &   72.40 \\
146 & $  11.5\pm  2.0$ & $   1.37\pm  0.08$ & $  72.49/ 81$ & $  59.8\pm  1.6$ &    5.31 \\
158 & $  28.7\pm 11.0$ & $   3.61\pm  0.96$ & $  19.33/ 13$ & $  16.0\pm  1.7$ &   35.50 \\
159 & $  17.4\pm 17.0$ & $   2.13\pm  1.10$ & $   6.88/  3$ & $   2.1\pm  0.5$ &   38.10 \\
160 & $   4.0\pm  0.0$ & $   0.20\pm  0.02$ & $  11.89/  7$ & $   2.0\pm  0.3$ &   37.70 \\
161 & $  38.3\pm 15.0$ & $   0.57\pm  0.14$ & $  10.64/  9$ & $   3.4\pm  1.1$ &   71.70 \\
172 & $  10.0\pm  6.3$ & $   1.21\pm  0.37$ & $   4.37/  7$ & $   5.1\pm  0.8$ &    2.58 \\
\hline 
\multicolumn{6}{l}{$^a$Power law photon index, $\Gamma$, except for source numbers 160 and 161} \\
\multicolumn{6}{l}{ where spectral parameter is blackbody temperature, $kT$, in keV.}
\end{tabular}
} 
\end{center}

\clearpage


\end{document}